\documentclass[12pt]{article}
\usepackage{a4wide,amssymb,cite}
\pdfoutput=1

\usepackage{a4wide,amssymb,graphicx}
\usepackage{epsfig}
\usepackage[usenames,dvipsnames]{color}
\usepackage{slashed}
\usepackage[compat=1.1.0]{tikz-feynman}

\newcommand{\be}{\begin{equation}}
\newcommand{\ee}{\end{equation}}
\newcommand{\bea}{\begin{eqnarray}}
\newcommand{\eea}{\end{eqnarray}}

\def\circa#1{\,\raise.3ex\hbox{$#1$\kern-.75em\lower1ex\hbox{$\sim$}}\,}

\begin{document}

\begin{titlepage}
%
%

%\rightline{CERN-PH-TH/2018-xxx}

%

\begin{centering}
\vspace{1cm}
{\Large {\bf Lepto-Quark Portal Dark Matter}} \\

\vspace{1.5cm}

{\bf Soo-Min Choi$^{1\sharp}$, Yoo-Jin Kang$^{1*}$, Hyun Min Lee$^{1,2\dagger}$ and Tae-Gyu Ro$^{1\ddagger}$}
%\\
\vspace{.5cm}

{\it $^1$Department of Physics, Chung-Ang University, Seoul 06974, Korea.} \vspace{.2cm}
\\
{\it $^2$School of Physics, Korea Institute for Advanced Study, Seoul 02455, Korea.}

\end{centering}
\vspace{2cm}

\begin{abstract}
\noindent
We consider the extension of the Standard Model with scalar leptoquarks as a portal to dark matter (DM), motivated by the recent anomalies in semi-leptonic $B$-meson decays. Taking singlet and triplet scalar leptoquarks as the best scenarios for explaining $B$-meson anomalies, we discuss the phenomenological constraints from rare meson decays, muon $(g-2)_\mu$, and leptoquark searches at the Large Hadron Collider (LHC).  Introducing leptoquark couplings to scalar dark matter, we find that the DM annihilations into a pair of leptoquarks open a wide parameter space, being compatible with XENON1T bound, and show that there is an interesting interplay between  LHC leptoquark searches and distinct signatures from cascade annihilations of dark matter.

\end{abstract}

\vspace{3cm}
  \begin{flushleft}
    $^\sharp$Email: soominchoi90@gmail.com \\    
    $^*$Email: yoojinkang91@gmail.com \\
     $^\dagger$Email: hminlee@cau.ac.kr   \\
     $^\ddagger$Email: shxorb234@gmail.com
  \end{flushleft}

\end{titlepage}

\section{Introduction}

% B-meson anomalies
Recently, there have been intriguing anomalies in the semi-leptonic decays of $B$-mesons at BaBar, Belle and LHCb experiments, which are based on the observables of testing Lepton Flavor Universality(LFU), i.e. $R_{K^{(*)}}$ \cite{RK,RKs,P5} and $R_{D^{(*)}}$ \cite{babar,belle,lhcb}. Thus, it is plausible that LFU might be violated due to new physics in the neutral and charged currents associated with muon and tau leptons, respectively. 
Currently, experimental values of $R_{K^{(*)}}$ and $R_{D^{(*)}}$ turn out to be deviated from  the SM expectations at about $4\sigma$ level per each. However, we still need to understand the hadronic uncertainties in angular distributions of related $B$-meson decays \cite{qcd} and the results are to be confirmed at LHCb with more data and Belle II \cite{more}. 
Nonetheless, it is important to study the consequences of new physics in direct searches at the LHC and other precision and indirect observables.

%Dark matter 
Dark matter (DM) is known to occupy about $85\%$ of the total matter density in the Universe, and there are a variety of evidences for the existence of dark matter such as galaxy rotation curves, gravitational lensing, large scale structure, etc. The Weakly Interacting Massive Particles (WIMPs) paradigm has driven forces for searching particle dark matter with non-gravitational interactions beyond the Standard Model (SM) for more than three decades. Various direct detection experiments \cite{xenon1t,panda,cdms,lux} have put stringent bounds on the cross section of DM-nucleon elastic scattering, and forthcoming XENON-nT  and large-scale experiments such as DARWIN \cite{darwin} and LZ \cite{LZ} will push the limits  further to the neutrino floor where there are irreducible backgrounds due to neutrino coherent scattering. 
In particular, Higgs-portal type models for dark matter have been strongly constrained, apart from the resonance region or the heavy DM masses.

%LQ
Leptoquark models \cite{LQs,LQS1} have been revived recently because they can provide an economic way of accommodating the aforementioned $B$-meson anomalies \cite{leptoquarks,RK-models,LQ-fit,LQ-early,nomura,kenji,watanabe} and can be tested at the LHC. 
Leptoquarks carry extra Yukawa-type couplings to the SM fermions, providing a source for violating LFU. 
Furthermore, leptoquark scalars or vectors could be originated from unified models of forces \cite{unification}, in analogy to colored triplet Higgs scalars or $X,Y$ gauge bosons in the minimal $SU(5)$ unification. 
The best scenarios for explaining the $B$-meson anomalies \cite{RK-models,LQ-fit} are: one $SU(2)_L$-singlet scalar leptoquark $S_1$  for $R_{D^{(*)}}$, and one $SU(2)_L$-triplet scalar leptoquark $S_3$ for $R_{K^{(*)}}$, or one $SU(2)_L$-singlet vector leptoquark for both $B$-meson anomalies. Leptoquark scenarios are phenomenologically rich, because the muon $(g-2)_\mu$ anomalies can be also explained by leptoquark couplings and various LHC searches can be reinterpreted to bound the leptoquark models.

%DM
In this article, we consider a leptoquark-portal model for dark matter where scalar dark matter communicates with the SM through the quartic couplings of scalar leptoquarks, $S_1$ and $S_3$. 
We show that sizable leptoquark couplings to dark matter lead to new annihilation channels of dark matter into a pair of leptoquarks, opening a wide parameter space where the correct relic density can be explained, being compatible with the direct detection bounds from XENON1T. 
Moreover, we also discuss that the cascade annihilations of dark matter can lead to distinct signatures for cosmic ray observation, in correlation to leptoquark searches at the LHC. 
We argue that our models with scalar leptoquarks are consistent with the current bounds from rare meson decays, mixings and lepton flavor violation, whereas the loop corrections of leptoquarks to DM-nucleon couplings and Higgs couplings can be negligible in most of the parameter space of our interest.

%Outline 
 The paper is organized as follows. 
 We first give a brief overview on the $R_{K^{(*)}}$ and $R_{D^{(*)}}$ anomalies and the necessary corrections to the effective Hamiltonians. 
 Then, in models with scalar leptoquarks, we derive the effective interactions for the semi-leptonic $B$-meson decays and discuss the conditions for $B$-meson anomalies and various constraints from rare meson decays, mixings, muon $(g-2)_\mu$ and leptoquark searches at the LHC. 
 Next we describe leptoquark-portal models for dark matter and consider various constraints on the models, coming from the relic density, direct and indirect detection of dark matter and Higgs data.
There are two appendices dealing with the details on effective Hamiltonians for $B$-meson decays and effective interactions for dark matter and Higgs due to leptoquarks, respectively. 
Finally, conclusions are drawn.

\section{Overview on $R_{K^{(*)}}$ and $R_{D^{(*)}}$ anomalies}

In this section, we give a brief overview on the status of the $B$-meson anomalies and the interpretations in terms of the effective Hamiltonians in the SM.

The reported value of $R_K={\cal B}(B\rightarrow K\mu^+\mu^-)/{\cal
  B}(B\rightarrow Ke^+e^-)$ ~\cite{RK} is
\begin{equation}
  R_K=0.745^{+0.090}_{-0.074}({\rm stat})\pm 0.036({\rm syst}), \quad 1\,{\rm GeV}^2<q^2 <6\,{\rm GeV}^2,
\end{equation}
which deviates from the SM prediction by $2.6\sigma$.
On the other hand for vector $B$-mesons, $R_{K^*}={\cal
  B}(B\rightarrow K^*\mu^+\mu^-)/{\cal B}(B\rightarrow
K^*e^+e^-)$~\cite{RKs} is
\bea
  R_{K^*}&=& 0.66^{+0.11}_{-0.07}({\rm stat})\pm 0.03({\rm syst}), \quad 0.045\,{\rm GeV}^2<q^2 <1.1\,{\rm GeV}^2,\nonumber\\
  R_{K^*}&=& 0.69^{+0.11}_{-0.07}({\rm stat})\pm 0.05({\rm syst}), \quad 1.1\,{\rm GeV}^2<q^2 <6.0\,{\rm GeV}^2,
\eea
which again differs from the SM prediction by 2.1--$2.3\sigma$ and
2.4--$2.5\sigma$, depending on the energy bins. The deviation in $R_{K^*}$ is supported by the reduction in the angular distribution of $B\rightarrow K^*\mu^+\mu^-$, the so called $P'_5$ variable~\cite{P5}.

The effective Hamiltonian for $b\rightarrow s  \mu^+ \mu^-$ is given by
\bea
\Delta {\cal H}_{{\rm eff},{\bar b}\rightarrow {\bar s}\mu^+ \mu^-} = -\frac{4G_F}{\sqrt{2}}  \,V^*_{ts} V_{tb}\,\frac{\alpha_{em}}{4\pi}\, (C^{\mu}_9 {\cal O}^\mu_9+C^{\mu}_{10} {\cal O}^\mu_{10}+C^{\prime\mu}_9{\cal O}^{\prime\mu}_9+C^{\prime\mu}_{10}{\cal O}^{\prime\mu}_{10} )+{\rm h.c.} \label{RKeff}
\eea
where $ {\cal O}^\mu_9 \equiv ({\bar s}\gamma^\mu P_L b) ({\bar \mu}\gamma_\mu \mu)$, $ {\cal O}^\mu_{10} \equiv ({\bar s}\gamma^\mu P_L b) ({\bar \mu}\gamma_\mu\gamma^5 \mu)$, $ {\cal O}^{\prime\mu}_9 \equiv ({\bar s}\gamma^\mu P_R b) ({\bar \mu}\gamma_\mu \mu)$ and $ {\cal O}^{\prime\mu}_{10} \equiv ({\bar s}\gamma^\mu P_R b) ({\bar \mu}\gamma_\mu\gamma^5 \mu)$, and $\alpha_{\rm em}$ is the electromagnetic coupling. In the SM, the Wilson coefficients are given by $C^{\mu,\rm SM}_9(m_b)=-C^{\mu,\rm SM}_{10}(m_b)=4.27$ and $C^{\prime\mu, {\rm SM}}_9(m_b)\approx-C^{\prime\mu,{\rm SM}}_{10}(m_b)\approx 0$. 

For $C^{\mu,\rm NP}_{10}=C^{\prime\mu, {\rm NP}}_9=C^{\prime\mu,{\rm NP}}_{10}=0$, the best-fit value for new physics contribution is given by $C^{\mu, {\rm NP}}_9=-1.11$ \cite{crivellin}, (while taking $[-1.28,-0.94]$ and $[-1.45,-0.75]$ within $1\sigma$ and $2\sigma$ errors), to explain the $R_{K^{(*)}}$ anomalies.
On the other hand, for $C^{\mu, {\rm NP}}_9=-C^{\mu,\rm NP}_{10}$ and others being zero, the best-fit value for new physics contribution is given by $C^{\mu, {\rm NP}}_9=-0.62$ \cite{crivellin}, (while taking $[-0.75,-0.49]$ and $[-0.88,-0.37]$ within $1\sigma$ and $2\sigma$ errors).

Taking the results of BaBar \cite{babar}, Belle \cite{belle} and LHCb \cite{lhcb} for $R_D={\cal B}(B\rightarrow D\tau\nu)/{\cal  B}(B\rightarrow Dl\nu)$ and $R_{D^*}={\cal B}(B\rightarrow D^*\tau\nu)/{\cal  B}(B\rightarrow D^*l\nu)$ with $l=e,\mu$ for BaBar and Belle and $l=\mu$ for LHCb, the Heavy Flavor Averaging Group \cite{hflav} reported the experimental world averages as follows,
\bea
R^{\rm exp}_D &=& 0.403\pm 0.040\pm 0.024, \\
R^{\rm exp}_{D^*}&=& 0.310\pm 0.015\pm 0.008.
\eea
On the other hand, taking into account the lattice calculation of $R_D$, which is $R_D=0.299\pm 0.011$ \cite{RD-lattice}, and the uncertainties in $R_{D^*}$ in various groups \cite{RD-SM,RD-SMetc}, we take the SM predictions for these ratios as follows,
\bea
R^{\rm SM}_D &=& 0.299\pm 0.011, \\
R^{\rm SM}_{D^*}&=& 0.260\pm 0.010.
\eea
Then, the combined derivation between the measurements and the SM predictions for $R_D$ and $R_{D^*}$ is about $4.1\sigma$. 
We quote the best fit values for $R_D$ and $R_{D^*}$ including the new physics contributions \cite{RD-bestfit},
\bea
\frac{R_D}{R^{\rm SM}_D}=\frac{R_{D^*}}{R^{\rm SM}_{D^*}}=1.21\pm 0.06. \label{RDratio}
\eea 

The effective Hamiltonian for $b\rightarrow c  \tau \nu$ in the SM is given by
\be
{\cal H}_{\rm eff}=\frac{4G_F}{\sqrt{2}} V_{cb} C_{cb}\, ({\bar c}\gamma^\mu P_L b) ({\bar\tau}\gamma_\mu P_L \nu_\tau) +{\rm h.c.} \label{RDeff}
\ee
where  $C_{cb}=1$ in the SM with $V_{cb}\approx 0.04$.  The new physics contribution may contain the dimension-6 four-fermion vector operators, ${\cal O}_{V_{R,L}}=({\bar c}\gamma^\mu P_{R,L} b) ({\bar\tau}\gamma_\mu P_L \nu_\tau) $ and/or scalar operators, ${\cal O}_{S_{R,L}}=({\bar c} P_{R,L} b) ({\bar\tau} P_L \nu_\tau)$. Then, in order to explain the $R_{D^{(*)}}$ anomalies in eq.~(\ref{RDratio}), the Wilson coefficient for the new physics contribution should be  $\Delta C_{cb}=0.1$ from eq.~(\ref{RDeff}), while taking $[0.072,0.127]$ and $[0.044,0.153]$ within $1\sigma$ and $2\sigma$ errors.

\section{Leptoquarks for $B$-meson anomalies}

It is known that $SU(2)_L$ singlet and triplet scalar leptoquarks can explain $R_{D^{(*)}}$ and $R_{K^{(*)}}$ anomalies, respectively \cite{RK-models,LQ-fit}. (See also Ref.~\cite{leptoquarks,LQ-early,nomura,watanabe}.)
Thus, in this section, focusing on those scalar leptoquark models, we discuss the phenomenological constraints coming from the $B$-meson anomalies.

\subsection{Effective interactions from scalar leptoquarks}

We consider the Lagrangian for an $SU(2)_L$ singlet scalar leptoquark $S_1$ with $Y=+\frac{1}{3}$, and an $SU(2)_L$ triplet scalar leptoquark, $S_3\equiv \Phi_{ab}$ with $Y=+\frac{1}{3}$, as follows,
\bea
{\cal L}_{LQ}= {\cal L}_{S_1}+{\cal L}_{S_3}
\eea
\bea
{\cal L}_{S_1} &=&-\lambda_{ij} Q^a_{Li} (i\sigma^2)_{ab} \,S_1 L^b_{Lj} +{\rm h.c.}\nonumber \\
&=&-\lambda_{ij} \overline{(Q^C)^a_{Ri}}\, (i\sigma^2)_{ab} \,S_1 \, L^b_{Lj} +{\rm h.c.} \label{sLQ}
\eea
where $a,b$ are $SU(2)_L$ indices, $\sigma^2$ is the second Pauli matrix and $\psi^C=C{\bar\psi}^T$ is the charge conjugate with $C=i\gamma^0\gamma^2$,
and
\bea
{\cal L}_{S_3} &=&-\kappa_{ij} Q^a_{Li} \Phi_{ab} L^b_{Lj} +{\rm h.c.}\nonumber \\
&=&-\kappa_{ij} \overline{(Q^C)^a_{Ri}}\, \Phi_{ab}\, L^b_{Lj} +{\rm h.c.} \label{tLQ}
\eea
with
\be
\Phi_{ab}=\left( \begin{array}{cc} \sqrt{2}\phi_3 & -\phi_2 \\  -\phi_2  & -\sqrt{2} \phi_1 \end{array} \right)
\ee
where $(\phi_1,\phi_2,\phi_3)$ forms an isospin triplet with $T_3=+1,0,-1$ and $Q=+\frac{4}{3}, +\frac{1}{3}, -\frac{2}{3}$.  We note that our conventions are comparable to those in the literature by writing $\Phi=(i\sigma^2) ({\vec \sigma}\cdot {\vec S})$ where $\vec\sigma$ are Pauli matrices and $\vec S$ are complex scalar fields. 

Then, after integrating out the leptoquark scalars, we obtain the effective Lagrangian for the SM fermions in the following,
\bea
{\cal L}_{\rm eff} &=& \Bigg(\frac{1}{4m^2_{S_1}}\, \lambda_{ij}\lambda^*_{kl}+\frac{3}{4m^2_{S_3}}\, \kappa_{ij} \kappa^*_{kl} \Bigg) \Big({\bar Q}_{Lk}\gamma^\mu Q_{Li}\Big)\Big({\bar L}_{Ll}\gamma_\mu L_{Lj}\Big) \nonumber \\
&&+ \Bigg(-\frac{1}{4m^2_{S_1}}\, \lambda_{ij}\lambda^*_{kl}+\frac{1}{4m^2_{S_3}}\, \kappa_{ij} \kappa^*_{kl} \Bigg) \Big({\bar Q}_{Lk}\gamma^\mu\sigma^I Q_{Li}\Big)\Big({\bar L}_{Ll}\gamma_\mu \sigma^I L_{Lj}\Big)
\eea
where  $\sigma^I(I=1,2,3)$ are the Pauli matrices.
There, we find that there are both $SU(2)_L$ singlet and triplet $V-A$ operators.
As compared to the case with $U(2)$ flavor symmetry \cite{LQ-fit}, the effective interactions for either singlet or triplet leptoquark can be written as
\bea
{\cal L}_{\rm eff} &=& -\frac{1}{v^2}\, \lambda^q_{ki} \lambda^l_{lj} \Bigg[ C_S\Big({\bar Q}_{Lk}\gamma^\mu Q_{Li}\Big)\Big({\bar L}_{Ll}\gamma_\mu L_{Lj}\Big)+ C_T  \Big({\bar Q}_{Lk}\gamma^\mu\sigma^I Q_{Li}\Big)\Big({\bar L}_{Ll}\gamma_\mu \sigma^I L_{Lj}\Big)   \Bigg].
\eea
So, we obtain $C_S=-C_T$ for the singlet leptoquark and $C_S=3C_T$ for the triplet leptoquark. 
A fit to low-energy data including the $R_{K^{(*)}}$ and $R_{D^{(*)}}$ anomalies has been done with four free parameters, $C_T, C_S, \lambda^q_{sb}$ and $\lambda^l_{\mu\mu}$, under the assumption that the CKM matrix stems solely from the mixing between up-type quarks  \cite{LQ-fit}. As a result, the best-fit values are given by $C_S\approx C_T\approx 0.02$ for $|\lambda^q_{sb}|<5|V_{cb}|$ \cite{LQ-fit}.

\subsection{Singlet scalar  leptoquark}

After integrating out the leptoquark $S_1$, from the results in eq.~(\ref{S1-app}), we obtain the effective Hamiltonian relevant for $b\rightarrow c\tau{\bar\nu}_\tau$ as
\bea
{\cal H}^{S_1}_{b\rightarrow c\tau{\bar\nu}_\tau}
=-\frac{\lambda^*_{33}\lambda_{23}}{2 m^2_{S_1}}\, ({\bar b}_L \gamma^\mu c_L) ({\bar\nu}_{\tau L}\gamma_\mu \tau_L) +{\rm h.c.}\equiv  \frac{1}{\Lambda^2_D} \, ({\bar b}_L \gamma^\mu c_L) ({\bar\nu}_{\tau L}\gamma_\mu \tau_L) +{\rm h.c.}.
\eea
As a consequence, the singlet leptoquark gives rise to the effective operator for explaining the $R_{D^{(*)}}$ anomalies and and the effective cutoff scale is to be $\Lambda_D\sim 3.5\,{\rm TeV}$ \cite{cutoff}. Thus, for $m_{S_1}\gtrsim 1\,{\rm TeV}$, we need $\sqrt{\lambda^*_{33}\lambda_{23}}\gtrsim 0.4$.

\begin{figure}
  \begin{center}
    \includegraphics[height=0.40\textwidth]{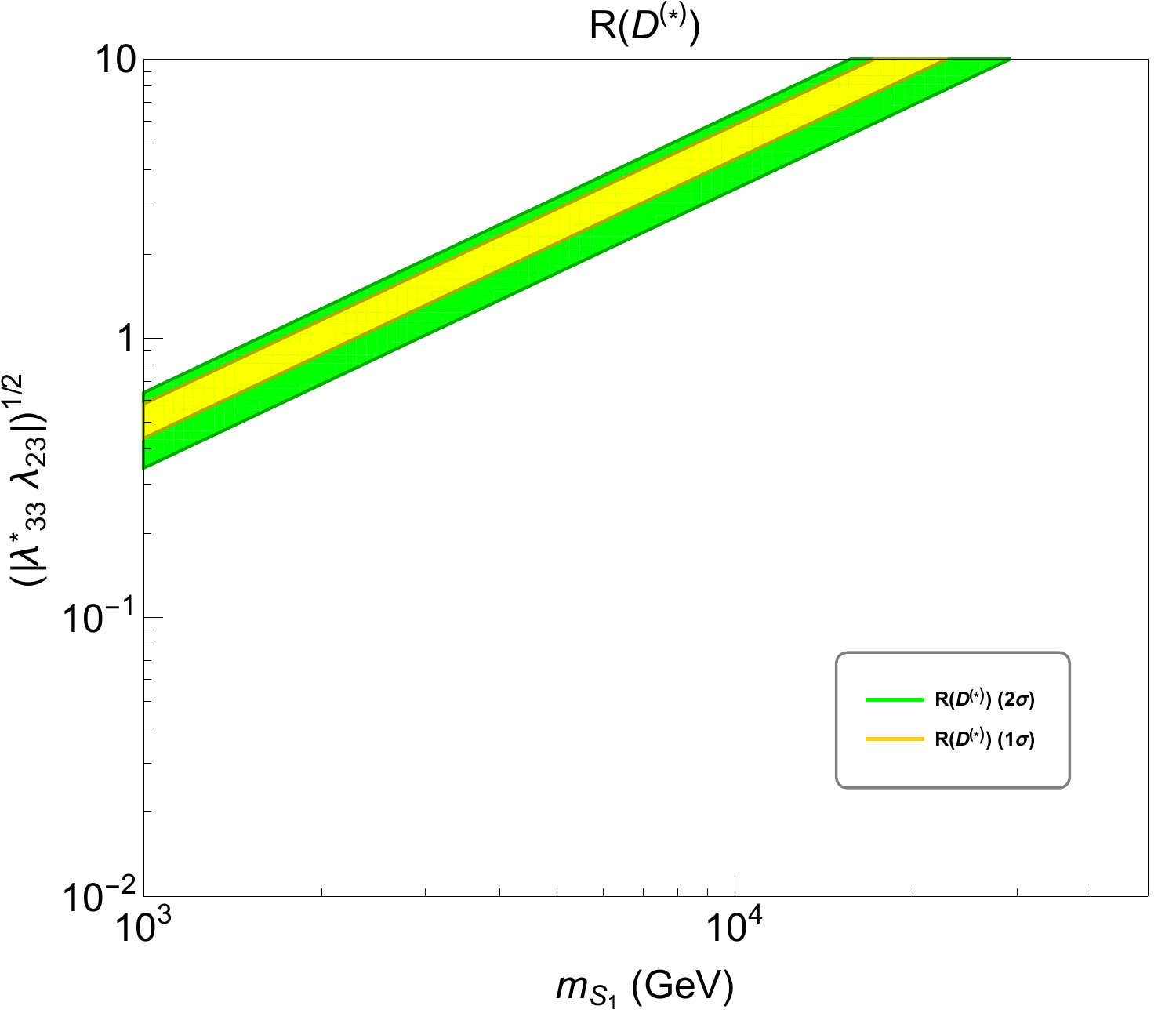}
     \includegraphics[height=0.40\textwidth]{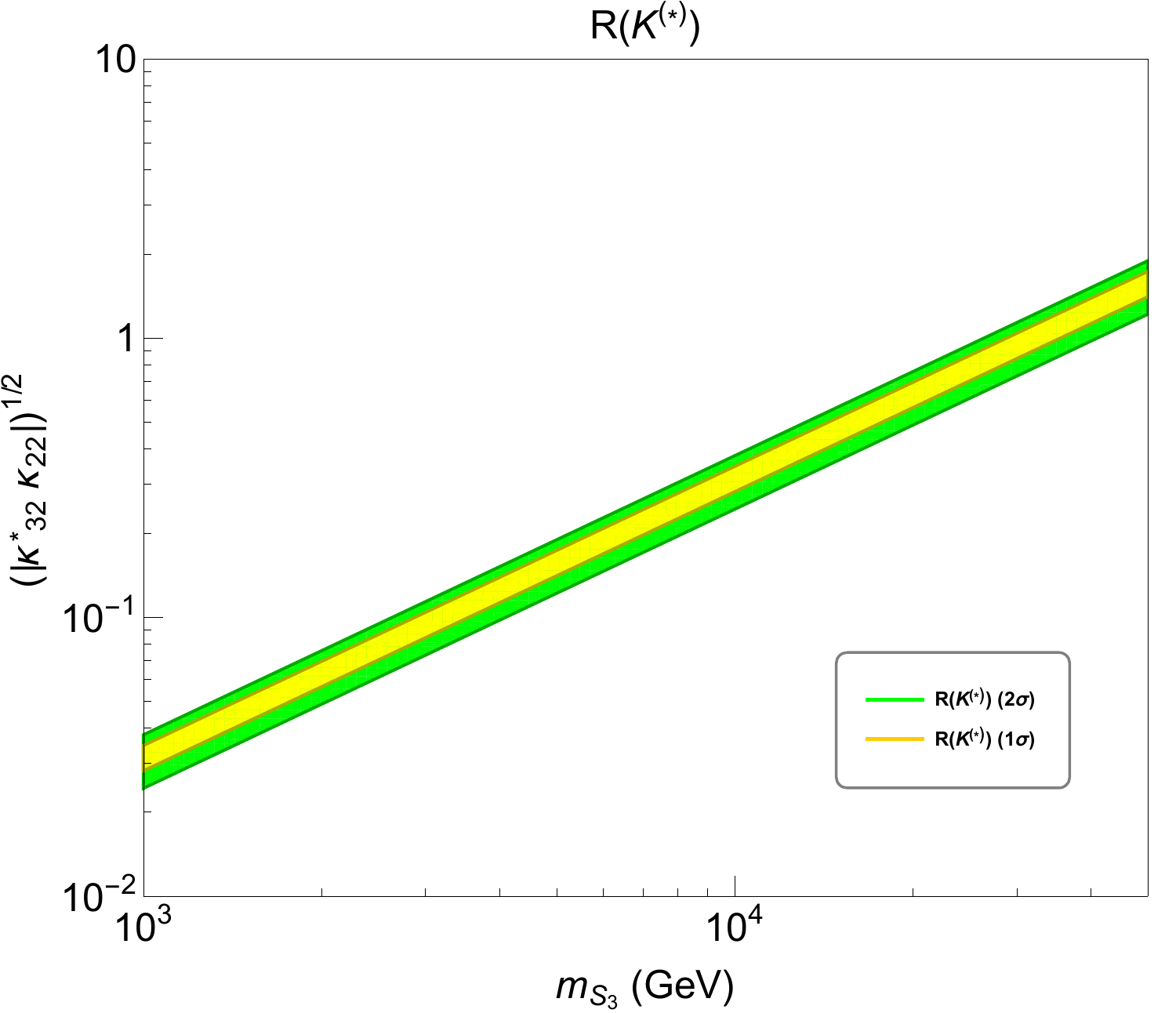}
      \end{center}
  \caption{Parameter space  for the leptoquark mass $m_{LQ}$ and the effective coupling $\lambda_{\rm eff}$, explaining the B-meson anomalies, in green(yellow) region at $2\sigma(1\sigma)$ level. We have taken $m_{LQ}=m_{S_1}$ and $\lambda_{\rm eff}=\sqrt{|\lambda^*_{33}\lambda_{23}|}$ for $R_{D^{(*)}}$ on left plot, and $m_{LQ}=m_{S_3}$ and $\lambda_{\rm eff}=\sqrt{|\kappa^*_{32}\kappa_{22}|}$ for $R_{K^{(*)}}$ on right plot.  }
  \label{B-anomalies}
\end{figure}

In the left plot of Fig.~\ref{B-anomalies}, we depict the parameter space for $m_{S_1}$ and the effective leptoquark coupling, $\lambda_{\rm eff}=\sqrt{|\lambda^*_{33}\lambda_{23}|}$, in which the $R_{D^{(*)}}$ anomalies can be explained within $2\sigma(1\sigma)$ errors  in green(yellow) region from the conditions below eq.~(\ref{RDeff}).

From the couplings of the singlet scalar leptoquark necessary for $R_{D^{(*)}}$ anomalies,
 \bea
{\cal L}_{S_1}&\supset& -\lambda_{33}\Big( \overline{(t^C)_{R}}\, S_1 \tau_{L} -\overline{(b^C)_{R}} \,S_1 \nu_{\tau L}   \Big)+{\rm h.c.} \nonumber \\
&&-\lambda_{23}\Big( \overline{(c^C)_{R}}\, S_1 \tau_{ L} -\overline{(s^C)_{R}} \,S_1 \nu_{\tau L}   \Big)+{\rm h.c.},
\eea
the decay modes of the singlet scalar leptoquark are given by $S_1\rightarrow {\bar t}{\bar\tau}, {\bar c}{\bar\tau}$ and $ S_1\rightarrow {\bar b}{\nu}_\tau, {\bar s}{\nu}_\tau$, which are summarized together with  the corresponding LHC bounds on leptoquark masses in Table~\ref{DBR}.

\begin{table}[h!]\small
\begin{center}
\begin{tabular}{|c||c|c||c|c|}
\hline
 {\rm LQs} & ${\rm BRs}$ & $m_{LQ,{\rm min}}$ &   ${\rm BRs}$  &  $m_{LQ,{\rm min}}$  \\ [0.5ex]
 \hline
$S_1$ & ${\rm B}({\bar t}{\bar \tau}/b\nu_\tau)=\frac{1}{2}\beta$ & 1.22\,{\rm TeV}($ b\nu_\tau)$ \cite{2bMET}   & ${\rm B}({\bar c}{\bar\tau}/s\nu_\tau)=\frac{1}{2}(1-\beta)$ & $950\,{\rm GeV}(\nu_\tau j)$ \cite{2jetMET} \\ [0.5ex]
\hline
$S_3(\phi_1)$ &  ${\rm B}({\bar b}{\bar\mu})=\gamma$ & $1.4\,{\rm TeV}$ \cite{2b2mu}  & ${\rm B}({\bar s}{\bar\mu})=1-\gamma$  &  $1.08\,{\rm TeV}\, ({\bar\mu} j)$ \cite{2mu2jet}  \\ [0.5ex]
\hline 
$S_3(\phi_2)$ &  ${\rm B}({\bar t}{\bar\mu}/{\bar b}{\bar \nu}_\mu)=\frac{1}{2}\gamma$ & $ 1.45\,{\rm TeV} \,({\bar t}{\bar\mu} )$ \cite{2t2mu}  &  ${\rm B}({\bar c}{\bar\mu}/{\bar s}{\bar \nu}_\mu)=\frac{1}{2}(1-\gamma)$  & $850\,{\rm GeV}\, ({\bar\mu}{\bar\nu}_\mu jj)$ \cite{mu2jetMET} \\ [0.5ex] 
\hline 
$S_3(\phi_3)$ & ${\rm B}({\bar t}{\bar\nu}_\mu)=\gamma$  & $1.12\,{\rm TeV}$  \cite{2tMET} & ${\rm B}({\bar c}{\bar\nu}_\mu)=1-\gamma$   & $950\,{\rm GeV}\, ({\bar\nu_\mu j})$ \cite{2jetMET}
\\ [0.5ex]
\hline 
\end{tabular}
\caption{Decay branching ratios of leptoquarks, and LHC bounds on leptoquark masses. Here, $\beta\equiv \lambda^2_{33}/(\lambda^2_{33}+\lambda^2_{23})$ and $\gamma\equiv  \kappa^2_{32}/(\kappa^2_{32}+ \kappa^2_{22})$. Most LHC bounds are given for $B=1$, except  in Ref.~\cite{mu2jetMET} where $B({\bar c}{\bar\mu})=B({\bar s}{\bar \nu}_\mu)=0.5$ was taken. }
\label{DBR}
\end{center}
\end{table}

\subsection{Triplet scalar leptoquark}

After integrating out the leptoquark $\phi_1$ with $Q=+\frac{4}{3}$,  from the results in eq.~(\ref{S3-app}),  we also obtain the effective Hamiltonian relevant for $b\rightarrow s\mu^+\mu^-$ as
\bea
{\cal H}^{S_3}_{b\rightarrow s\mu^+\mu^-}
=-\frac{\kappa^*_{32}\kappa_{22}}{ m^2_{\phi_1}}\, ({\bar b}_L \gamma^\mu s_L) ({\bar\mu}_L\gamma_\mu \mu_L) +{\rm h.c.}\equiv \frac{1}{\Lambda^2_K}\, ({\bar b}_L \gamma^\mu s_L) ({\bar\mu}_L\gamma_\mu \mu_L) +{\rm h.c.}.
\eea
As a consequence, the triplet leptoquark gives rise to the effective operator of the $(V-A)$ form for the quark current, that is, $C^{\mu,\rm NP}_9=-C^{\mu,\rm NP}_{10}\neq 0$, as favored by the $R_{K^{(*)}}$ anomalies, and the effective cutoff scale is to be $\Lambda_K\sim 30\,{\rm TeV}$ \cite{cutoff}. The result is in contrast to the case for $Z'$ models with family-dependent charges such as $Q'=x(B_3-L_3)+y(L_\mu-L_\tau)$ with $x, y$ being arbitrary parameters  where $C^{\mu,\rm NP}_9\neq 0$ and $C^{\mu,\rm NP}_{10}=0$ \cite{Zprime}. Then, for $m_{\phi_1}\gtrsim 1\,{\rm TeV}$, we need $\sqrt{\kappa^*_{32}\kappa_{22}}\gtrsim 0.03$.  
Therefore, we can combine scalar leptoquarks, $S_1$ and $S_3$, to explain $R_{D^{(*)}}$ and $R_{K^{(*)}}$ anomalies, respectively. 

In the right plot of Fig.~\ref{B-anomalies}, we depict the parameter space for $m_{S_3}$ and the effective leptoquark coupling, $\lambda_{\rm eff}=\sqrt{|\kappa^*_{32}\kappa_{22}|}$, in which the $R_{K^{(*)}}$ anomalies can be explained within $2\sigma(1\sigma)$ errors in green(yellow) region from the conditions below eq.~(\ref{RKeff}).

Likewise as for the singlet scalar leptoquark, from the triplet leptoquark couplings necessary for $R_{K^{(*)}}$ anomalies, 
\bea
{\cal L}_{S_3} &\supset&-\kappa_{32}\Big({\sqrt{2}}\, {\overline{(t^C)_{R}}}\, \phi_3\nu_{\mu L} - \overline{(t^C)_{R}} \,\phi_2 \mu_{ L}- \overline{(b^C)_{R}}\, \phi_2\nu_{\mu L} -{\sqrt{2}}\, {\overline{(b^C)_{R}}
} \,\phi_1 \mu_{ L}   \Big)+{\rm h.c.} \nonumber \\
&&-\kappa_{22}\Big( {\sqrt{2}}\,\overline{(c^C)_{R}}\, \phi_3\nu_{\mu L} - \overline{(c^C)_{R}} \,\phi_2 \mu_{ L}- \overline{(s^C)_{R}}\, \phi_2\nu_{\mu L} -{\sqrt{2}}\, \overline{(s^C)_{R}}\, \phi_1\mu_{ L}   \Big)+{\rm h.c.},
\eea
the decay modes of the singlet scalar leptoquark are given by $\phi_1\rightarrow {\bar b} {\bar\mu}, {\bar s}{\bar\mu}$, $\phi_2\rightarrow {\bar t}{\bar\mu},  {\bar c}{\bar\mu}, {\bar b} {\bar\nu}_\mu,  {\bar s} {\bar\nu}_\mu$, and $\phi_3\rightarrow {\bar t} {\bar\nu}_\mu, {\bar c} {\bar\nu}_\mu$.
As will be discussed in the next section, the bounds from $B\rightarrow K\nu{\bar \nu}$ could require $\kappa_{33}$ and $\kappa_{23}$ to be sizable. In this case, the decay modes containing ${\bar\tau}$ or ${\bar\nu}_\tau$ are relevant too. The decay branching ratios of the triplet leptoquark and the corresponding LHC bounds on the mass of triplet scalar leptoquark are also summarized in Table~\ref{DBR}.

\section{Constraints on leptoquarks}

We discuss the constraints on scalar leptoquark models, due to other rare meson decays, muon $(g-2)_\mu$, lepton flavor violation as well as the LHC searches.
The constraints discussed in this section can give rise to important implications for the indirect signatures of DM annihilation into a leptoquark pair in the later discussion.

\subsection{Rare meson decays and mixing}

In leptoquark models explaining the B-meson anomalies, there is no $B-{\bar B}$ mixing at tree level, but instead it appears at one-loop level. Therefore, the resulting new contribution to the $B_s-{\bar B}_s$ mixing is about $1\%$ level \cite{leptoquarks}, which can be ignored.

Both singlet and triplet leptoquarks contribute to $B\rightarrow K^{(*)}\nu {\bar \nu}$ at tree level, so their couplings are severely constrained in this case \cite{leptoquarks,LQ-fit}. 
The effective Hamiltonian relevant for ${\bar b}\rightarrow {\bar s}\nu {\bar \nu}$ \cite{BKnunu} is
\bea
{\cal H}_{{\bar b}\rightarrow {\bar s}\nu {\bar \nu}}= -\frac{\sqrt{2}\alpha_{\rm em} G_F}{\pi}\, V_{tb} V^*_{ts} \sum_l C^l_L  ({\bar b}\gamma^\mu P_L s)({\bar \nu}_{l}\gamma_\mu P_L\nu_{l}) 
\eea
where $C^l_L=C^{\rm SM}_L+C^{l,{\rm NP}}_\nu$. Here, the SM contribution $C^{\rm SM}_L$ is given by $C^{\rm SM}_L=-X_t/s^2_W$
where $s_W\equiv \sin\theta_W$ and $X_t=1.469\pm 0.017$.
From the result in eq.~(\ref{S1-bnunu-app}), the scalar leptoquarks leads to additional contributions to the effective Hamiltonian for $B\rightarrow K\nu {\bar \nu}$ as
\bea
C^{l,{\rm NP}}_\nu= -\left(  \frac{\lambda^*_{3i}\lambda_{2j}}{2m^2_{S_1}}+\frac{\kappa^*_{3i} \kappa_{2j}}{2m^2_{\phi_2}}\right) \, \frac{\pi}{ \sqrt{2}\alpha_{\rm em} G_FV_{tb} V^*_{ts}}. 
\eea
Therefore, the ratio of the branching ratios are given by
\bea
R_{K^{(*)}\nu}&\equiv& \frac{B(B\rightarrow K^{(*)}\nu{\bar\nu})}{B(B\rightarrow K^{(*)}\nu{\bar\nu})\Big|_{\rm SM}} \nonumber \\
&=& \frac{2}{3}+\frac{1}{3} \frac{|C^{\rm SM}_L+C^{l,{\rm NP}}_\nu|^2}{|C^{\rm SM}_L|^2}.
\eea
Comparing the experimental bounds on $B(B\rightarrow K^{(*)}\nu{\bar\nu})$ \cite{BKnunu-exp} given by
\bea
B(B\rightarrow K\nu{\bar\nu})<1.6\times 10^{-5}, \quad\quad B(B\rightarrow K^{*}\nu{\bar\nu})< 2.7\times 10^{-5},
\eea
to the SM values \cite{BKnunu-SM} given by 
\bea
B(B\rightarrow K\nu{\bar\nu})\Big|_{\rm SM}&=&(3.98\pm 0.43\pm 0.19)\times 10^{-6}, \nonumber   \\
\quad  B(B\rightarrow K^{*}\nu{\bar\nu})\Big|_{\rm SM}&=&(9.19\pm 0.86\pm 0.50)\times 10^{-6},
\eea
and ignoring the imaginary part of $C^{l,{\rm NP}}_\nu$, we get the $R_{K^*\nu}$ bound as
\bea
-10.1<{\rm Re}(C^{l,{\rm NP}}_\nu)<22.8.
\eea

Taking into account $\kappa_{32}$ and $\kappa_{22}$, which are necessary for $B\rightarrow K^{(*)}\mu^+\mu^-$, the triplet scalar leptoquark contributes only to $B\rightarrow K^{(*)}\nu_\mu{\bar\nu}_\mu$.
In this case, as the triplet leptoquark contribution to $C^{\mu,{\rm NP}}_\nu$ is about the same as $C^{\mu,{\rm NP}}_9=-0.61$,  it  satisfies the $R_{K^*\nu}$ bound on its own easily. 

On the other hand, the singlet leptoquark with nonzero $\lambda_{33}$ and $\lambda_{23}$, which are necessary for $B\rightarrow D^{(*)}\tau{\bar\nu}_\tau$, contribute significantly to $B\rightarrow K^{(*)}\nu_\tau {\bar \nu}_\tau$.
Therefore, we need to cancel the singlet scalar leptoquark contributions to $B\rightarrow K^{(*)}\nu_\tau {\bar\nu}_\tau$, by imposing that 
\bea
 \frac{\lambda^*_{33}\lambda_{23}}{2m^2_{S_1}}+\frac{\kappa^*_{33} \kappa_{23}}{2m^2_{\phi_2}}\approx 0.
\eea
Ignoring the mass splitting generated within the triplet scalar leptoquark due to potential higher dimensional operators after electroweak symmetry breaking, we get $m_{\phi_1}=m_{\phi_2}=m_{\phi_3}\equiv m_{S_3}$. Then, in order to cancel the contributions to  $B\rightarrow K^{(*)}\nu_\tau{\bar\nu}_\tau$ or $B\rightarrow K^{(*)}\nu_{\mu,\tau}{\bar\nu}_{\tau,\mu}$, the necessary conditions for the additional couplings are
\bea
|\kappa^*_{33} \kappa_{23}|&\approx &|\lambda^*_{33} \lambda_{23}|\, \Big(\frac{m^2_{S_3}}{m^2_{S_1}}\Big), \\
|\lambda^*_{32}\lambda_{23}| &\approx &|\kappa^*_{32}\kappa_{23}|\, \Big(\frac{m^2_{S_1}}{m^2_{S_3}}\Big).  \label{lam32}
\eea
Therefore, for $m_{S_3}\sim m_{S_1}$, the additional  couplings for the triplet leptoquark, $\kappa_{23}$ and $\kappa_{33}$,  must satisfy $\sqrt{|\kappa^*_{33}\kappa_{23}|}\approx\sqrt{ |\lambda^*_{33} \lambda_{23}|}\gtrsim 0.4$, because $\sqrt{\lambda^*_{33}\lambda_{23}}\gtrsim 0.4$ to explain the $R_{D^{(*)}}$ anomalies.
On the other hand,  for $m_{S_3}\sim m_{S_1}$, the additional coupling for the singlet leptoquark, $\lambda_{32}$ must satisfy $\sqrt{|\lambda^*_{32} \lambda_{23}|}\approx\sqrt{|\kappa^*_{32}\kappa_{23}|}$, up to the conditions, $\sqrt{\lambda^*_{33}\lambda_{23}}\gtrsim 0.4$ and $\sqrt{|\kappa^*_{32}\kappa_{22}|}\gtrsim 0.03$, for explaining $R_{D^{(*)}}$ and  $R_{K^{(*)}}$ anomalies, respectively. Then,  it is easy to get a sizable $\lambda_{32}$ coupling in order to explain the deviation in $(g-2)_\mu$ as will be discussed later.

In summary, taking account of bounds from $B\rightarrow K^{(*)}\nu{\bar\nu}$, the necessary flavor structure for leptoquark couplings is given by the following,
\bea
\lambda=\left(\begin{array}{ccc}  0  & 0 & 0 \\   0 &  0 & \lambda_{23} \\ 0 & \lambda_{32} & \lambda_{33}  \end{array} \right), \quad \kappa=\left(\begin{array}{ccc}  0  & 0 & 0 \\   0 &  \kappa_{22} & \kappa_{23} \\ 0 & \kappa_{32} & \kappa_{33}  \end{array} \right).
\eea
If the extra couplings for $B\rightarrow K^{(*)}\nu{\bar\nu}$ are sizable, namely, $\lambda_{32}\sim \lambda_{23}, \lambda_{33}$ for the singlet leptoquark, and $\kappa_{23}, \kappa_{33}\gtrsim \kappa_{22}, \kappa_{32}$ for the triplet leptoquark, the decay branching ratios of leptoquarks are changed, so that the LHC searches for leptoquarks as well as the indirect searches for leptoquark portal dark matter will be affected. In particular, we will discuss the impact of extra couplings on the signatures of DM annihilations into a leptoquark pair in detail in the later section. 

For the later discussion on $(g-2)_\mu$ in the next subsection, we illustrate some sets of consistent leptoquark couplings for $m_{S_3}\sim m_{S_1}\gtrsim 1\,{\rm TeV}$. For $\lambda_{32}=\lambda_{33}=1$ and $\kappa_{23}=0.1$, we find that $\lambda_{23}\gtrsim 0.16$, $\kappa_{32}\sim \kappa_{33}\gtrsim 1.6$ and $\kappa_{22}\gtrsim 5.6\times 10^{-4}$. In this case, we need a hierarchy of couplings, $\lambda_{32}=\lambda_{33}\gg\lambda_{23}$ and $\kappa_{32}\sim \kappa_{33}\gg\kappa_{23}\gg\kappa_{22}$. Instead, choosing $\lambda_{32}=\lambda_{33}=0.1$ and $\kappa_{23}=1$, we obtain that $\lambda_{23}\gtrsim 1.6$, $\kappa_{32}\sim 0.16$, $\kappa_{33}\gtrsim 0.16$ and $\kappa_{22}\gtrsim 5.6\times 10^{-3}$. Then, we need a hierarchy of couplings, $\lambda_{23}\gg \lambda_{32}=\lambda_{33}$ and $\kappa_{23}\gg \kappa_{32}\sim\kappa_{33}\gg\kappa_{22}$.

\subsection{$(g-2)_\mu$}

 \begin{figure}
  \begin{center}
    \includegraphics[height=0.40\textwidth]{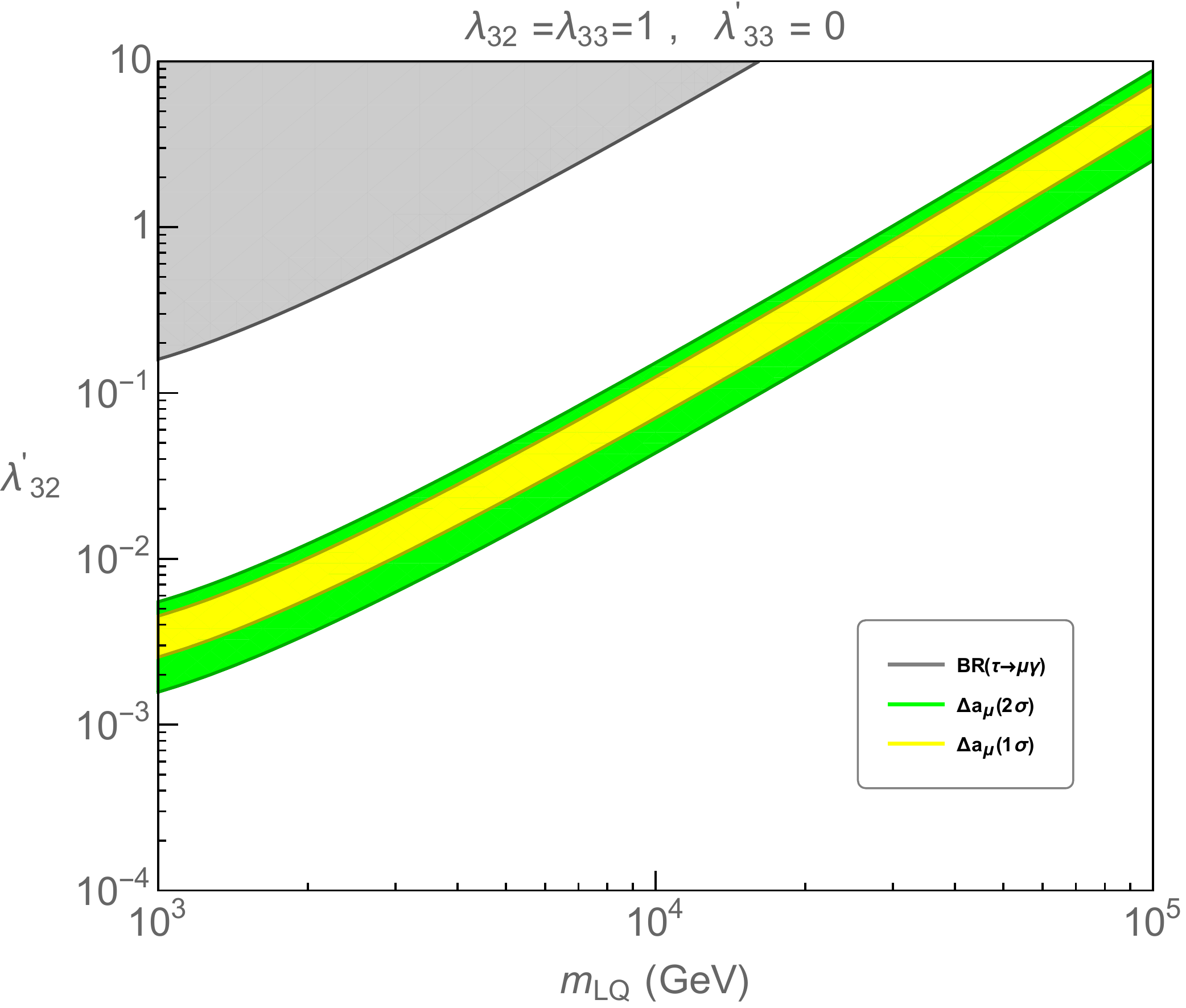}
     \includegraphics[height=0.40\textwidth]{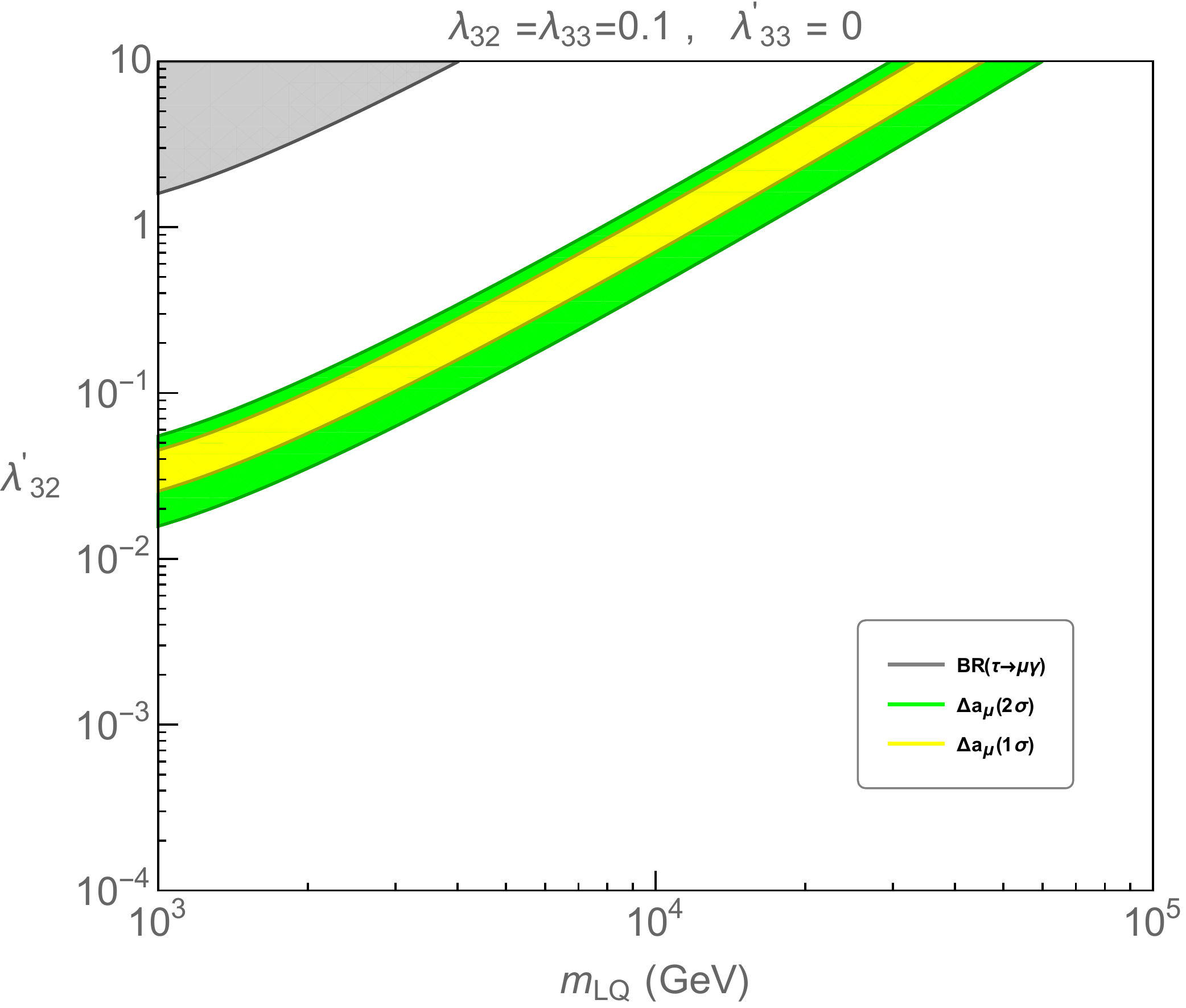}
      \end{center}
  \caption{Parameter space  for $m_{LQ}=m_{S_1}$ and $\lambda'_{32}$ allowed by $(g-2)_\mu$, in green(yellow) region, at $2\sigma(1\sigma)$ level. The gray region is excluded by the bound on ${\rm BR}(\tau\rightarrow\mu\gamma)$. We have fixed $\lambda_{32}=\lambda_{33}=1(0.1)$ on left(right) plot and $\lambda'_{33}=0$ in both plots.}
  \label{mg2}
\end{figure}

For the singlet scalar leptoquark, the relevant Yukawa couplings for $(g-2)_\mu$ with an additional Yukawa coupling, are given as follows,
\bea
{\cal L}_{S_1}\supset - \lambda_{ij} \overline{(Q^C)^a_{Ri}} (i\sigma^2)_{ab}S_1 L^b_{jL} - \lambda'_{ij} \overline{(u^C)_{Li}}S_1 e_{jR}  +{\rm h.c.} 
\eea
Then, the chirality-enhanced effect from the top quark contributes most \cite{leptoquarks}, as follows,
\bea
a^{S_1}_\mu= \frac{m_\mu }{4\pi^2}\,{\rm Re}[C^{22}_R]
\eea
with
\bea
C^{ij}_R\equiv -\frac{N_c}{12m_{S_1}^2} \,m_t \lambda_{3i} \lambda^{\prime *}_{3j} \Big(7+4\log \Big(\frac{m^2_t}{m^2_{S_1}} \Big) \Big).
\eea
The deviation of the anomalous magnetic moment of muon between experiment and SM
values is given \cite{amu,pdg} by
\be
\Delta a_\mu = a^{\rm exp} -a^{\rm SM} = 288(80)\times 10^{-11},
\ee
which is a $3.6\sigma$ discrepancy from the SM \cite{pdg}.
We note that as discussed in eq.~(\ref{lam32}), the extra couplings for the triplet leptoquark, $\kappa_{23}$ and $\kappa_{33}$, allow for a sizable $\lambda_{32}$, leading to a large deviation in $(g-2)_\mu$ without a conflict to the bound from $B(B\rightarrow K^{(*)}\nu{\bar \nu})$.     

On the other hand, the additional coupling also contributes to the branching ratio of $\tau\rightarrow \mu\gamma$ as follows,
\bea
{\rm BR}(\tau\rightarrow \mu\gamma)= \frac{\alpha m^3_\tau}{256\pi^4}\, \tau_\tau \Big(|C^{23}_R|^2+|C^{23}_L|^2 \Big)
\eea
where $C^{ij}_L=C^{ij}_R( \lambda_{3i}\rightarrow \lambda^{\prime }_{3i}, \lambda'_{3j}\rightarrow \lambda_{3j} )$ and the lifetime of tau is given by $\tau_\tau=(290.3\pm 0.5)\times 10^{-15}\, {\rm s}$ \cite{pdg}. 
The current experimental bound is given \cite{taumu} by 
\be
{\rm BR}(\tau\rightarrow\mu\gamma)<4.4\times 10^{-8}.
 \ee
 
 In Fig.~\ref{mg2}, we show the parameter space for the singlet scalar leptoquark mass $m_{LQ}$  and the extra leptoquark coupling $\lambda'_{32}$, where the $(g-2)_\mu$ anomaly can be explained, in green(yellow) region at $2\sigma(1\sigma)$ level. The gray region is excluded by the bound on $B(\tau\rightarrow \mu\gamma)$. We have taken $\lambda_{32}=\lambda_{33}=1(0.1)$ on left(right) plot and $\lambda'_{33}=0$. Therefore, for $m_{LQ}\lesssim 10-50\,{\rm TeV}$ under perturbativity and leptoquark couplings less than unity, the $(g-2)_\mu$ anomaly can be explained in our model, being compatible with $B(\tau\rightarrow \mu\gamma)$.

\subsection{Leptoquark searches}

There are two main production channels for leptoquarks at the LHC, one is pair production via gluon fusion and the other is single production via gluon-quark fusion \cite{LQS1,LQS2}.

In the case of $R_{K^{(*)}}$ anomalies, the triplet scalar leptoquark ($\phi_1$) couples to $b/s,\mu$. The other components of the triplet leptoquark couple to $b/s,\nu_\mu$ and $t/c,\mu$ for $\phi_2$ and $t/c,\nu_\mu$ for $\phi_3$.
On the other hand, in the case of $R_{D^{(*)}}$ anomalies, the singlet scalar leptoquark ($S_1$) couples to $b/s,\nu_\tau$ and $t/c,\tau$.
When the leptoquark pair production via gluon fusion is dominant, the current limits on leptoquark masses listed in Table 1 apply.  The current LHC bounds on leptoquarks depend on decay modes, but the leptoquark masses are constrained to be greater than about  $1\,{\rm TeV}$ in most cases. 

When the Yukawa couplings, $\phi_1$-$b/s$-$\mu$, $S_1$-$b$-$\nu_\tau$ and $S_1$-$c$-$\tau$ couplings, present in models explaining the $B$-anomalies, are  sizable, the leptoquarks can be singly produced by $b/s/c$ quark fusions with gluons. 
For instance, in the case of $\phi_1$, the relevant production/decay channels are $pp\rightarrow \phi^*_1 \phi_1=b{\bar b}(s{\bar s})\mu^+\mu^-$ and  $pp\rightarrow \phi_1 \mu^+\rightarrow b(s)\mu^+\mu^- $ \cite{LQS1}.

\section{Leptoquarks and scalar dark matter}

We introduce a scalar dark matter that have direct interactions to scalar leptoquarks  and the SM Higgs doublet $H$ by quartic couplings. Thus,  this is the minimal dark matter model without a need of extra mediator particle.
In this section, we regard scalar leptoquarks as portals to scalar dark matter and discuss the impacts of leptoquarks on direct and indirect detection of dark matter as well as Higgs data.

We can also consider leptoquark-portal models for fermion or vector dark matter too.
But, in this case, there is a need of mediator particles \cite{LQ-DM} and/or non-renormalizable interactions \cite{LQ-DM2}, leading to more parameters in the model, so this case is postponed to a future publication for comparison \cite{TBP}.

\subsection{Annihilation cross sections for scalar dark matter}

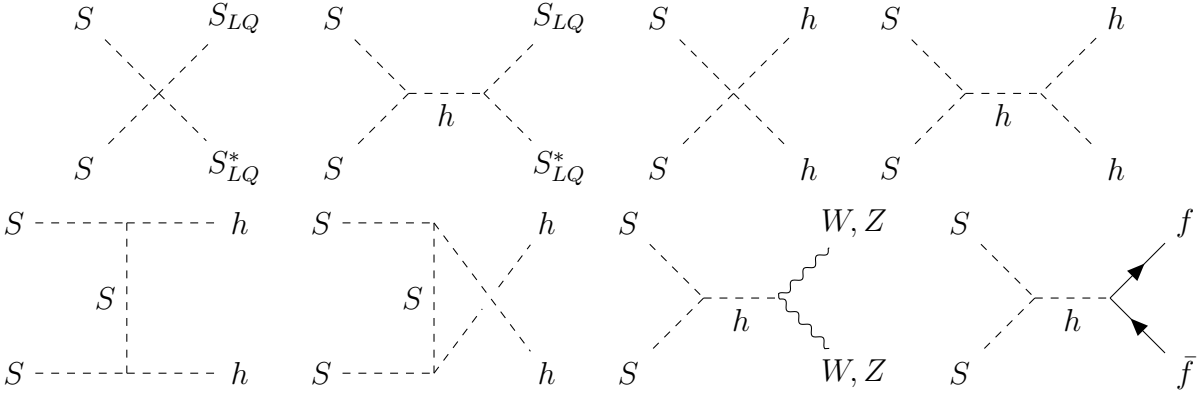
\begin{figure}[!t]
\begin{center}
\begin{tikzpicture}[baseline=(a)]
\begin{feynman}
\vertex (a) at (0, 0);
\vertex (b) at (-1, 1) {\(S\)};
\vertex (c) at (-1, -1) {\(S\)};
\vertex (d) at (1, 1) {\(S_{LQ}\)};
\vertex (e) at (1, -1) {\(S^*_{LQ}\)};
\diagram* {
(b) -- [scalar] (a) -- [scalar] (c),
(d) -- [scalar] (a) -- [scalar] (e),
};
\end{feynman}
\end{tikzpicture}
\quad
\begin{tikzpicture}[baseline=(a)]
\begin{feynman}
\vertex (a) at (0, 0);
\vertex (b) at (-1, 1) {\(S\)};
\vertex (c) at (-1, -1) {\(S\)};
\vertex (d) at (1, 0);
\vertex (e) at (2, 1) {\(S_{LQ}\)};
\vertex (f) at (2, -1) {\(S^*_{LQ}\)};
\diagram* {
(b) -- [scalar] (a) -- [scalar] (c),
(a) -- [scalar, edge label'={\(h\)}] (d),
(e) -- [scalar] (d) -- [scalar] (f),
};
\end{feynman}
\end{tikzpicture}
\quad
\begin{tikzpicture}[baseline=(a)]
\begin{feynman}
\vertex (a) at (0, 0);
\vertex (b) at (-1, 1) {\(S\)};
\vertex (c) at (-1, -1) {\(S\)};
\vertex (d) at (1, 1) {\(h\)};
\vertex (e) at (1, -1) {\(h\)};
\diagram* {
(b) -- [scalar] (a) -- [scalar] (c),
(d) -- [scalar] (a) -- [scalar] (e),
};
\end{feynman}
\end{tikzpicture}
\quad
\begin{tikzpicture}[baseline=(a)]
\begin{feynman}
\vertex (a) at (0, 0);
\vertex (b) at (-1, 1) {\(S\)};
\vertex (c) at (-1, -1) {\(S\)};
\vertex (d) at (1, 0);
\vertex (e) at (2, 1) {\(h\)};
\vertex (f) at (2, -1) {\(h\)};
\diagram* {
(b) -- [scalar] (a) -- [scalar] (c),
(a) -- [scalar, edge label'={\(h\)}] (d),
(e) -- [scalar] (d) -- [scalar] (f),
};
\end{feynman}
\end{tikzpicture}
\quad
\begin{tikzpicture}[baseline=(a)]
\begin{feynman}
\vertex (a) at (0, 0);
\vertex (b) at (0, 1);
\vertex (c) at (0, -1);
\vertex (d) at (-1.5, 1) {\(S\)};
\vertex (e) at (-1.5, -1) {\(S\)};
\vertex (f) at (1.5, 1) {\(h\)};
\vertex (g) at (1.5, -1) {\(h\)};
\diagram* {
(d) -- [scalar] (b) -- [scalar] (f),
(b) -- [scalar, edge label'={\(S\)}] (c),
(e) -- [scalar] (c) -- [scalar] (g),
};
\end{feynman}
\end{tikzpicture}
\quad
\begin{tikzpicture}[baseline=(a)]
\begin{feynman}
\vertex (a) at (0, 0);
\vertex (b) at (0, 1);
\vertex (c) at (0, -1);
\vertex (d) at (-1.5, 1) {\(S\)};
\vertex (e) at (-1.5, -1) {\(S\)};
\vertex (f) at (1.5, 1) {\(h\)};
\vertex (g) at (1.5, -1) {\(h\)};
\node (h) at (0.75, 0);
\diagram* {
(d) -- [scalar] (b) -- [scalar] (g),
(b) -- [scalar, edge label'={\(S\)}] (c),
(e) -- [scalar] (c) -- [scalar] (h) -- [scalar] (f),
};
\end{feynman}
\end{tikzpicture}
\quad
\begin{tikzpicture}[baseline=(a)]
\begin{feynman}
\vertex (a) at (0, 0);
\vertex (b) at (-1, 1) {\(S\)};
\vertex (c) at (-1, -1) {\(S\)};
\vertex (d) at (1, 0);
\vertex (e) at (2, 1) {\(W, Z\)};
\vertex (f) at (2, -1) {\(W, Z\)};
\diagram* {
(b) -- [scalar] (a) -- [scalar] (c),
(a) -- [scalar, edge label'={\(h\)}] (d),
(e) -- [photon] (d) -- [photon] (f),
};
\end{feynman}
\end{tikzpicture}
\quad
\begin{tikzpicture}[baseline=(a)]
\begin{feynman}
\vertex (a) at (0, 0);
\vertex (b) at (-1, 1) {\(S\)};
\vertex (c) at (-1, -1) {\(S\)};
\vertex (d) at (1, 0);
\vertex (e) at (2, 1) {\(f\)};
\vertex (f) at (2, -1) {\(\bar{f}\)};
\diagram* {
(b) -- [scalar] (a) -- [scalar] (c),
(a) -- [scalar, edge label'={\(h\)}] (d),
(f) -- [fermion] (d) -- [fermion] (e),
};
\end{feynman}
\end{tikzpicture}
\end{center}
\caption{Feynman diagrams  for  annihilations of scalar dark matter at tree level.}
\label{SDMFeyn}
\end{figure}

We consider a  scalar leptoquark $S_{LQ}=S_{1,3}$ and a singlet real scalar dark matter $S$.
Then,  the most general renormalizable Lagrangian consistent with $S\rightarrow -S$ is
\bea
{\cal L}_S&=&|D_\mu S_{LQ}|^2-m^2_{LQ} |S_{LQ}|^2+ \frac{1}{2}(\partial_\mu S)^2 - \frac{1}{2} m^2_S S^2  \nonumber \\
&&- \frac{1}{4} \lambda_1 S^4-\lambda_2 |S_{LQ}|^4- \frac{1}{2}\lambda_3 S^2|S_{LQ}|^2 - \frac{1}{2} \lambda_4 S^2 |H|^2-\lambda_5 |H|^2 |S_{LQ}|^2.
\eea
The above Lagrangian generalizes the Higgs portal interactions to those for leptoquarks.

After electroweak symmetry breaking with $H=(0,v+h)^T/\sqrt{2}$, the new interactions  relevant for $SS\rightarrow S_{LQ} S^*_{LQ}, hh$ are
\bea
{\cal L}_{\rm S, int}= -\frac{1}{2}\lambda_3 S^2|S_{LQ}|^2 -\frac{1}{4}\lambda_4 S^2 (2 v h+ h^2)-\frac{1}{2}\lambda_5 |S_{\rm LQ}|^2   (2 v h+ h^2).
\eea

Scalar dark matter $S$ annihilates through three channels at tree level, $SS\rightarrow f{\bar f}$ with $f$ being the SM fermions, $SS\rightarrow hh$, with $h$ being the SM Higgs boson,  $SS\rightarrow VV$ with $V$ being electroweak gauge bosons, and $SS\rightarrow S_{LQ} S^*_{LQ}$ for $m_{LQ}<m_S$. Depending on the quartic couplings, a heavy scalar dark matter may annihilate into a pair of leptoquarks dominantly, leaving the signatures in both anti-proton and positron from cosmic rays, due to the decay products of leptoquarks, as will be discussed later.

For $m_{LQ}>m_S$, $SS\rightarrow S_{LQ} S^*_{LQ}$ channels are kinematically closed, so instead leptoquark loops make corrections to $SS\rightarrow VV$ with $V$ being electroweak gauge bosons and contribute to new annihilations such as $SS\rightarrow gg, Z\gamma,\gamma\gamma$. 
In this case, depending on the relative contributions of  $SS\rightarrow f{\bar f}, hh, VV$ channels, the loop-induced annihilation channels can be relevant.

We obtain the effective interactions between scalar dark matter and the SM gauge bosons due to leptoquarks with $m_{LQ}>m_S$, as follows,
\bea
{\cal L}_{S,{\rm eff}}=D_3 \, S^2\, G_{\mu\nu}G^{\mu\nu}+D_2\, S^2\, W_{\mu\nu}W^{\mu\nu}+  D_1\, S^2\, F_{Y\mu\nu} F^{Y\mu\nu}
\eea
The details on the above effective interactions are given in Appendix B.
Then, in the basis of mass eigenstates, the above effective interactions become
\bea
{\cal L}_{S,{\rm eff}}&=&D_{gg} \, S^2 \, G_{\mu\nu}G^{\mu\nu}+D_{WW} \, S^2 \, W^+_{\mu\nu}W^{-\mu\nu}+D_{ZZ} \, S^2 \, Z_{\mu\nu}Z^{\mu\nu} \nonumber \\
&&+D_{Z\gamma} \, S^2\, Z_{\mu\nu}F^{\mu\nu}+ D_{\gamma\gamma} \, S^2\, F_{\mu\nu} F^{\mu\nu}
\eea
where 
\bea
D_{gg}&=&D_3, \\
D_{WW}&=& 2D_2, \\
D_{ZZ}&=& D_1\sin^2\theta_W+ D_2 \cos^2\theta_W, \\
D_{Z\gamma}&=& (D_2-D_1) \sin(2\theta_W), \\
D_{\gamma\gamma} &=&D_1 \cos^2\theta_W + D_2 \sin^2\theta_W.
\eea

First, the tree-level annihilation cross sections are
\bea
(\sigma v_{\rm rel})_{SS\rightarrow S_{LQ} S^*_{LQ}}= \frac{N_c N_{LQ}}{32 \pi m_S^2}\sqrt{1-\frac{m_{ LQ}^2}{m_S^2}}\,\Bigg( \lambda_3 + \frac{\lambda_4\lambda_5 v^2 }{4m_S^2-m_h^2} \Bigg)^2,
\eea
\bea
(\sigma v_{\rm rel})_{SS\rightarrow hh}=\frac{\lambda_4^2}{64\pi m_S^2}\sqrt{1-\frac{m_h^2}{m_S^2}} \Bigg( 1 + \frac{3 m_h^2}{4m_S^2-m_h^2}-\frac{2\lambda_4 v^2 }{2m_S^2-m_h^2} \Bigg)^{2},
\eea
\bea
( \sigma v_{\rm rel} )_{SS \rightarrow f\bar{f}} = \frac{N_c \lambda_4^2}{4\pi} \frac{m_f^2}{ (4m_S^2-m_h^2)^2}\Bigg( 1-\frac{m_f^2}{m_S^2}\Bigg)^{3/2},
\eea
with $f$ being all the SM fermions satisfying $m_f<m_S$.
Here, we note that $N_c=3$ is the number of colors and $N_{LQ}=1,3$ for $S_{LQ}=S_1, S_3$, respectively.

On the other hand, for $m_{LQ}>m_S$,  instead of $SS\rightarrow S_{LQ} S^*_{LQ}$, we need to consider the loop-induced annihilation cross sections \cite{cline,loops} for $SS\rightarrow gg, \gamma\gamma, Z\gamma$,  as follows,
\bea
(\sigma v_{\rm rel})_{SS\rightarrow gg}&=& \frac{64D^2_{gg}m^2_S}{\pi}, \\
(\sigma v_{\rm rel})_{SS\rightarrow \gamma\gamma}&=& \frac{ 8D^2_{\gamma\gamma}m^2_S}{\pi }, \\
(\sigma v_{\rm rel})_{SS\rightarrow Z\gamma} &=&  
 \frac{ 4D^2_{Z\gamma}m^2_S}{\pi}\left( 1-\frac{m^2_Z}{4m^2_S} \right)^3.
 \eea
Adding loop corrections of leptoquarks to tree-level contributions coming from the Higgs portal coupling $\lambda_4$, we also obtain the annihilation cross sections for  $SS\rightarrow WW, ZZ$, respectively,
 \bea
  ( \sigma v_{\rm rel})_{SS\rightarrow WW}&=&\Bigg[\frac{\lambda_4^2 m_S^2}{2\pi(m_h^2-4m_S^2)^2}\bigg(1-\frac{m_W^2}{m_S^2}+\frac{3m_W^4}{4m_S^4} \bigg)+\frac{4|D_{WW}|^2m_S^2}{\pi}\bigg(1-\frac{m_W^2}{m_S^2}+\frac{3m_W^4}{8m_S^4} \bigg) \nonumber \\
&&+\frac{3\lambda_4 {\rm Re}[D_{WW}]m_W^2}{2\pi (m_h^2-4m_S^2)}\bigg(2-\frac{m_W^2}{m_S^2}\bigg)\Bigg]\sqrt{1-\frac{m_W^2}{m_S^2}}, \\
 ( \sigma v_{\rm rel})_{SS\rightarrow ZZ} &=& \Bigg[\frac{\lambda_4^2 m_S^2}{4\pi(m_h^2-4m_S^2)^2}\bigg(1-\frac{m_Z^2}{m_S^2}+\frac{3m_Z^4}{4m_S^4} \bigg)+\frac{8|D_{ZZ}|^2 m_S^2}{\pi}\bigg(1-\frac{m_Z^2}{m_S^2}+\frac{3m_Z^4}{8m_S^4} \bigg) \nonumber \\\nonumber \\
&&+\frac{3\lambda_4 {\rm Re}[D_{ZZ}]m_Z^2}{2\pi (m_h^2-4m_S^2)}\bigg(2-\frac{m_Z^2}{m_S^2}\bigg)\Bigg]\sqrt{1-\frac{m_Z^2}{m_S^2}}.
 \eea

 \begin{figure}
  \begin{center}
    \includegraphics[height=0.35\textwidth]{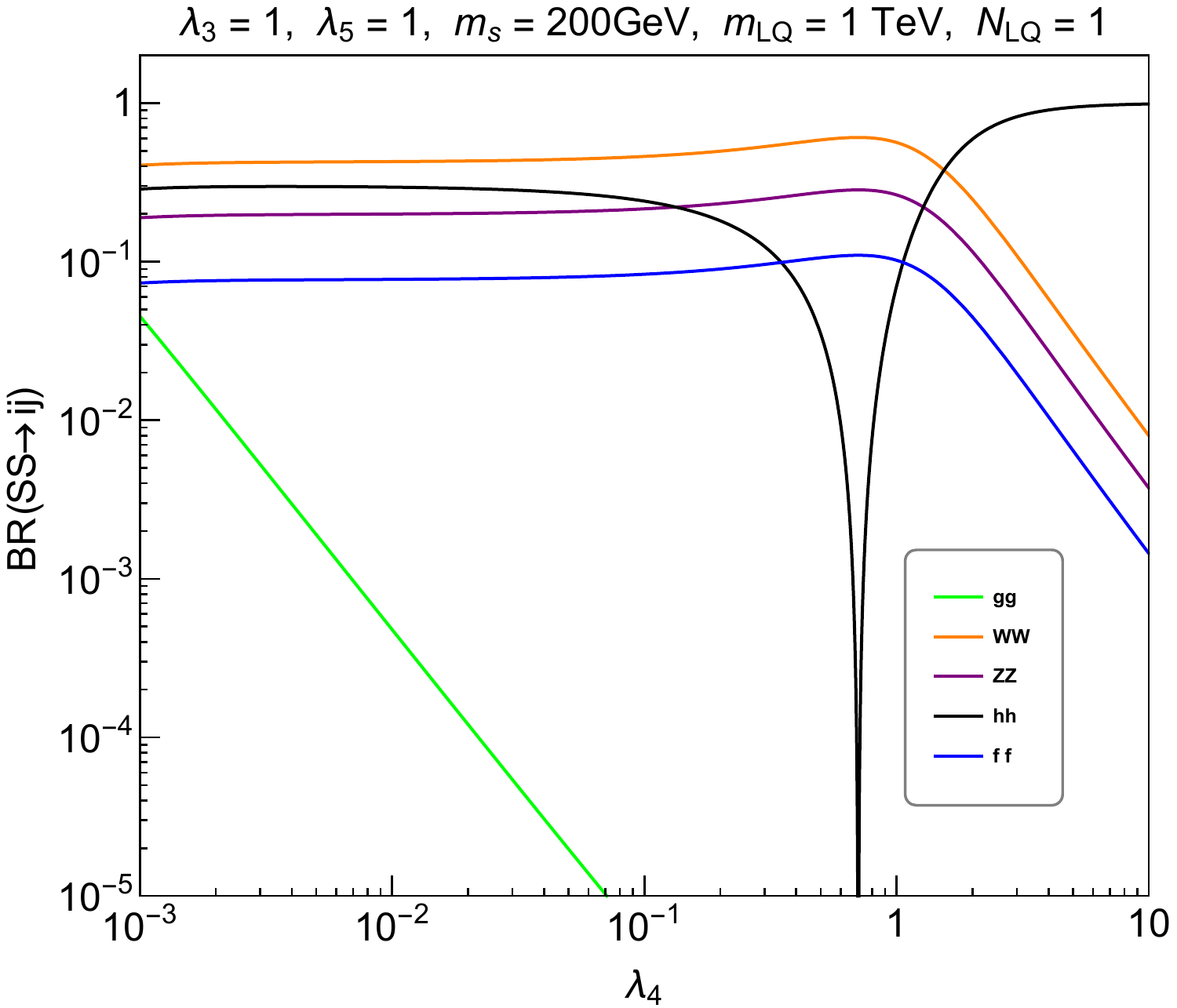}
        \includegraphics[height=0.35\textwidth]{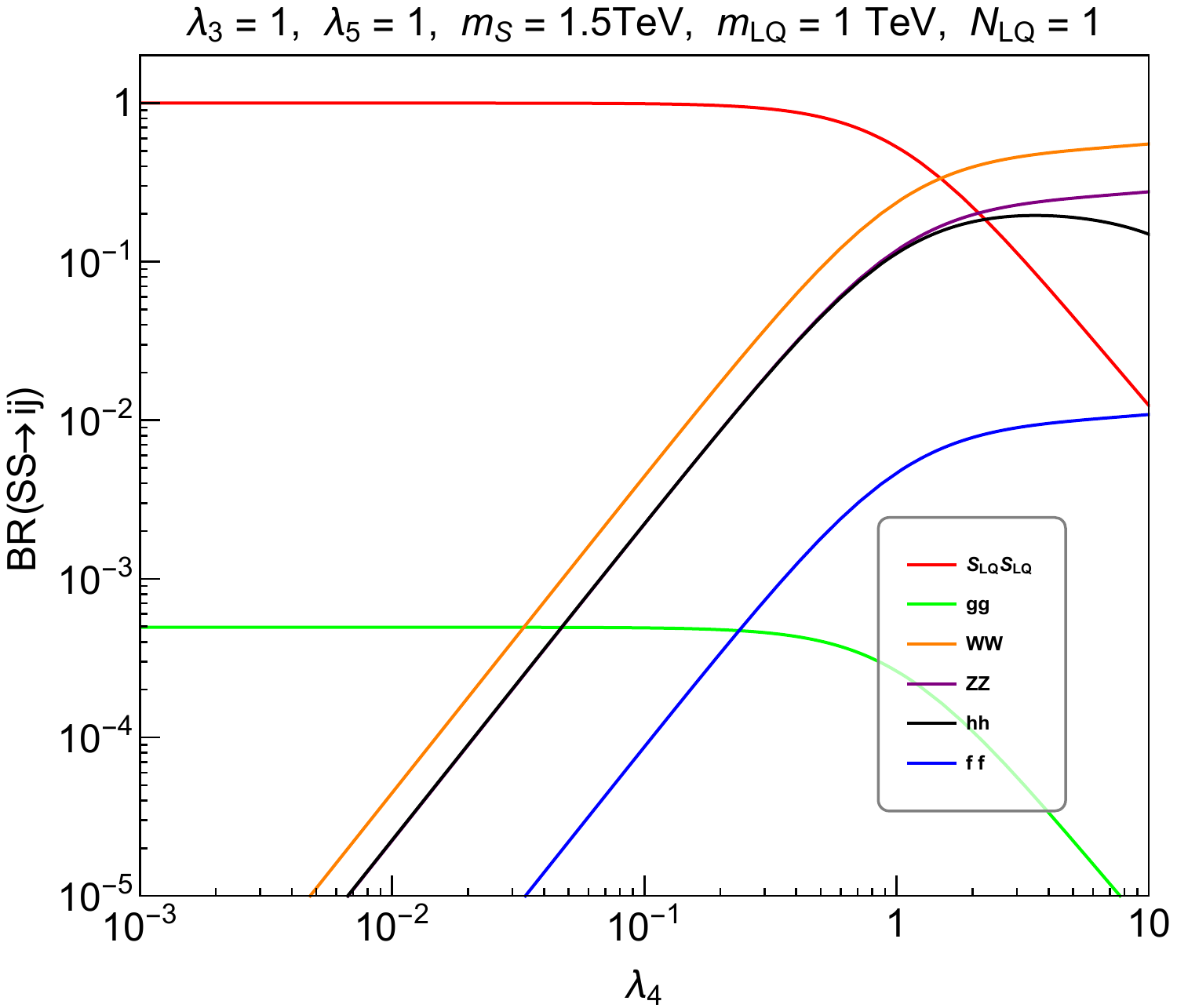} \vspace{0.5cm} \\
         \includegraphics[height=0.35\textwidth]{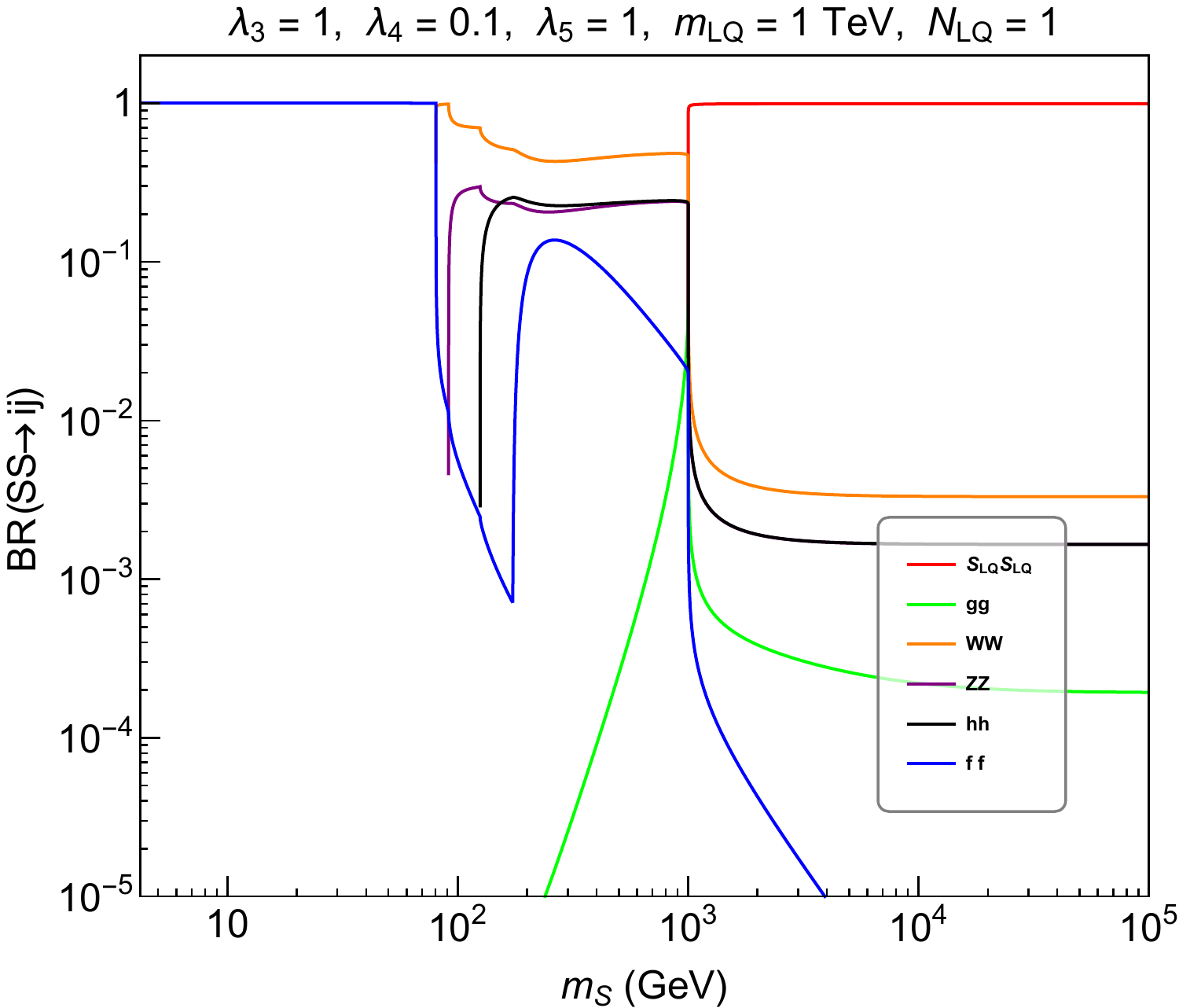}
        \includegraphics[height=0.35\textwidth]{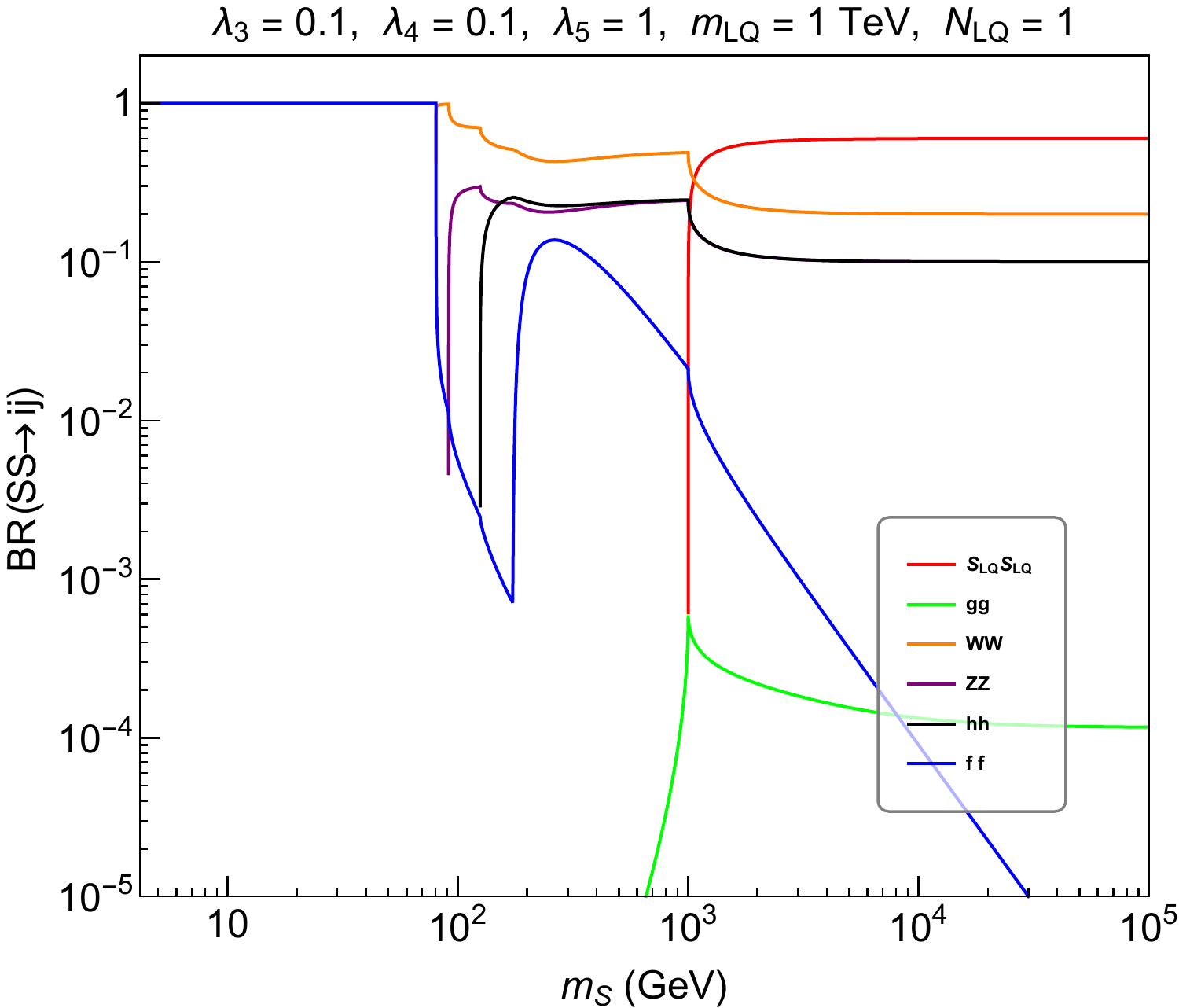}
  \end{center}
  \caption{Branching ratios of annihilation cross sections for dark matter as a function of $\lambda_4$ (upper panel) or $m_S$ (lower panel), in models with singlet scalar leptoquark. Branching ratios for $WW$({\color{orange} orange}), $ZZ$ ({\color{purple}purple}), $gg$ ({\color{green}green}), $hh$ (black), $f{\bar f}$ ({\color{blue}blue}),  and $S_{LQ} S^*_{LQ}$ ({\color{red}red}) channels are shown. }
  \label{BRa}
\end{figure}

 \begin{figure}
  \begin{center}
    \includegraphics[height=0.35\textwidth]{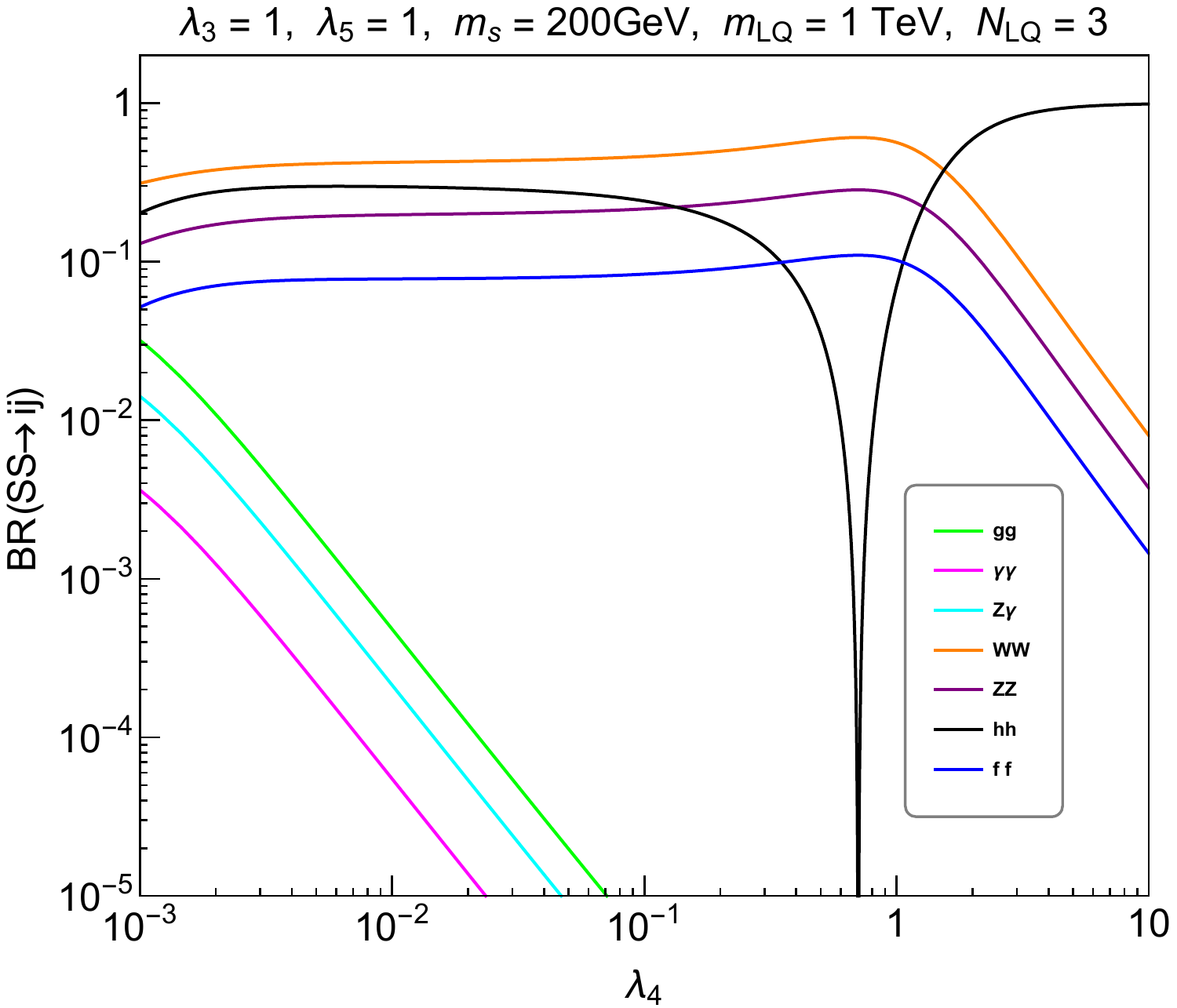}
        \includegraphics[height=0.35\textwidth]{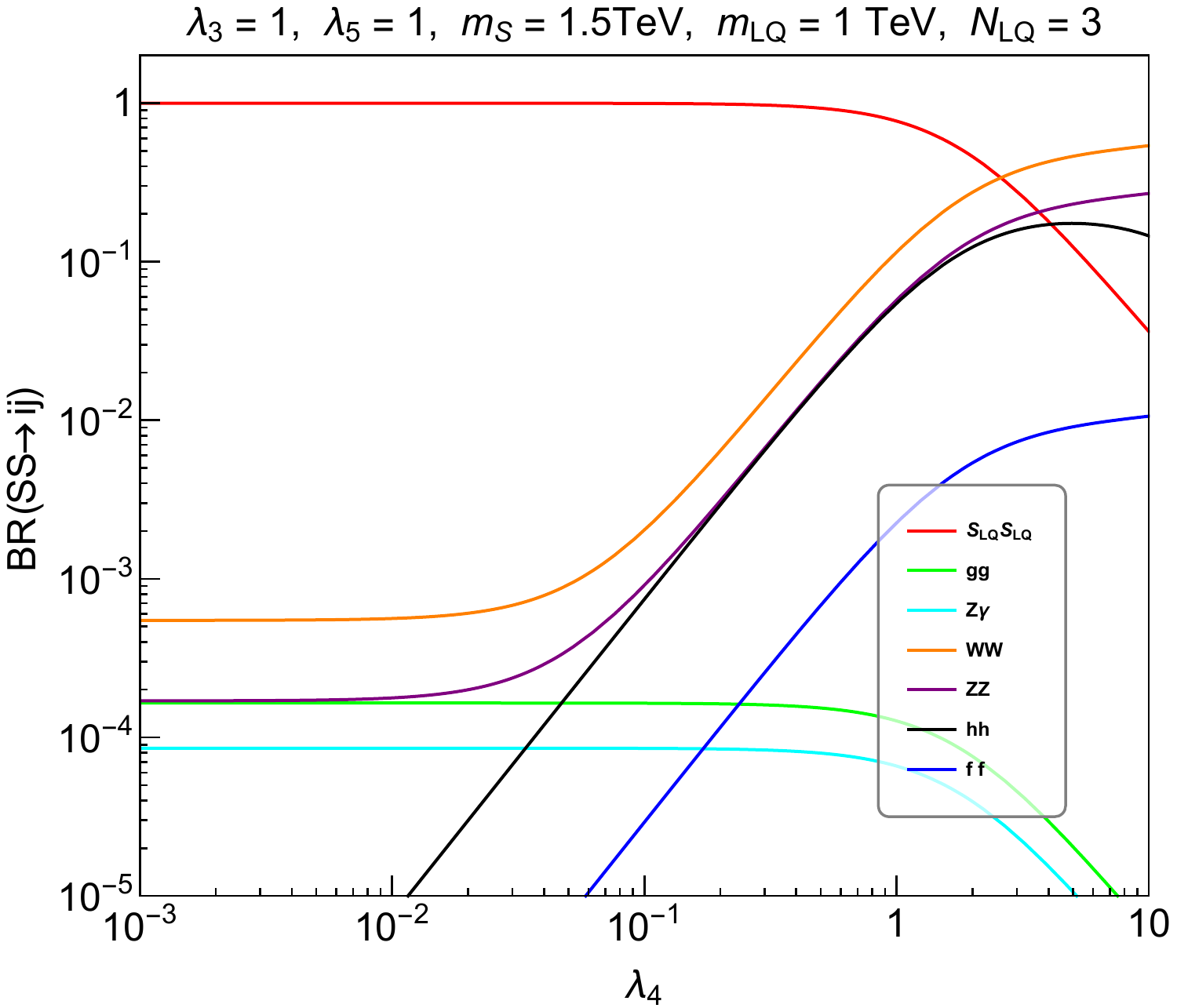} \vspace{0.5cm} \\
         \includegraphics[height=0.35\textwidth]{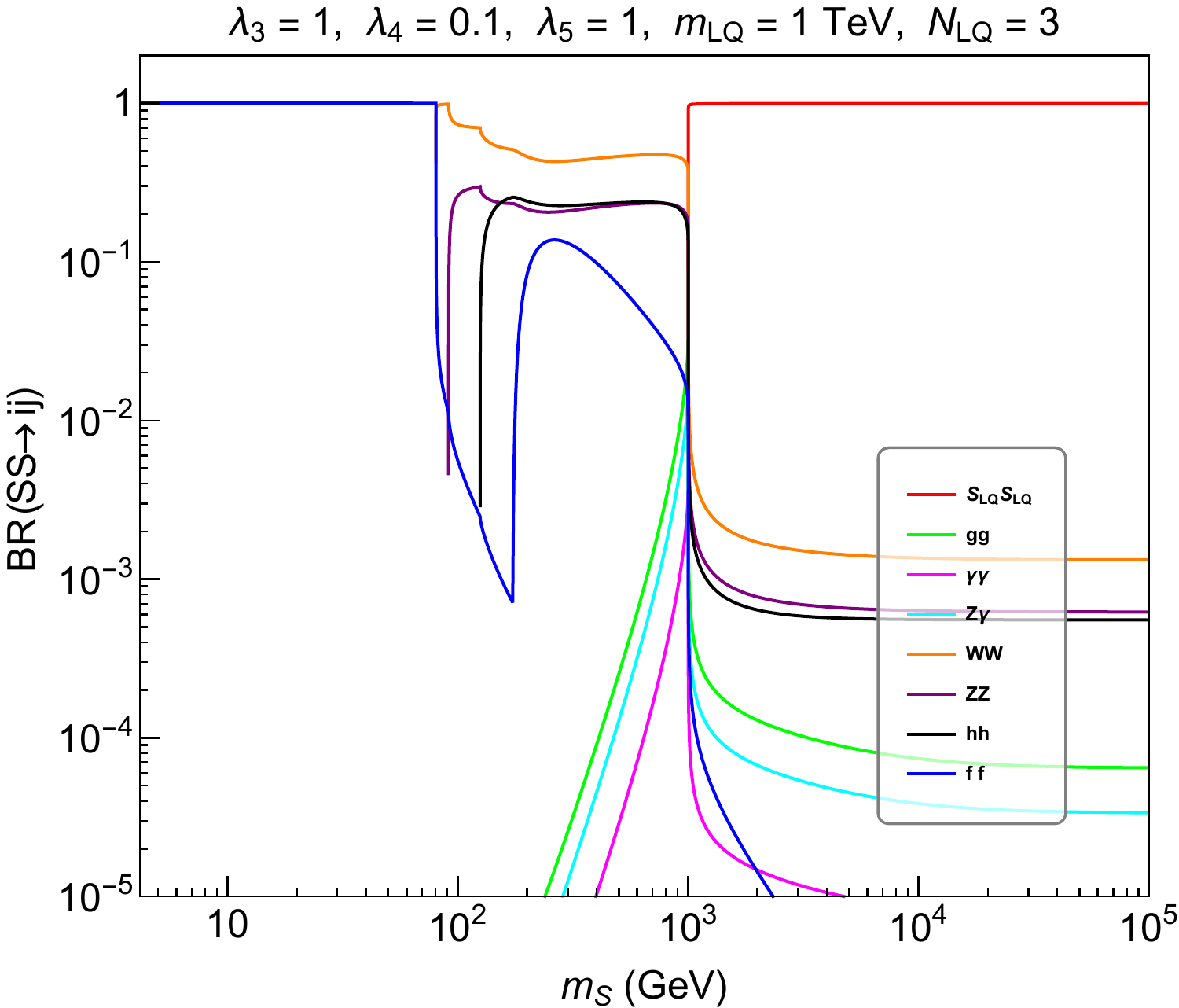}
        \includegraphics[height=0.35\textwidth]{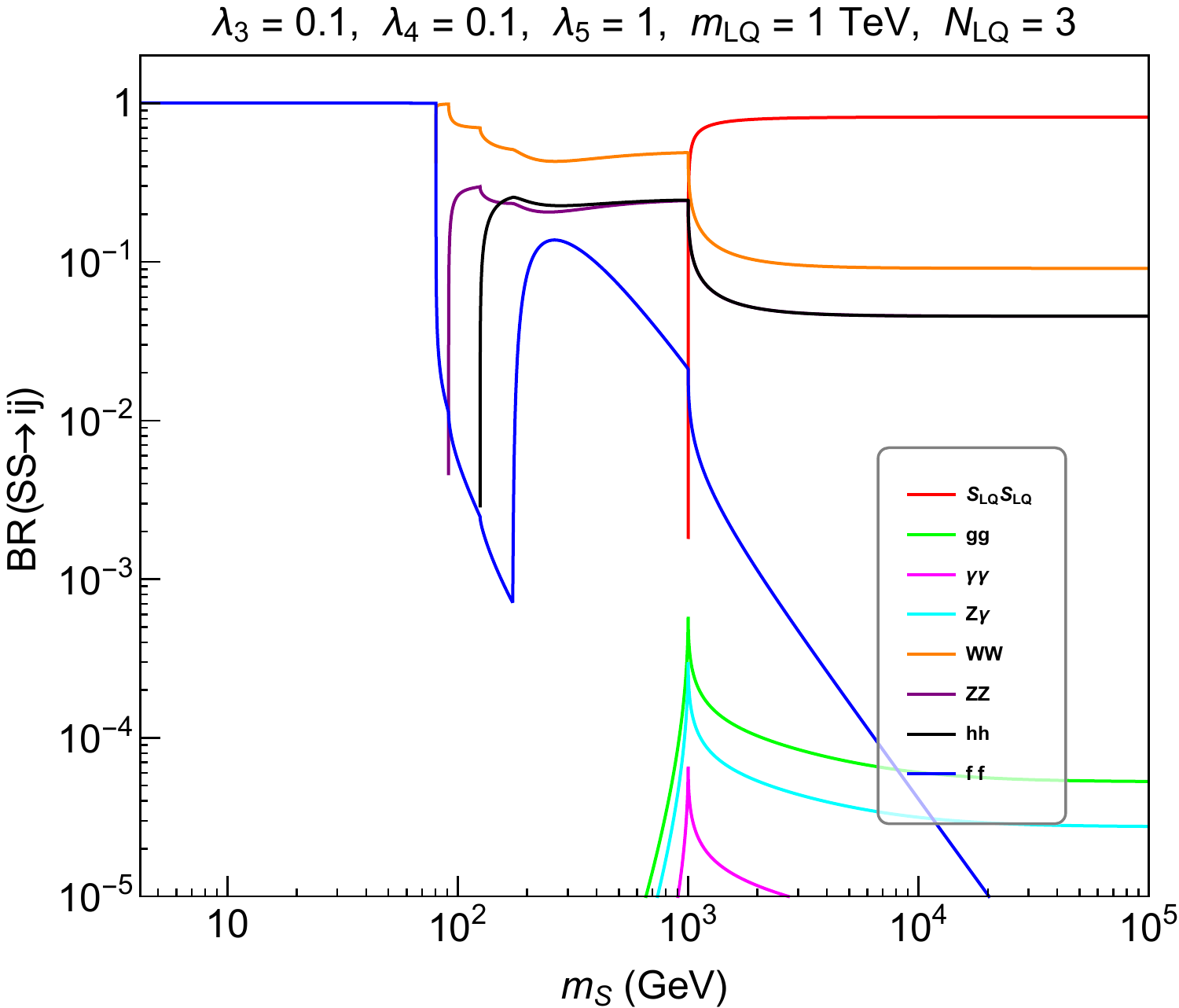}
  \end{center}
  \caption{The same as in Fig.~\ref{BRa}, but in models with triplet scalar leptoquark. }
  \label{BRb}
\end{figure}

In Figs.~\ref{BRa} and ~\ref{BRb}, we show the branching ratios of annihilation cross sections of dark matter, ${\rm BR}(SS\rightarrow ij)$, as a function of $\lambda_4$ in the upper panel and $m_S$ in the lower panel, for singlet and triplet scalar leptoquarks, respectively.

For light dark matter with $m_S<m_{LQ}$, we find that the tree-level annihilation processes such as $WW, hh, f{\bar f}, ZZ$ are dominant and the loop-induced processes due to leptoquarks are suppressed, except the $gg$ channel, which can be as large as $1-10\%$ of the total annihilation cross section, depending on whether the scalar leptoquark is singlet or triplet. In the case of triplet scalar leptoquark, the $Z\gamma, \gamma\gamma$  channels can be as large as $1\%$ or $0.1\%$ of the total annihilation cross section, so they could be probed by Fermi-LAT \cite{Fermiline} or HESS \cite{HESSline} line searches.
On the other hand, for heavy dark matter with $m_S>m_{LQ}$, the $S_{LQ}S^*_{LQ}$ channel becomes dominant while the other tree-level processes are negligible as far as $|\lambda_3|\gtrsim |\lambda_4|$.

\subsection{Direct detection bounds}

 \begin{figure}
  \begin{center}
    \includegraphics[height=0.38\textwidth]{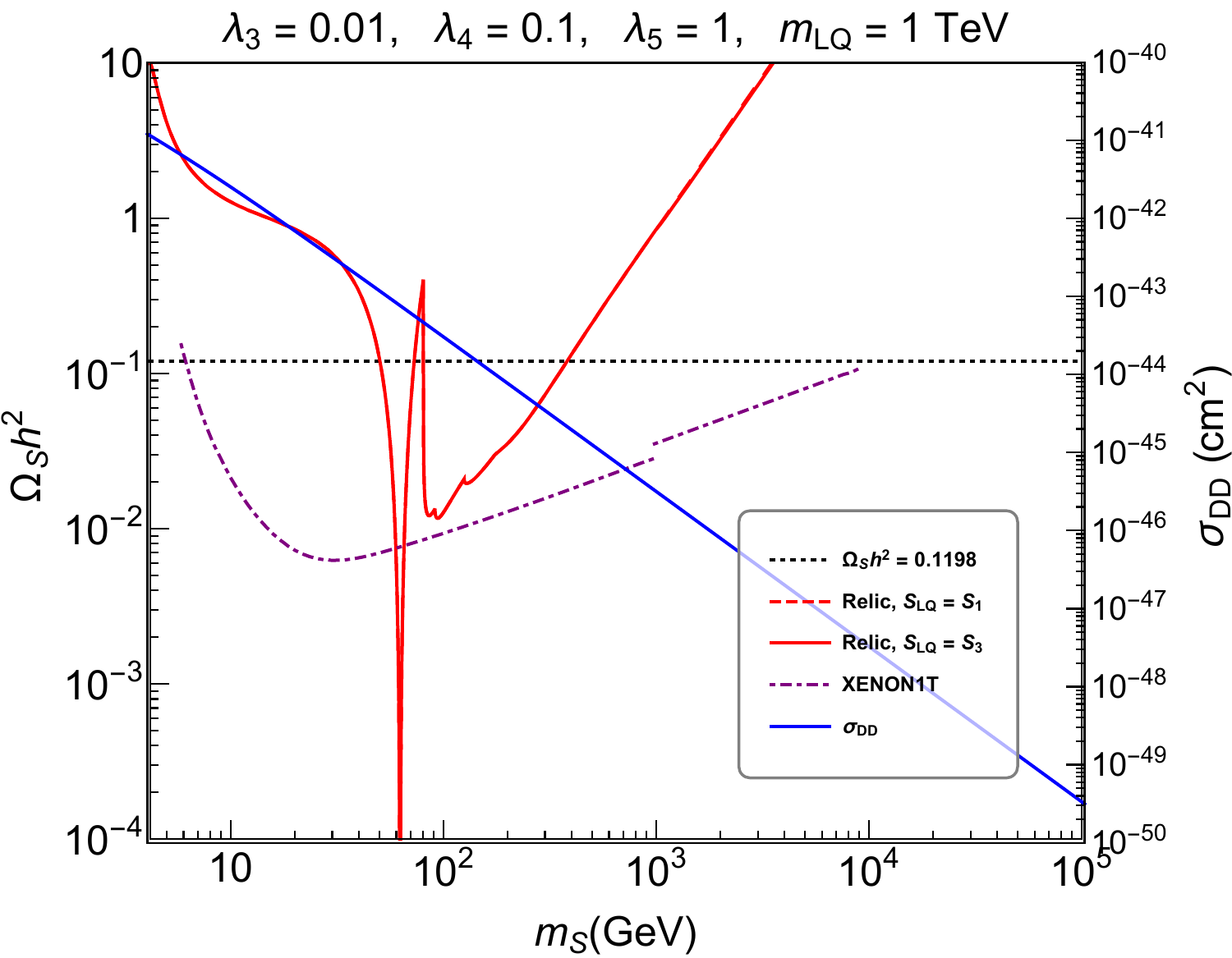}
        \includegraphics[height=0.38\textwidth]{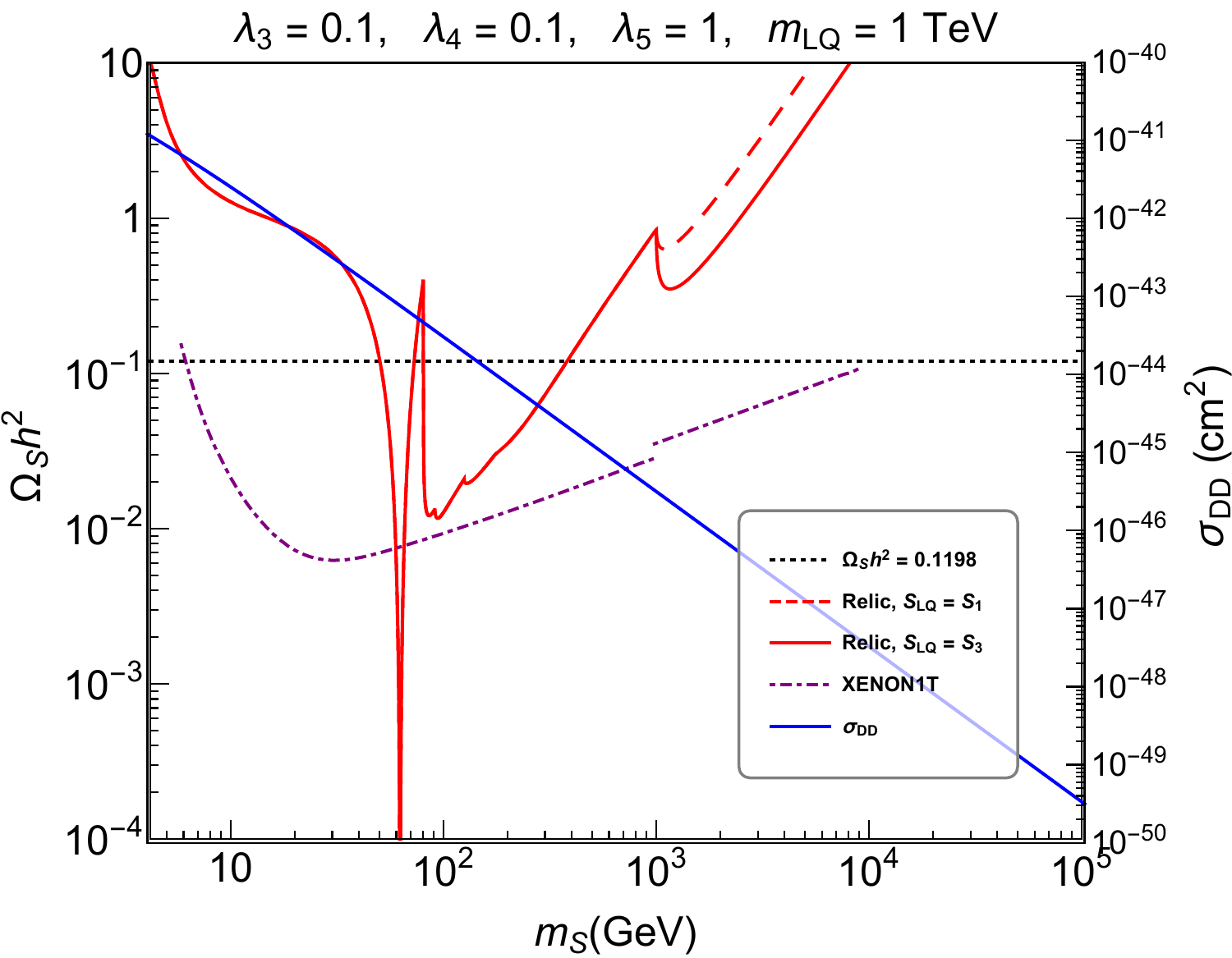}  \vspace{.3cm} \\
           \includegraphics[height=0.38\textwidth]{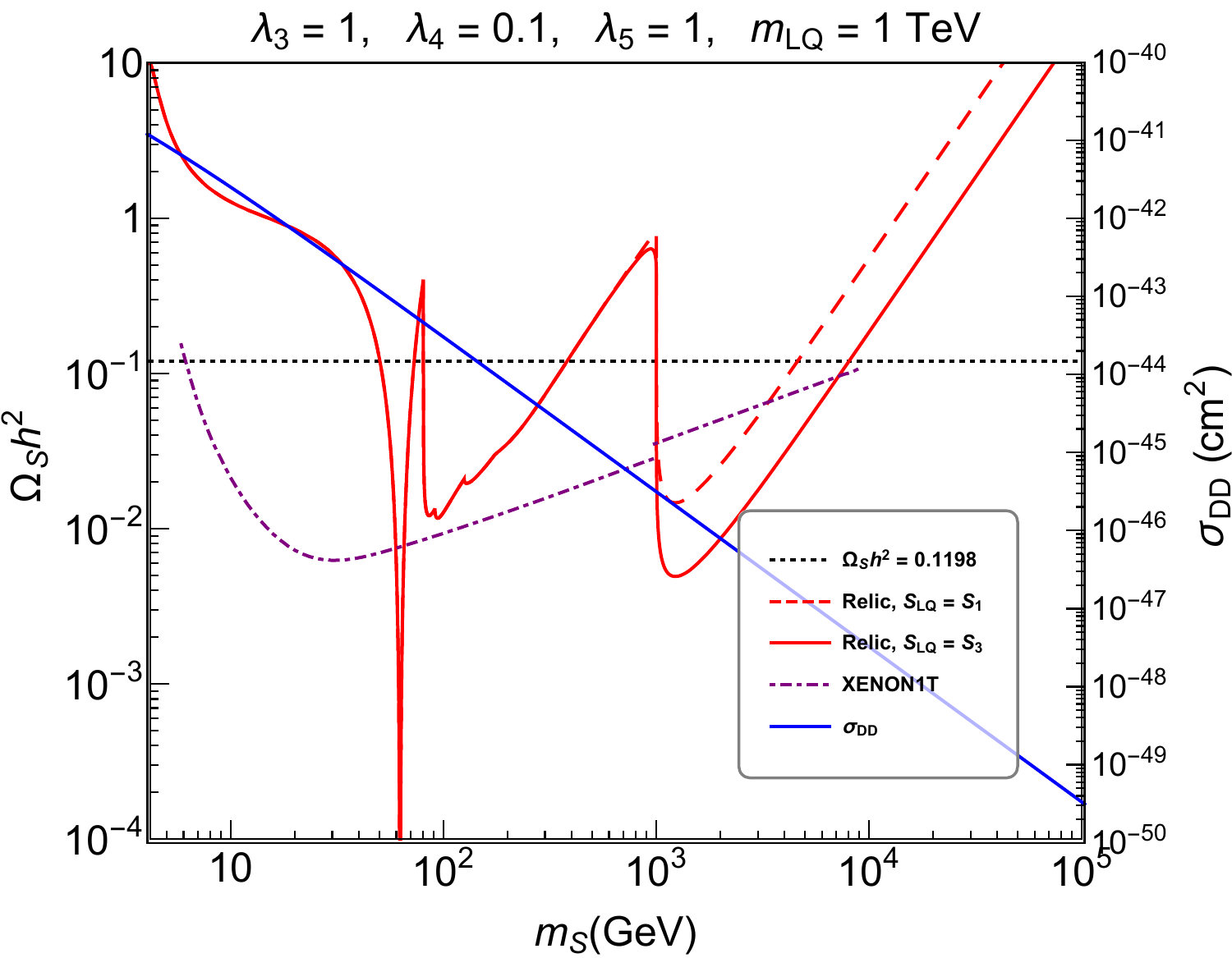}
  \end{center}
  \caption{Dark matter relic density as a function of $m_S$ in red solid (dashed) lines for triplet (singlet) scalar leptoquarks. DM-nucleon scattering cross section and XENON1T bound are shown in blue line and purple dot-dashed line, respectively.  $\lambda_3= 0.01,0.1,1$ are taken from the top left plot clockwise, and $\lambda_4=0.1$, $\lambda_5=1$ and $m_{LQ}=1\,{\rm TeV}$ are taken for all plots. }
  \label{relic}
\end{figure}

For scalar dark matter, the effective DM-quark interaction is induced due to the SM Higgs exchange at tree level, as follows,
\bea
{\cal L}_{{\rm eff}, Sq{\bar q}}= \frac{\lambda_4 m_q}{m^2_h}\, S^2 {\bar q} q. 
\eea
Moreover, taking a small momentum transfer for the DM-nucleon scattering in eq.~(\ref{effgluon}), the effective interactions between scalar dark matter and gluons, generated by loop corrections with leptoquarks, become
\bea
{\cal L}_{{\rm eff}, Sgg}= \frac{\alpha_S\lambda_4}{96\pi m^2_{\rm LQ}}\, l_3(S_{LQ})\, S^2 G_{\mu\nu}G^{\mu\nu}
\eea
where $l_3(S_{LQ})$ is the Dynkin index of a leptoquark $S_{LQ}$ under $SU(3)_C$.
Then, the spin-independent cross section for DM-nucleon elastic scattering
is given by
\bea
\sigma_{S-N}=\frac{\mu^2_N}{\pi m^2_S A^2} \Big( Z f_p + (A-Z) f_n \Big)^2 
\eea
where $Z, A-Z$ are the numbers of protons and neutrons in the detector nucleus, $\mu_N=m_N m_S/(m_N+m_S)$ is the reduced mass of DM-nucleon system, and
\bea
f_{p,n}=\frac{\lambda_4 m_{p,n} }{m^2_h}\Bigg(  \sum_{q=u,d,s} f^{p,n}_{Tq} +\frac{2}{9} f^{p,n}_{TG}\Bigg) -\frac{\lambda_3m_{p,n}}{108 m^2_{LQ}}\,l_3(S_{LQ})\, f^{p,n}_{TG}
\eea
with $f^{p,n}_{TG}=1-\sum_{q=u,d,s} f^{p,n}_{Tq}$.
Here, the mass fractions are  $f^p_{T_u}=0.023$, $f^p_{T_d}=0.032$ and $f^p_{T_s}=0.020$ for a proton and $f^n_{T_u}=0.017$, $f^n_{T_d}=0.041$ and $f^n_{T_s}=0.020$ for a neutron \cite{hisano}. 
Therefore, the quartic coupling $\lambda_4$  between scalar dark matter and SM Higgs is strongly constrained by direct detection experiments such as XENON1T \cite{xenon1t}.
Consequently, tree-level annihilations of scalar dark matter into $hh, f{\bar f}, WW, ZZ$ are constrained, while the leptoquark-induced annihilations at tree or loop levels can be relevant.

In Fig.~\ref{relic}, we show the DM relic density as a function of DM mass in red solid(dashed) lines for triplet(singlet) scalar leptoquarks. We also show the DM-nucleon scattering cross section in blue lines as can be read from the right vertical axis, and the XENON1T bound in purple dot-dashed lines. 
We find that the extra annihilation of dark matter into a pair of leptoquarks opens a new parameter space at $m_S>m_{LQ}$ due to a sizable leptoquark portal coupling, $\lambda_3$, avoiding the direct detection bound from XENON1T.

\subsection{Indirect detection bounds}

\begin{figure}
  \begin{center}
    \includegraphics[height=0.40\textwidth]{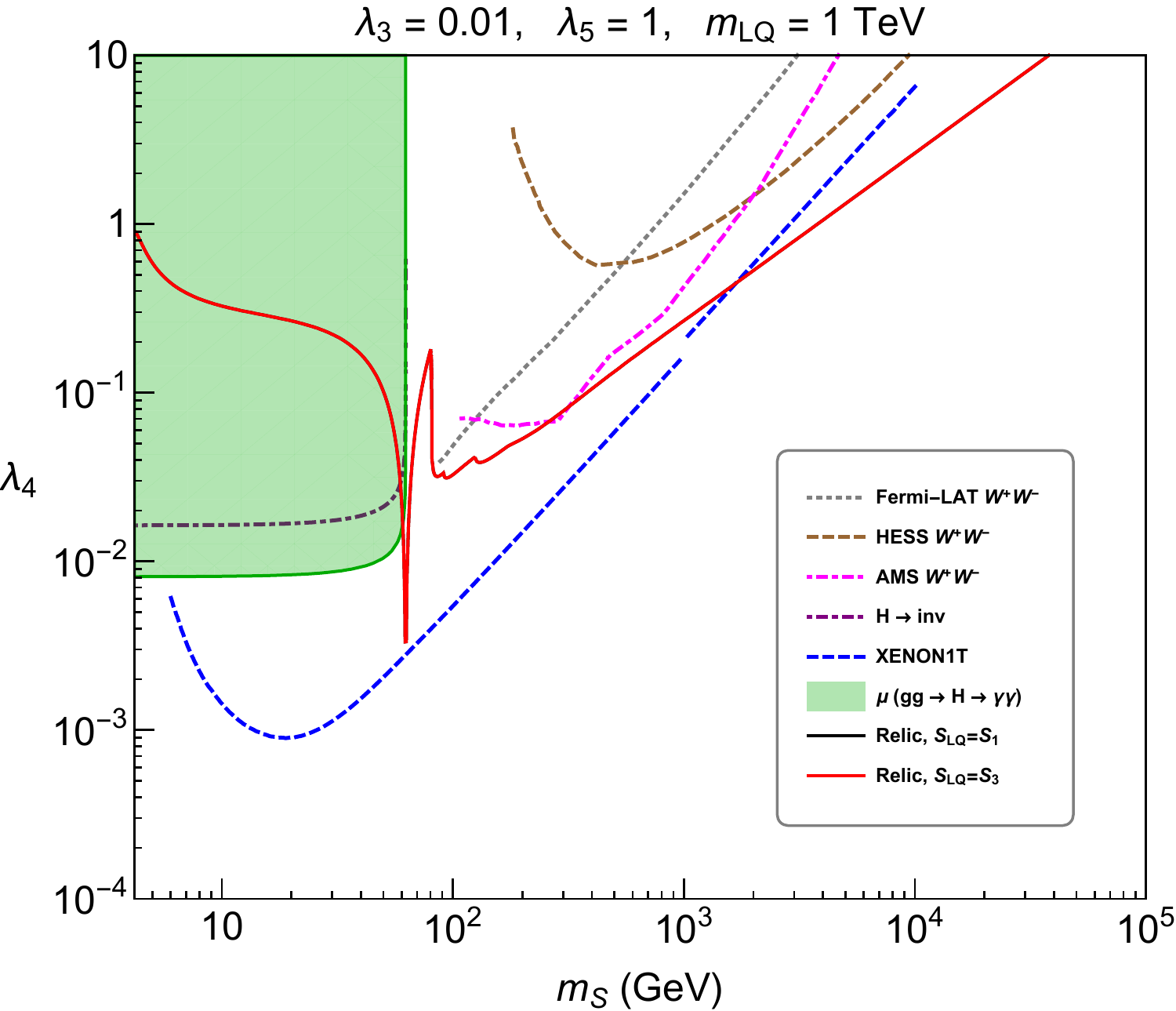}
     \includegraphics[height=0.40\textwidth]{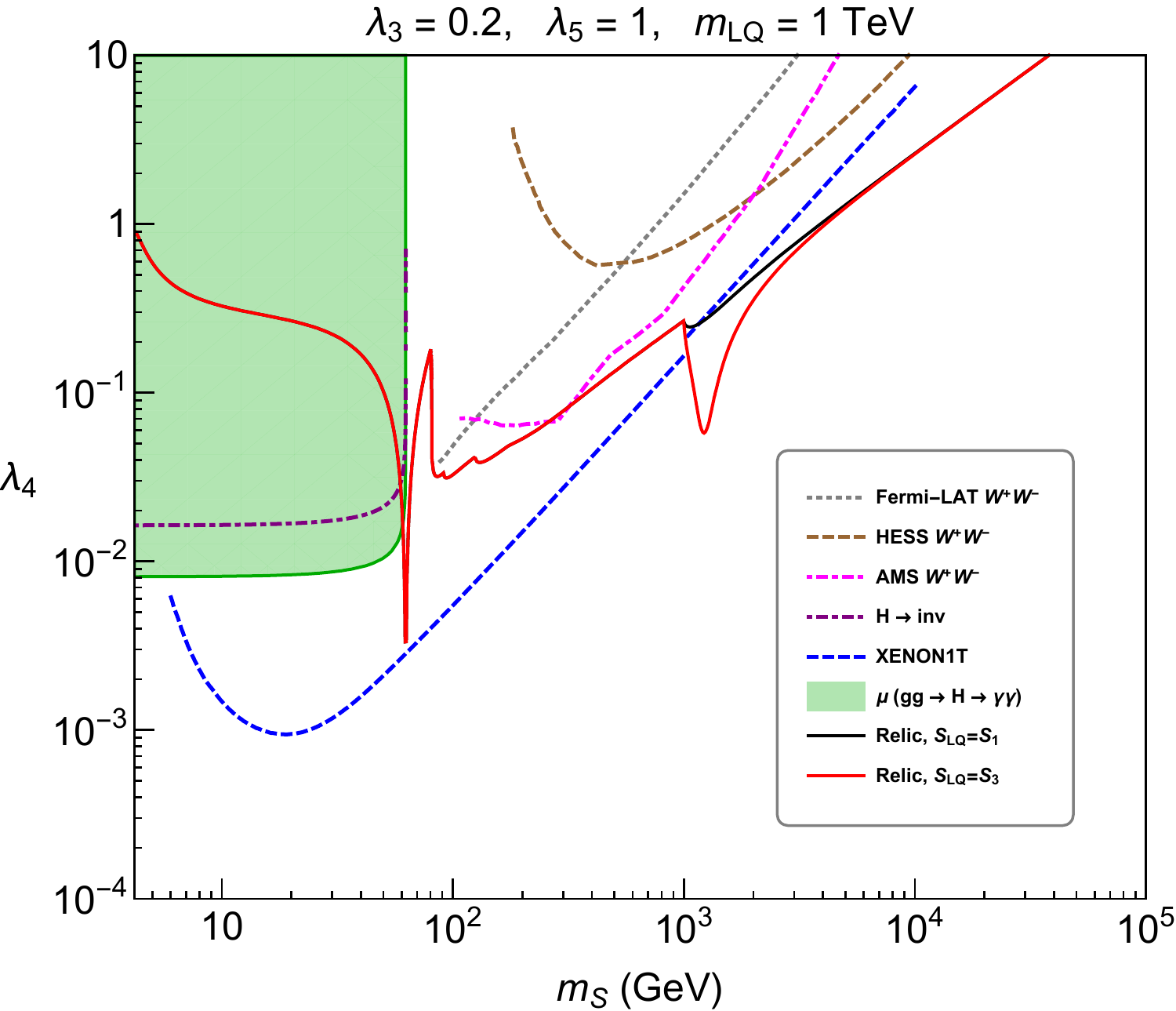}  \vspace{0.5cm} \\
      \includegraphics[height=0.40\textwidth]{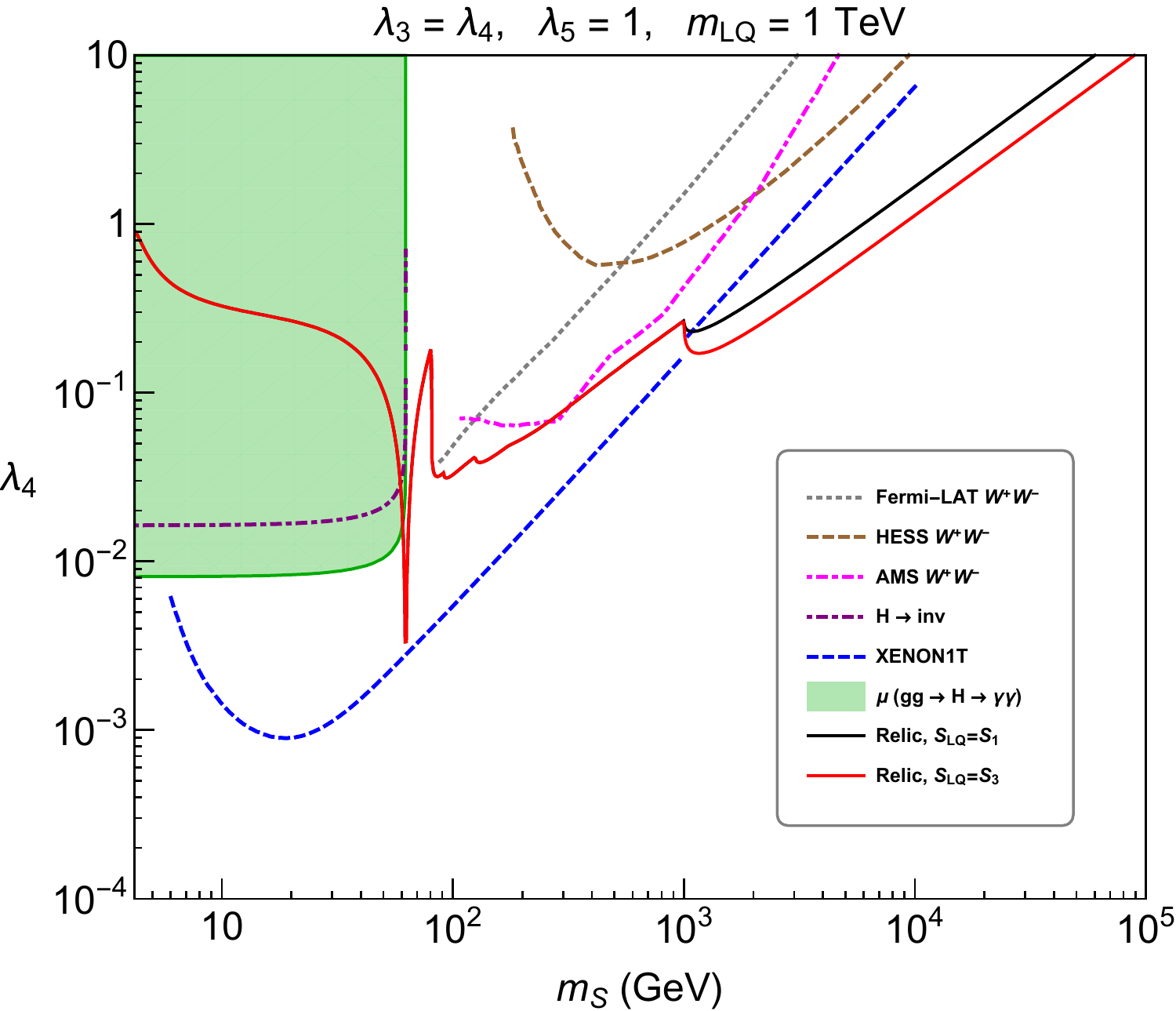}
        \includegraphics[height=0.40\textwidth]{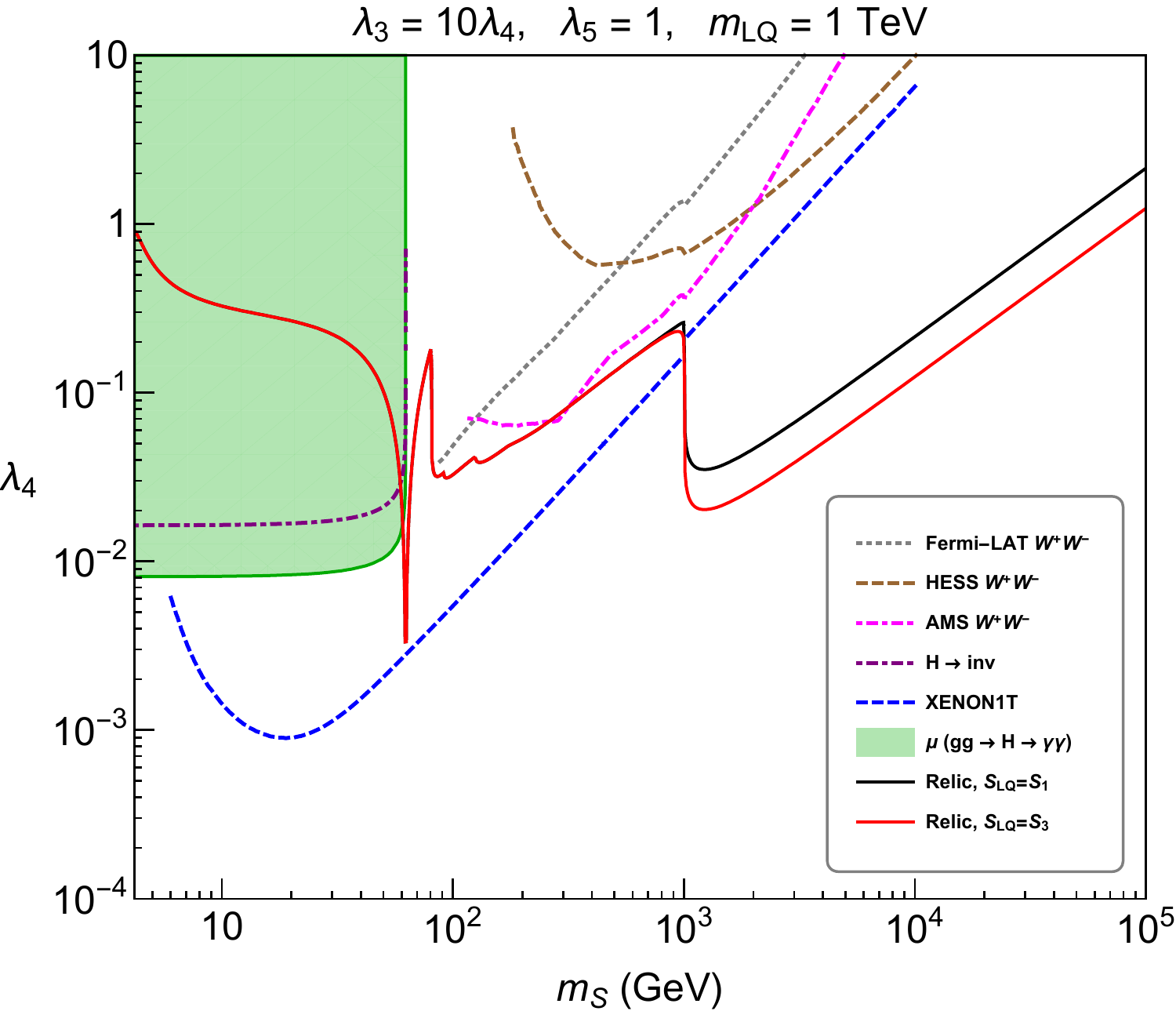}
  \end{center}
  \caption{Relic density for scalar dark matter and various bounds in the parameter space, $\lambda_4$ vs $m_S$. The correct relic density can be obtained along the black and red solid lines, for models with singlet and triplet scalar leptoquarks, respectively. XENON1T bounds are shown in blue dashed lines. Indirect detection constraints from gamma-ray searches in Fermi-LAT (gray dotted) and HESS (brown dashed), and antiproton search in AMS-02 (pink dot-dashed) are overlaid. The bound from Higgs invisible decay is shown in purple dot-dashed line and the green regions are excluded by visible decays such as the Higgs diphoton signal strength.  }
  \label{DD1}
\end{figure}

For relatively light scalar dark matter with $m_S<m_{LQ}$, the DM annihilation cross sections into $hh,WW,ZZ,t{\bar t},b{\bar b}$ are dominant. In this case, Fermi-LAT dwarf galaxies \cite{Fermi} and HESS  gamma-rays \cite{HESS} and AMS-02 antiprotons \cite{AMS} can constrain the model.

In Fig.~\ref{DD1}, we depict the parameter space in $\lambda_4$ vs $m_S$ in black and red solid lines, satisfying the correct relic density for models with singlet and triplet scalar leptoquarks, respectively.
In the same plots, we superimpose the indirect detection bounds on the DM annihilations into a $WW$ pair from Fermi-LAT and HESS gamma-ray and AMS-02 anti-proton searches, and include the direct detection bounds from XENON1T. Moreover, the region with $m_S<m_h/2$ can be also constrained by Higgs data such as Higgs invisible decay and the signal strength for $gg\rightarrow h\rightarrow \gamma\gamma$, as will be discussed in the next subsection.

As a result, the Higgs data as well as indirect detection constrains the region with light and weak-scale dark matter, but the XENON1T experiment constrains most, ruling out most of the DM masses below $m_S=1\,{\rm TeV}$, except the resonance region near $m_S=m_h/2$. However, we find that the correct relic density can be obtained for a small value of $\lambda_4$ due to the contribution of DM annihilation channels into a leptoquark pair with a sizable leptoquark-portal coupling $\lambda_3$ for $m_S>m_{LQ}$. Therefore, there is a wide parameter space above $m_S=1\,{\rm TeV}$ that is consistent with the XENON1T bound.

We remark the indirect signatures of dark matter and the leptoquark decays in the case of heavy scalar dark matter. 
For  $m_S>m_{LQ}$, dark matter can annihilate into a pair of leptoquarks, each of which decays into a pair of quark and lepton in cascade. In the case with leptoquark couplings to explain the $B$-meson anomalies, the branching ratios of final products of DM annihilations are shown in Table \ref{ABR}, depending on the decay branching ratios of leptoquarks.
In the case where the extra leptoquark couplings, $\lambda_{32}$, $\kappa_{23}$ and $\kappa_{33}$, introduced for accommodating $B\rightarrow K^{(*)}{\bar\nu}\nu$ bounds and/or the $(g-2)_\mu$ excess, are dominant, as discussed in Section 4.1, we also show  the corresponding branching ratios of final products of DM annihilations in Table \ref{ABR2}.

\begin{table}[h!]\small
\begin{center}
\begin{tabular}{|c||c|c|c|}
\hline
 {\rm LQs} & ${\rm BRs}$ &  ${\rm BRs}$  & ${\rm BRs}$ \\ [0.5ex]
 \hline
$S_1S^*_1$ & ${\rm B}(|{\bar t}{\bar \tau}+b \nu_\tau|^2)$  & ${\rm B}(|{\bar c}{\bar\tau}+ s\nu_\tau|^2)$ & ${\rm B}(({\bar t}{\bar \tau}+b\nu_\tau)^*({\bar c}{\bar\tau}+s\nu_\tau)+{\rm h.c.})$   \\ [0.5ex]
 &  $=\beta^2$ & $=(1-\beta)^2$ &  $=2\beta(1-\beta)$  \\ [0.5ex]
 \hline
$\phi_1 \phi^*_1$ &  ${\rm B}({\bar b}b{\bar\mu} \mu)$ & ${\rm B}({\bar s}s{\bar\mu} \mu)$ & ${\rm B}({\bar b}s{\bar\mu}\mu+{\rm h.c.})$   \\ [0.5ex]
 &  $=\gamma^2$ & $=(1-\gamma)^2$ &  $=2\gamma(1-\gamma)$  \\ [0.5ex]
  \hline
$\phi_2\phi^*_2$ &  ${\rm B}(|{\bar t}{\bar\mu}+b \nu_\mu|^2)$ & ${\rm B}(|{\bar c}{\bar\mu}+s \nu_\mu|^2)$  &  ${\rm B}(({\bar t}{\bar\mu}+b \nu_\mu)^*({\bar c}{\bar\mu}+s\nu_\mu)+{\rm h.c.})$   \\ [0.5ex] 
 &  $=\gamma^2$ & $=(1-\gamma)^2$ &  $=2\gamma(1-\gamma)$  \\ [0.5ex]
 \hline
$\phi_3\phi^*_3$ & ${\rm B}({\bar t}t {\bar\nu}_\mu  \nu_\mu)$ & ${\rm B}({\bar c}c{\bar\nu}_\mu \nu_\mu)$  & ${\rm B}({\bar t}c{\bar\nu}_\mu \nu_\mu+{\rm h.c.})$
\\ [0.5ex]
 &  $=\gamma^2$ & $=(1-\gamma)^2$ &  $=2\gamma(1-\gamma)$  \\ [0.5ex]
\hline 
\end{tabular}
\caption{Branching ratios of products of DM annihilations into leptoquarks. Here, $\beta\equiv \lambda^2_{33}/(\lambda^2_{33}+\lambda^2_{23})$ and $\gamma\equiv  \kappa^2_{32}/(\kappa^2_{32}+ \kappa^2_{22})$. }
\label{ABR}
\end{center}
\end{table}

In particular, for $m_S\gtrsim m_{LQ}$, a leptoquark pair is produced with almost zero velocities, so each leptoquark decays into a pair of quark and lepton such as ${\bar q}\,{\bar l}$ or $q'\, l'$, back-to-back. In this case, a pair of two quarks ($q'{\bar q}$) or a pair of leptons ($l'{\bar l}$) carry about the energy of DM mass, so we take them as if they are produced from the direct annihilations of dark matter with mass $m_S/2$ and impose the indirect detection bounds on the annihilation cross section.
But, if $m_S\gg m_{LQ}$, leptoquarks produced from the DM annihilations are boosted so the full energy spectra for quarks or leptons carry the energy spectra of wide box rather than a monochromatic energy. In this case, we need to take more care before imposing the indirect detection bounds. Henceforth, ignoring the boost effects of leptoquarks, in particular, for $m_{S}\gtrsim m_{LQ}$, we discuss the indirect detection bounds for the direct annihilations of dark matter to cascade annihilations.

\begin{figure}
  \begin{center}
    \includegraphics[height=0.40\textwidth]{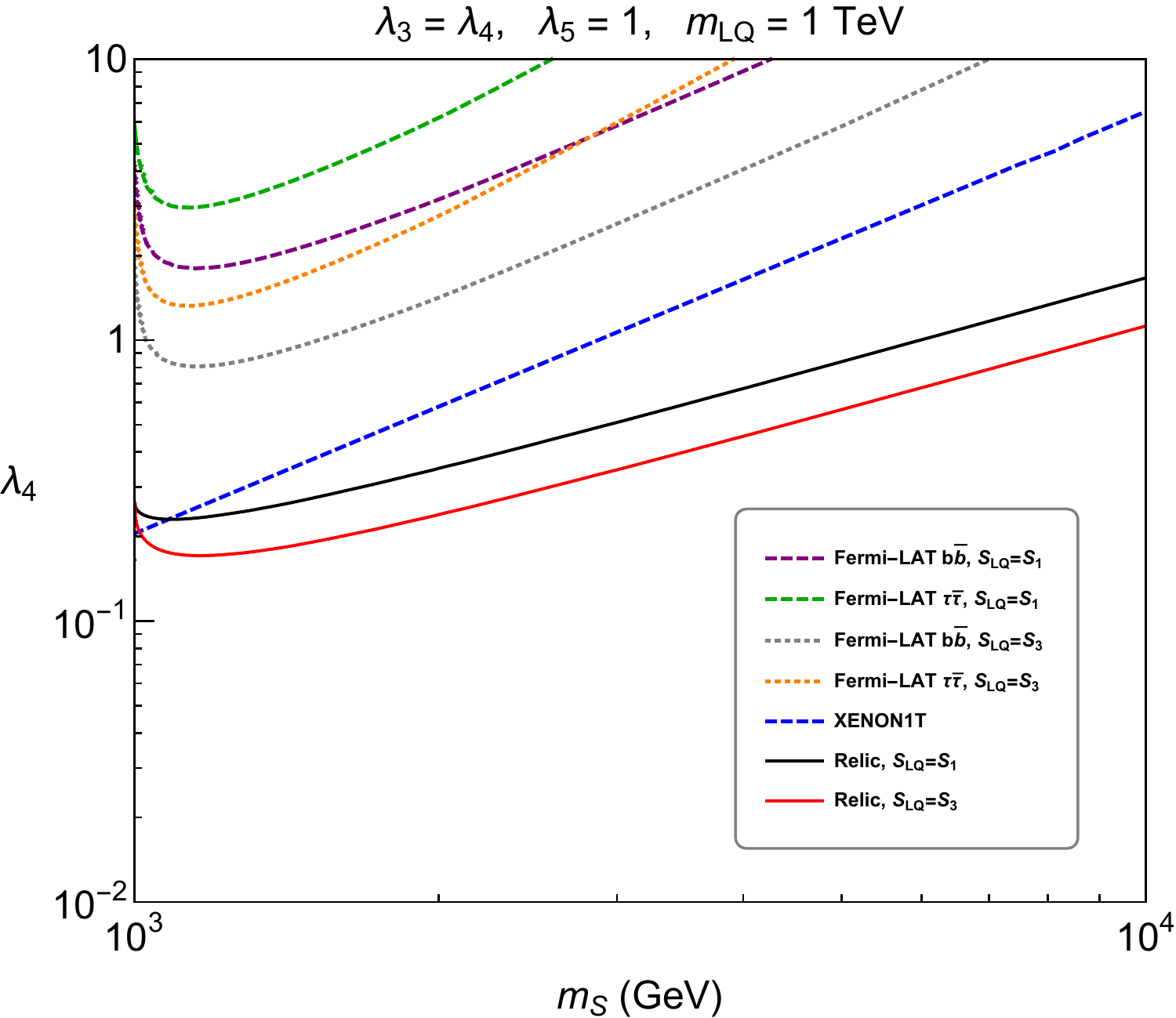}
     \includegraphics[height=0.40\textwidth]{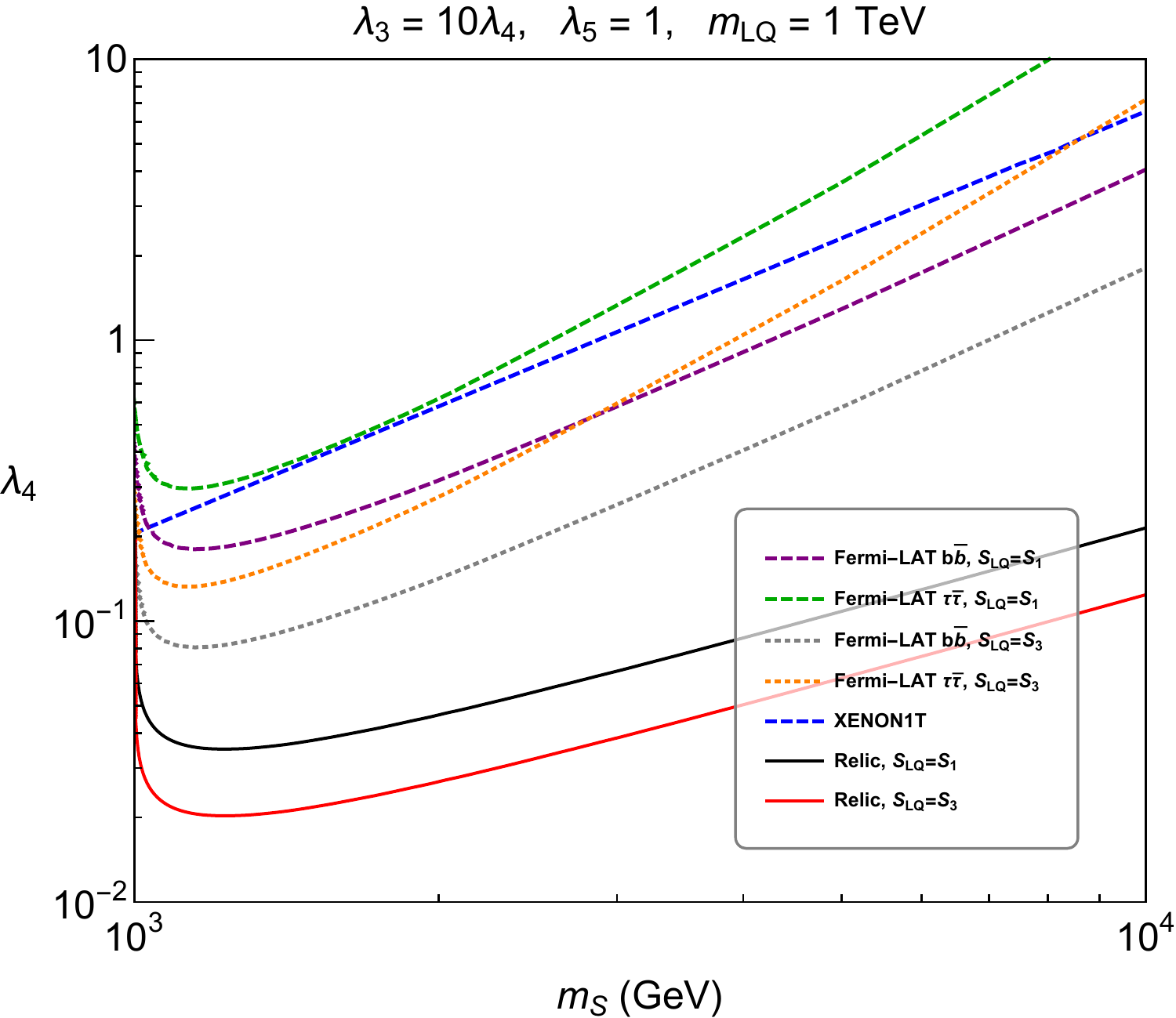}  
  \end{center}
  \caption{Relic density for scalar dark matter and various bounds in the parameter space, $\lambda_4$ vs $m_S$. The correct relic density can be obtained along the black and red solid lines, for models with singlet and triplet scalar leptoquarks, respectively. XENON1T bounds are shown in blue dashed lines. Fermi-LAT gamma-ray constraints on $b{\bar b}$ and $\tau{\bar\tau}$ coming from cascade annihilations are also shown in purple dashed (gray dotted) and green dashed (orange dotted) lines for singlet (triplet) leptoquarks.   }
  \label{DD2}
\end{figure}

First, we consider the case in Table \ref{ABR} with leptoquark couplings necessary to explain the $B$-meson anomalies.
In this case, for a singlet leptoquark with $\lambda_{33}\gg \lambda_{23}$ or $\beta\approx 1$, we get the branching ratios of products of DM annihilations into leptoquarks as ${\rm B}({\bar t}t 
\,{\bar\tau}\tau): {\rm B}({\bar b}b\, {\bar \nu}_\tau \nu_\tau): {\rm B}({\bar t} b \, {\bar\tau}\nu_\tau+{\rm h.c.})=\frac{1}{2}:\frac{1}{2}:1$. Then, we can impose the Fermi-LAT diffuse gamma-ray constraints from ${\bar b}b$ and ${\bar\tau}\tau$ \cite{Fermi} on $\frac{1}{4}\langle\sigma v\rangle_{SS\rightarrow S_{LQ}S^*_{LQ}}$.  Similarly, for a triplet leptoquark with $\kappa_{32}\gg \kappa_{22}$ or $\gamma\approx 1$, we get ${\rm B}({\bar b}b\,{\bar\mu}\mu):{\rm B}({\bar t}t\, {\bar\mu}\mu):{\rm B}({\bar b}b\, {\bar\nu}_\mu \nu_\mu) : {\rm B}({\bar t}b\, {\bar\mu}\nu_\mu+{\rm h.c.}):{\rm B}({\bar t}t\, {\bar\nu}_\mu \nu_\mu)=1:\frac{1}{4}:\frac{1}{4}:\frac{1}{2}:1$. In this case, we can impose the Fermi-LAT  bounds for ${\bar b}b$ and ${\bar\mu}\mu$ \cite{Fermi} on $\frac{5}{12}\langle\sigma v\rangle_{SS\rightarrow S_{LQ}S^*_{LQ}}$ too. 
In general, positron, anti-proton and gamma-ray constraints are equally relevant for leptoquark-portal dark matter. 

In Fig.~\ref{DD2}, in the parameter space in $\lambda_4$ vs $m_S$, in addition to the correct relic density conditions for models with singlet and triplet scalar leptoquarks, respectively, in black and red solid lines and the direct detection bounds from XENON1T, we superimpose the Fermi-LAT constraints from $b{\bar b}$ and $\tau{\bar\tau}$ on the products of DM cascade annihilations into a leptoquark pair. Here, we assume that each leptoquark decays into a pair of quark and lepton, according to Table 2 with $\beta\approx 1$ and $\gamma\approx 1$.  Then, as explained in the caption of Fig.~\ref{DD2}, the resulting Fermi-LAT bounds are shown to constrain the parameter space as strong as or stronger than the XENON1T bounds, depending on the value of leptoquark-portal coupling $\lambda_3$.

\begin{table}[h!]\small
\begin{center}
\begin{tabular}{|c||c|c|c|}
\hline
 {\rm LQs} & ${\rm BRs}$ &  ${\rm BRs}$  & ${\rm BRs}$ \\ [0.5ex]
 \hline
$S_1S^*_1$ & ${\rm B}({\bar t}t {\bar\mu}\mu)=\frac{1}{4}$  & ${\rm B}({\bar b}b{\bar\nu}_\mu \nu_\mu)=\frac{1}{4}$ & ${\rm B}({\bar t}b {\bar\mu}\nu_\mu+{\rm h.c.})=\frac{1}{2}$   \\ [0.5ex]
 \hline
$\phi_1 \phi^*_1$ &  ${\rm B}({\bar b}b{\bar\tau} \tau)=\gamma^{\prime 2}$ & ${\rm B}({\bar s}s{\bar\tau} \tau)=(1-\gamma')^2$ & ${\rm B}({\bar b}s{\bar\tau}\tau+{\rm h.c.})=2\gamma'(1-\gamma')$   \\ [0.5ex]
  \hline
$\phi_2\phi^*_2$ &  ${\rm B}(|{\bar t}{\bar\tau}+b\nu_\tau|^2)$ & ${\rm B}(|{\bar c}{\bar\tau}+s \nu_\tau|^2)$  &  ${\rm B}(({\bar t}{\bar\tau}+b \nu_\tau)^*({\bar c}{\bar\tau}+s \nu_\tau)+{\rm h.c.})$   \\ [0.5ex] 
 &  $=\gamma^{\prime 2}$ & $=(1-\gamma')^2$ &  $=2\gamma'(1-\gamma')$  \\ [0.5ex]
 \hline
$\phi_3\phi^*_3$ & ${\rm B}({\bar t}t {\bar\nu}_\tau  \nu_\tau)=\gamma^{\prime 2}$ & ${\rm B}({\bar c}c{\bar\nu}_\tau \nu_\tau)=(1-\gamma')^2$  & ${\rm B}({\bar t}c{\bar\nu}_\tau \nu_\tau+{\rm h.c.})=2\gamma'(1-\gamma')$
\\ [0.5ex]
\hline 
\end{tabular}
\caption{Branching ratios of products of DM annihilations into leptoquarks, for the dominance of the extra couplings, $\lambda_{32}$, $\kappa_{23}$ and $\kappa_{33}$. Here, $\gamma'\equiv  \kappa^2_{33}/(\kappa^2_{23}+ \kappa^2_{33})$. }
\label{ABR2}
\end{center}
\end{table}

Next, we consider the case  in Table \ref{ABR2} where the extra leptoquark couplings introduced for accommodating $B\rightarrow K^{(*)}{\bar\nu}\nu$ bounds and/or the $(g-2)_\mu$ excess are dominant. 
In this case, for a singlet leptoquark, we get  the branching ratios of products of DM annihilations into leptoquarks as ${\rm B}({\bar t}t 
\,{\bar\mu}\mu): {\rm B}({\bar b}b\, {\bar \nu}_\mu \nu_\mu): {\rm B}({\bar t} b \, {\bar\mu}  \nu_\mu+{\rm h.c.})=\frac{1}{2}:\frac{1}{2}:1$. Then, we can impose the Fermi-LAT diffuse gamma-ray constraints from ${\bar b}b$ and ${\bar\mu}\mu$ \cite{Fermi}  on $\frac{1}{4}\langle\sigma v\rangle_{SS\rightarrow S_{LQ}S^*_{LQ}}$ as for the case in Table 2.  Similarly, for a triplet leptoquark with $\kappa_{33}\gg \kappa_{23}$ as in the first benchmark point discussed in the last paragraph in Section 4.1 or $\gamma'\approx 1$, we get ${\rm B}({\bar b}b\,{\bar\tau}\tau):{\rm B}({\bar t}t\, {\bar\tau}\tau):{\rm B}({\bar b}b\, {\bar\nu}_\tau \nu_\tau) : {\rm B}({\bar t}b\, {\bar\tau}\nu_\tau+{\rm h.c.}):{\rm B}({\bar t}t\, {\bar\nu}_\tau \nu_\tau)=1:\frac{1}{4}:\frac{1}{4}:\frac{1}{2}:1$. In this case,  the similar Fermi-LAT  bounds for ${\bar b}b$ and ${\bar \tau}\tau$  \cite{Fermi} can be also imposed on $\frac{5}{12}\langle\sigma v\rangle_{SS\rightarrow S_{LQ}S^*_{LQ}}$  as for the case in Table 2.

In summary, the leptoquark-portal couplings lead to potentially distinct signatures with quarks and leptons mixed from the cascade annihilations of dark matter, as compared to the case with direct annihilations into a quark pair or a lepton pair. Our lepto-quark portal scenario is different from the Higgs portal scenario with additional $SU(2)_L$ singlet or triplet scalars, because the final states in the cascade DM annihilations contain quarks and leptons together due to the leptoquark decays in our case.
In other words, the region with $m_S>m_{LQ}$ can be constrained by indirect detection experiments too. 
The more general cases that the boost effects of leptoquarks cannot be ignored will be discussed in a future work.

\subsection{Higgs data}

The decay rate of the Higgs boson  into a pair of dark matter particles is
\bea
\Gamma(h\rightarrow SS)= \frac{\lambda^2_4 v^2}{32\pi m_h}\,\sqrt{1-\frac{4m^2_S}{m^2_h}}.
\eea
The decay rate of  the Higgs boson into a diphoton or a digluon is also modified due to leptoquarks, as given in eqs.~(\ref{Hgg}) and (\ref{Hgluglu}).
Corrections to $h\rightarrow WW,ZZ$ are small because they are already present at tree level in the SM, so we can ignore them. Then, the total Higgs decay width is modified to
\bea
\Gamma_h\approx \Gamma_{h, {\rm SM}}+\Gamma(h\rightarrow SS)
\eea
where $ \Gamma_{h, {\rm SM}}=4\,{\rm MeV}$ in the SM.
The bound from invisible Higgs decay, ${\rm BR}(h\rightarrow SS)<0.24$, leads to the following condition \cite{Higgsinv},
\bea
{\rm BR}(h\rightarrow SS)=\frac{\Gamma(h\rightarrow SS)}{\Gamma_h}<0.24.
\eea

The diphoton signal strength for gluon-fusion production is given by
\bea
\mu_{\gamma\gamma}= R_{gg}\,R_{\gamma\gamma}
\eea
where 
\bea
R_{gg}=\frac{\sigma(gg\rightarrow h)}{\sigma(gg\rightarrow h)_{\rm SM}}=\frac{
\Gamma(h\rightarrow gg)}{\Gamma_h \cdot {\rm BR}(h\rightarrow gg)_{\rm SM}},  \quad R_{\gamma\gamma}=\frac{\Gamma(h\rightarrow \gamma\gamma)}{\Gamma_h\cdot  {\rm BR}(h\rightarrow \gamma\gamma)_{\rm SM}}.
\eea
The other visible decays, $h\rightarrow ij$, are similarly modified to $\mu_{ij}=R_{gg}R_{ij}$, through the modified total decay width of Higgs boson, with $R_{ij}={\rm BR}(h\rightarrow ij)/{\rm BR}(h\rightarrow ij)_{\rm SM}=\Gamma_{h,{\rm SM}}/\Gamma_h$.  
The measurements of $gg\rightarrow h\rightarrow \gamma\gamma$ show $\mu_{\gamma\gamma}=1.10^{+0.23}_{-0.22}$ from the combined fit of LHC $7\,{\rm TeV}+8\,{\rm TeV}$ data \cite{Combi}, and  $\mu_{\gamma\gamma}=0.81^{+0.19}_{-0.18}$ and $\mu_{\gamma\gamma}=1.10^{+0.20}_{-0.18}$ from the ATLAS and CMS $13\,{\rm TeV}$ data, respectively \cite{ATLAS,CMS}.

In our model, as far as $|\lambda_5|\lesssim 10$, the decay rate into a diphoton or a digluon can be ignored, but the diphoton signal strength is modified by the enhanced total decay width of Higgs boson due to the invisible decay mode. This result can be read from Fig.~\ref{DD1} in the purple dot-dashed lines the region above which is excluded by Higgs invisible decay and in the green region which is excluded by the Higgs diphoton signal strength.

\section{Conclusions}

We have presented leptoquark models where scalar leptoquarks not only lead to the effective operators necessary for the $B$-meson anomalies but also become a portal to scalar dark matter through quartic couplings. We showed that the  annihilations of dark matter into a leptoquark pair allow for a wide parameter space that is consistent with both the correct relic density and the XENON1T bound. These new annihilation channels lead to four-body final states  in cascade with quarks and leptons mixed, due to the leptoquark decays. 
Therefore, there is an interesting interplay between the cascade annihilations of dark matter and the leptoquark search channels at the LHC, which can be tested in the current and future experiments.

\section*{Acknowledgments}

The work is supported in part by Basic Science Research Program through the National Research Foundation of Korea (NRF) funded by the Ministry of Education, Science and Technology (NRF-2016R1A2B4008759). The work of TGR is supported in part by the Chung-Ang University Research Scholarship Grants in 2018.

\def\theequation{A.\arabic{equation}}

\setcounter{equation}{0}

\vskip0.8cm
\noindent
{\Large \bf Appendix A: Effective Hamiltonians for $B$-meson decays.} 
\vskip0.4cm
\noindent

From eq.~(\ref{sLQ}), we obtain the relevant Yukawa couplings for the singlet scalar leptoquark $S_1$ in components,
\bea
{\cal L}_{S_1} &=&-\lambda_{3j}\Big( \overline{(t^C)_{R}}\, S_1 l_{j L} -\overline{(b^C)_{R}} \,S_1 \nu_{j L}   \Big)+{\rm h.c.} \nonumber \\
&&-\lambda_{2j}\Big( \overline{(c^C)_{R}}\, S_1 l_{j L} -\overline{(s^C)_{R}} \,S_1 \nu_{j L}   \Big)+{\rm h.c.} +\cdots.
\eea
Then, after integrating out the leptoquark $S_1$, we obtain the effective Hamiltonian relevant for $b\rightarrow c\tau{\bar\nu}_\tau$ as
\bea
{\cal H}^{S_1}_{b\rightarrow c\tau{\bar\nu}_\tau}&=&-\frac{\lambda^*_{33}\lambda_{23}}{m^2_{S_1}}\, (\overline{(c^C)_{R}}\tau_L) ({\bar\nu}_{\tau L} (b^C)_{R})+{\rm h.c.} \nonumber \\
&=&-\frac{\lambda^*_{33}\lambda_{23}}{2 m^2_{S_1}}\, ( \overline{(c^C)_{R}}\gamma^\mu (b^C)_{R})({\bar\nu}_{\tau L}\gamma_\mu \tau_L) +{\rm h.c.} \nonumber \\
&=&-\frac{\lambda^*_{33}\lambda_{23}}{2 m^2_{S_1}}\, ({\bar b}_L \gamma^\mu c_L) ({\bar\nu}_{\tau L}\gamma_\mu \tau_L) +{\rm h.c.}. \label{S1-app}
\eea
where use is made of Fierz identity in the second line.

In particular, in MSSM, down-type squarks (${\tilde b}^*_{Rk}$) \cite{RPV} belong to singlet scalar leptoquarks. 
We introduce the R-parity violating (RPV) superpotential as follows,
\bea
W\supset \lambda'_{ijk} L_i Q_j D^c_k,  \label{sup}
\eea
resulting in the component field Lagrangian for  doublet scalar leptoquarks $S_2\equiv {\tilde u}_{Lk}$ with $Y=+\frac{1}{6}$ or singlet scalar leptoquarks $S_1={\tilde b}^*_{Rk}$ with $Y=+\frac{1}{3}$ as
\bea
{\cal L}_{RPV}=-\lambda'_{ijk} L_i {\tilde Q}_j d^c_k+{\rm h.c.} +\cdots.
\eea
Picking up the necessary terms for $R_{K^{(*)}}$ and $R_{D^{(*)}}$ anomalies, we get, in terms of two component spinors,
\bea
{\cal L}_{RPV}&=&-\lambda'_{jk3} l_{jL} {\tilde u}_{L k}  b^c -\lambda^{\prime }_{jk2} { l}_{jL}{\tilde u}_{Lk}  s^c \nonumber \\
&&-\lambda'_{j3k}  \nu_{j L} b_{L} {\tilde b}^*_{Rk} -\lambda^{\prime }_{j2k}  { l}_{jL} c_L {\tilde b}^*_{Rk}+{\rm h.c.}+\cdots. 
\eea
Then, after integrating out the up-type squarks, ${\tilde u}_{Lk}$, and down-type squarks, ${\tilde b}^*_{Rk}$, we obtain the effective Hamiltonian for the semi-leptonic B-decays in terms of four-component spinors, as follows,
\bea
{\cal H}^{RPV}_{\rm eff}&=&-\frac{\lambda'_{2k3}\lambda^{\prime *}_{2k2}}{m^2_{{\tilde u}_{Lk}}}\, ({\bar b}_R \mu_{L}) ({\bar \mu}_{L} s_R)-\frac{\lambda'_{32k} \lambda^{\prime *}_{33k}}{m^2_{{\tilde d}_{Rk}}} (\overline{(c^C)_{R}}\tau_L) ({\bar\nu}_{\tau L} (b^C)_{R})+{\rm h.c.} \nonumber \\
&=&-\frac{\lambda'_{2k3}\lambda^{\prime *}_{2k2}}{2m^2_{{\tilde u}_{Lk}}}\,({\bar b}_R\gamma^\mu s_R)({\bar \mu}_{L}\gamma_\mu \mu_{L})-\frac{\lambda'_{32k} \lambda^{\prime *}_{33k}}{2m^2_{{\tilde d}_{Rk}}} ({\bar b}_L \gamma^\mu c_L) ({\bar\nu}_{\tau L}\gamma_\mu \tau_L)+{\rm h.c.}.
\eea 
Therefore, the effective Hamiltonian  for the $b$-to-$s$ transition is of the $(V+A)$ form, which was originally proposed to explain $R_K$ anomalies \cite{RK-early,SLee} but is not consistent with $R_{K^*}$ anomalies as it favors $(V-A)$ form. On the other hand, the effective Hamiltonian  for the $b$-to-$c$ transition is consistent with the $R_{D^{(*)}}$ anomalies \cite{RPV2,RD-bestfit}.

From eq.~(\ref{tLQ}),  we obtain the relevant Yukawa couplings for  the triplet leptoquark $S_3$  in components,
\bea
{\cal L}_{S_3} &=&-\kappa_{3j}\Big({\sqrt{2}}\, {\overline{(t^C)_{R}}}\, \phi_3\nu_{j L} - \overline{(t^C)_{R}} \,\phi_2l_{j L}- \overline{(b^C)_{R}}\, \phi_2\nu_{j L} -{\sqrt{2}}\, {\overline{(b^C)_{R}}
} \,\phi_1 l_{j L}   \Big)+{\rm h.c.} \nonumber \\
&&-\kappa_{2j}\Big( {\sqrt{2}}\,\overline{(c^C)_{R}}\, \phi_3\nu_{j L} - \overline{(c^C)_{R}} \,\phi_2 l_{j L}- \overline{(s^C)_{R}}\, \phi_2\nu_{j L} -{\sqrt{2}}\, \overline{(s^C)_{R}}\, \phi_1l_{j L}   \Big)+{\rm h.c.} +\cdots.
\eea
Then, after integrating out the leptoquark $\phi_1$ with $Q=+\frac{4}{3}$, we obtain the effective Hamiltonian relevant for $b\rightarrow s\mu^+\mu^-$ as
\bea
{\cal H}^{S_3}_{b\rightarrow s\mu^+\mu^-}&=&-\frac{2\kappa^*_{32}\kappa_{22}}{m^2_{\phi_1}}\, (\overline{(s^C)_{R}}\mu_L) ({\bar\mu}_L (b^C)_{R})+{\rm h.c.} \nonumber \\
&=&-\frac{\kappa^*_{32}\kappa_{22}}{ m^2_{\phi_1}}\, ( \overline{(s^C)_{R}}\gamma^\mu (b^C)_{R})({\bar\mu}_L\gamma_\mu \mu_L)+{\rm h.c.} \nonumber \\
&=&-\frac{\kappa^*_{32}\kappa_{22}}{ m^2_{\phi_1}}\, ({\bar b}_L \gamma^\mu s_L) ({\bar\mu}_L\gamma_\mu \mu_L) +{\rm h.c.}. \label{S3-app}
\eea
Here, we note that use is made of the Fierz identity in the second line and $ \overline{(s^C)_{R}}\gamma^\mu (b^C)_{R}=b^\dagger_L{\bar\sigma}^\mu s_L={\bar b}_L \gamma^\mu s_L$ is used in the third line.

The Yukawa couplings for the singlet scalar leptoquark also lead to effective Hamiltonian for $b\rightarrow s\nu_i {\bar \nu}_j$ as follows,
\bea
{\cal H}^{S_1}_{b\rightarrow s\nu_i {\bar \nu}_j}&=&\frac{\lambda^*_{3i}\lambda_{2j}}{m^2_{S_1}} (\overline{(s^c)_{R}}\nu_{jL}) ({\bar\nu}_{iL} (b^c)_{R})+{\rm h.c.}  \nonumber \\
&=&\frac{\lambda^*_{3i}\lambda_{2j}}{2m^2_{S_1}} (\overline{(s^c)_{R}}\gamma^\mu (b^c)_R) ({\bar\nu}_{iL}\gamma_\mu \nu_{jL})+{\rm h.c.} \nonumber \\
&=& \frac{\lambda^*_{3i}\lambda_{2j}}{2m^2_{S_1}}({\bar b}_L\gamma^\mu s_L)({\bar \nu}_{iL}\gamma_\mu \nu_{jL}) +{\rm h.c.}.   \label{S1-bnunu-app}
\eea
A similar effective interactions can be obtained for the triplet scalar leptoquark, as discussed in the text.

\def\theequation{B.\arabic{equation}}

\setcounter{equation}{0}

\vskip0.8cm
\noindent
{\Large \bf Appendix B: Effective interactions  for dark matter and Higgs boson due to leptoquark loops.} 
\vskip0.4cm
\noindent

For heavy leptoquarks, we the effective interactions between scalar dark matter and SM gauge bosons, induced by leptoquarks, as follows,
\bea
{\cal L}_{S,{\rm eff}}=D_3 \, S^2\, G_{\mu\nu}G^{\mu\nu}+D_2\, S^2\, W_{\mu\nu}W^{\mu\nu}+  D_1\, S^2\, F_{Y\mu\nu} F^{Y\mu\nu}
\eea
where
\bea
D_3&=& \frac{\alpha_S\lambda_3}{32\pi m^2_{LQ}}\, N_{LQ} l_3(S_{LQ}) A_0(y), \label{effgluon} \\
D_2&=& \frac{\alpha\lambda_3}{32\pi m^2_{LQ}}\, N_c l_2(S_{LQ}) A_0(y), \\
D_1 &=& \frac{\alpha_Y\lambda_3}{32\pi m^2_{LQ}}\, N_c N_{LQ} Y^2_{LQ} A_0(y)
\eea
with
\bea
A_0(y)=-y^{-2} [y-f(y)],
\eea
\bea
f(y)= \left\{ \begin{array}{cc} {\rm arcsin}^2\sqrt{y},\quad y\leq 1, \\ -\frac{1}{4} \Big[\ln \frac{1+\sqrt{1-y^{-1}}}{1-\sqrt{1-y^{-1}}}-i\pi \Big],\quad y>1, \end{array} \right.
\eea
and $y\equiv m^2_S/m^2_{LQ}$.
Here, $l_{2,3}(S_{LQ})$ are the Dynkin indices of $S_{LQ}$ under $SU(2)_L$ and $SU(3)_c$, respectively, i.e. $l_3(S_{1,3})=\frac{1}{2}$,  $l_2(S_1)=0$, and $l_2(S_3)=2$, and $N_{LQ}=1,3$ for $S_{LQ}=S_1, S_3$, respectively. 

Moreover, leptoquark couplings to the SM Higgs can modify the decay rates of Higgs boson into a diphoton or a digluon, as follows,
\bea
\Gamma(h\rightarrow \gamma\gamma)&=& \frac{G_F \alpha^2_{\rm em} m^3_h}{128\sqrt{2}\pi^3}\Bigg|\sum_f N_c Q^2_f A_{1/2}(x_f)+A_1(x_W) \nonumber \\
&& \quad +N_c \, g_{LQ} \sum_{i=1,\cdots,N_{LQ}} Q^2_{LQ} A_0(x_{LQ}) \Bigg|^2,  \label{Hgg}\\
\Gamma(h\rightarrow gg)&=&  \frac{G_F \alpha^2_s m^3_h}{36\sqrt{2}\pi^3}\left|\frac{3}{4}\sum_f  A_{1/2}(x_f)+\frac{3}{4}N_{LQ}\,  g_{LQ}\,  A_0(x_{LQ}) \right|^2 \label{Hgluglu}
\eea
where $g_{LQ}\equiv  \lambda_5 v^2/(2m^2_{LQ})$,  $x_i=m^2_h/(4m^2_i)$ and the loop functions are
\bea
A_{1/2}(x)&=& 2x^{-2}[x+(x-1)f(x)], \\
A_1(x)&=& -x^{-2}[2x^2+3x+3(2x-1)f(x)].
\eea


\begin{thebibliography}{999}


%RK

\bibitem{RK}
%\cite{Aaij:2014ora}
%\bibitem{Aaij:2014ora}
  R.~Aaij {\it et al.} [LHCb Collaboration],
  %``Test of lepton universality using $B^{+}\rightarrow K^{+}\ell^{+}\ell^{-}$ decays,''
  Phys.\ Rev.\ Lett.\  {\bf 113} (2014) 151601
  doi:10.1103/PhysRevLett.113.151601
  [arXiv:1406.6482 [hep-ex]].
  %%CITATION = doi:10.1103/PhysRevLett.113.151601;%%

\bibitem{RKs}
S. Bifani (2017), Seminar at CERN, URL: https://indico.cern.ch/event/580620/;
%\cite{Bifani:2017gyn}
%\bibitem{Bifani:2017gyn}
  S.~Bifani [LHCb Collaboration],
  %``Status of New Physics searches with $b \to s \ell^{+}\ell^{-}$ transitions @ LHCb,''
  arXiv:1705.02693 [hep-ex];
  %%CITATION = ARXIV:1705.02693;%%
%\cite{Aaij:2017vbb}
%\bibitem{Aaij:2017vbb}
  R.~Aaij {\it et al.} [LHCb Collaboration],
  %``Test of lepton universality with $B^{0} \rightarrow K^{*0}\ell^{+}\ell^{-}$ decays,''
  JHEP {\bf 1708} (2017) 055
  doi:10.1007/JHEP08(2017)055
  [arXiv:1705.05802 [hep-ex]].
  %%CITATION = doi:10.1007/JHEP08(2017)055;%%

\bibitem{P5}
%\cite{Aaij:2013qta}
%\bibitem{Aaij:2013qta}
  R.~Aaij {\it et al.} [LHCb Collaboration],
  %``Measurement of Form-Factor-Independent Observables in the Decay $B^{0} \to K^{*0} \mu^+ \mu^-$,''
  Phys.\ Rev.\ Lett.\  {\bf 111} (2013) 191801
  doi:10.1103/PhysRevLett.111.191801
  [arXiv:1308.1707 [hep-ex]];
  %%CITATION = doi:10.1103/PhysRevLett.111.191801;%%
%\cite{Aaij:2015oid}
%\bibitem{Aaij:2015oid}
  R.~Aaij {\it et al.} [LHCb Collaboration],
  %``Angular analysis of the $B^{0} \to K^{*0} \mu^{+} \mu^{-}$ decay using 3 fb$^{-1}$ of integrated luminosity,''
  JHEP {\bf 1602} (2016) 104
  doi:10.1007/JHEP02(2016)104
  [arXiv:1512.04442 [hep-ex]].
  %%CITATION = doi:10.1007/JHEP02(2016)104;%%


%RD

\bibitem{babar}
%\cite{Lees:2012xj}
%\bibitem{Lees:2012xj}
  J.~P.~Lees {\it et al.} [BaBar Collaboration],
  %``Evidence for an excess of $\bar{B} \to D^{(*)} \tau^-\bar{\nu}_\tau$ decays,''
  Phys.\ Rev.\ Lett.\  {\bf 109} (2012) 101802
  doi:10.1103/PhysRevLett.109.101802
  [arXiv:1205.5442 [hep-ex]];
  %%CITATION = doi:10.1103/PhysRevLett.109.101802;%%
%\cite{Lees:2013uzd}
%\bibitem{Lees:2013uzd}
  J.~P.~Lees {\it et al.} [BaBar Collaboration],
  %``Measurement of an Excess of $\bar{B} \to D^{(*)}\tau^- \bar{\nu}_\tau$ Decays and Implications for Charged Higgs Bosons,''
  Phys.\ Rev.\ D {\bf 88} (2013) no.7,  072012
  doi:10.1103/PhysRevD.88.072012
  [arXiv:1303.0571 [hep-ex]].
  
 \bibitem{belle}
  %%CITATION = doi:10.1103/PhysRevD.88.072012;%%
%\cite{Huschle:2015rga}
%\bibitem{Huschle:2015rga}
  M.~Huschle {\it et al.} [Belle Collaboration],
  %``Measurement of the branching ratio of $\bar{B} \to D^{(\ast)} \tau^- \bar{\nu}_\tau$ relative to $\bar{B} \to D^{(\ast)} \ell^- \bar{\nu}_\ell$ decays with hadronic tagging at Belle,''
  Phys.\ Rev.\ D {\bf 92} (2015) no.7,  072014
  doi:10.1103/PhysRevD.92.072014
  [arXiv:1507.03233 [hep-ex]];
  %%CITATION = doi:10.1103/PhysRevD.92.072014;%%
% \bibitem{RDexp3}
% %\AmhisHMA
% %\lref\AmhisHMA{
%   Y.~Amhis {\it et al.} [Heavy Flavor Averaging Group (HFAG)],
%   %``Averages of $b$-hadron, $c$-hadron, and $\tau$-lepton properties as of summer 2014,''
% [arXiv:1412.7515 [hep-ex]].
%\cite{Abdesselam:2016cgx}
%\bibitem{Abdesselam:2016cgx}
  A.~Abdesselam {\it et al.} [Belle Collaboration],
  %``Measurement of the branching ratio of $\bar{B}^0 \rightarrow D^{*+} \tau^- \bar{\nu}_{\tau}$ relative to $\bar{B}^0 \rightarrow D^{*+} \ell^- \bar{\nu}_{\ell}$ decays with a semileptonic tagging method,''
  arXiv:1603.06711 [hep-ex].
  %%CITATION = ARXIV:1603.06711;%%
  
\bibitem{lhcb}
%\cite{Aaij:2015yra}
%\bibitem{Aaij:2015yra}
  R.~Aaij {\it et al.} [LHCb Collaboration],
  %``Measurement of the ratio of branching fractions $\mathcal{B}(\bar{B}^0 \to D^{*+}\tau^{-}\bar{\nu}_{\tau})/\mathcal{B}(\bar{B}^0 \to D^{*+}\mu^{-}\bar{\nu}_{\mu})$,''
  Phys.\ Rev.\ Lett.\  {\bf 115} (2015) no.11,  111803  doi:10.1103/PhysRevLett.115.111803. [arXiv:1506.08614 [hep-ex]]; Erratum: [Phys.\ Rev.\ Lett.\  {\bf 115} (2015) no.15,  159901]
  doi:10.1103/PhysRevLett.115.159901.
  %%CITATION = doi:10.1103/PhysRevLett.115.159901, 10.1103/PhysRevLett.115.111803;%%


\bibitem{qcd}
%\cite{Straub:2015ica}
%\bibitem{Straub:2015ica}
  A.~Bharucha, D.~M.~Straub and R.~Zwicky,
  %``$B\to V\ell^+\ell^-$ in the Standard Model from light-cone sum rules,''
  JHEP {\bf 1608} (2016) 098
  doi:10.1007/JHEP08(2016)098
  [arXiv:1503.05534 [hep-ph]];
  %%CITATION = doi:10.1007/JHEP08(2016)098;%%
  %137 citations counted in INSPIRE as of 20 Jul 2017
%\cite{Ciuchini:2015qxb}
%\bibitem{Ciuchini:2015qxb}
  M.~Ciuchini, M.~Fedele, E.~Franco, S.~Mishima, A.~Paul, L.~Silvestrini and M.~Valli,
  %``$B\to K^* \ell^+ \ell^-$ decays at large recoil in the Standard Model: a theoretical reappraisal,''
  JHEP {\bf 1606} (2016) 116
  doi:10.1007/JHEP06(2016)116
  [arXiv:1512.07157 [hep-ph]];
  %%CITATION = doi:10.1007/JHEP06(2016)116;%%
  %80 citations counted in INSPIRE as of 20 Jul 2017
%\cite{Neshatpour:2017qvi}
%\bibitem{Neshatpour:2017qvi}
  S.~Neshatpour, V.~G.~Chobanova, T.~Hurth, F.~Mahmoudi and D.~Martinez Santos,
  %``Direct comparison of global fits to the B ->$ K^* $$\mu^+\mu^-$ data assuming hadronic corrections or new physics,''
  arXiv:1705.10730 [hep-ph];
  %%CITATION = ARXIV:1705.10730;%%
  %2 citations counted in INSPIRE as of 20 Jul 2017
  %\cite{Arbey:2018ics}
%\bibitem{Arbey:2018ics}
  A.~Arbey, T.~Hurth, F.~Mahmoudi and S.~Neshatpour,
  %``Hadronic and New Physics Contributions to $B \to K^* \ell^+ \ell^-$,''
  arXiv:1806.02791 [hep-ph].
  %%CITATION = ARXIV:1806.02791;%%
  %1 citations counted in INSPIRE as of 17 Jul 2018


\bibitem{more}
 %\cite{Albrecht:2017odf}
%\bibitem{Albrecht:2017odf}
  J.~Albrecht, F.~Bernlochner, M.~Kenzie, S.~Reichert, D.~Straub and A.~Tully,
  %``Future prospects for exploring present day anomalies in flavour physics measurements with Belle II and LHCb,''
  arXiv:1709.10308 [hep-ph].
  %%CITATION = ARXIV:1709.10308;%%
  %4 citations counted in INSPIRE as of 27 Feb 2018



% Direct detection

\bibitem{xenon1t}
%\cite{Aprile:2018dbl}
%\bibitem{Aprile:2018dbl}
  E.~Aprile {\it et al.} [XENON Collaboration],
  %``Dark Matter Search Results from a One Tonne$\times$Year Exposure of XENON1T,''
  arXiv:1805.12562 [astro-ph.CO];
  %%CITATION = ARXIV:1805.12562;%%
  %29 citations counted in INSPIRE as of 15 Jul 2018
  %\cite{Aprile:2017iyp}
%\bibitem{Aprile:2017iyp}
  E.~Aprile {\it et al.} [XENON Collaboration],
  %``First Dark Matter Search Results from the XENON1T Experiment,''
  Phys.\ Rev.\ Lett.\  {\bf 119} (2017) no.18,  181301
  doi:10.1103/PhysRevLett.119.181301
  [arXiv:1705.06655 [astro-ph.CO]].
  %%CITATION = doi:10.1103/PhysRevLett.119.181301;%%
  %429 citations counted in INSPIRE as of 17 Jul 2018
  
  
 \bibitem{panda} 
%\cite{Cui:2017nnn}
%\bibitem{Cui:2017nnn} 
  X.~Cui {\it et al.} [PandaX-II Collaboration],
  %``Dark Matter Results From 54-Ton-Day Exposure of PandaX-II Experiment,''
  Phys.\ Rev.\ Lett.\  {\bf 119}, no. 18, 181302 (2017)
  doi:10.1103/PhysRevLett.119.181302
  [arXiv:1708.06917 [astro-ph.CO]].
  %%CITATION = doi:10.1103/PhysRevLett.119.181302;%%
  %109 citations counted in INSPIRE as of 23 Feb 2018


\bibitem{cdms}
%\cite{Agnese:2017njq}
%\bibitem{Agnese:2017njq}
  R.~Agnese {\it et al.} [SuperCDMS Collaboration],
  %``Results from the Super Cryogenic Dark Matter Search Experiment at Soudan,''
  Phys.\ Rev.\ Lett.\  {\bf 120} (2018) no.6,  061802
  doi:10.1103/PhysRevLett.120.061802
  [arXiv:1708.08869 [hep-ex]].
  %%CITATION = doi:10.1103/PhysRevLett.120.061802;%%
  %8 citations counted in INSPIRE as of 04 Mar 2018

  
  \bibitem{lux}
  %\cite{Akerib:2016vxi}
  %\bibitem{Akerib:2016vxi} 
  D.~S.~Akerib {\it et al.} [LUX Collaboration],
  %``Results from a search for dark matter in the complete LUX exposure,''
  Phys.\ Rev.\ Lett.\  {\bf 118}, no. 2, 021303 (2017)
  doi:10.1103/PhysRevLett.118.021303
  [arXiv:1608.07648 [astro-ph.CO]].
  %%CITATION = doi:10.1103/PhysRevLett.118.021303;%%
  %547 citations counted in INSPIRE as of 23 Feb 2018 
  


\bibitem{darwin}
%\cite{Aalbers:2016jon}
%\bibitem{Aalbers:2016jon}
  J.~Aalbers {\it et al.} [DARWIN Collaboration],
  %``DARWIN: towards the ultimate dark matter detector,''
  JCAP {\bf 1611} (2016) 017
  doi:10.1088/1475-7516/2016/11/017
  [arXiv:1606.07001 [astro-ph.IM]].
  %%CITATION = doi:10.1088/1475-7516/2016/11/017;%%
  %130 citations counted in INSPIRE as of 16 Jul 2018


\bibitem{LZ}
%\cite{Akerib:2015cja}
%\bibitem{Akerib:2015cja}
  D.~S.~Akerib {\it et al.} [LZ Collaboration],
  %``LUX-ZEPLIN (LZ) Conceptual Design Report,''
  arXiv:1509.02910 [physics.ins-det].
  %%CITATION = ARXIV:1509.02910;%%
  %209 citations counted in INSPIRE as of 16 Jul 2018







%LQ models


\bibitem{LQs}
%\cite{Buchmuller:1986zs}
%\bibitem{Buchmuller:1986zs}
  W.~Buchmuller, R.~Ruckl and D.~Wyler,
  %``Leptoquarks in Lepton - Quark Collisions,''
  Phys.\ Lett.\ B {\bf 191} (1987) 442
   Erratum: [Phys.\ Lett.\ B {\bf 448} (1999) 320].
  doi:10.1016/S0370-2693(99)00014-3, 10.1016/0370-2693(87)90637-X
  %%CITATION = doi:10.1016/S0370-2693(99)00014-3, 10.1016/0370-2693(87)90637-X;%%
  %679 citations counted in INSPIRE as of 17 Jul 2018




\bibitem{LQS1}  
%\cite{Dorsner:2016wpm}
%\bibitem{Dorsner:2016wpm}
  I.~Dorsner, S.~Fajfer, A.~Greljo, J.~F.~Kamenik and N.~Kosnik,
  %``Physics of leptoquarks in precision experiments and at particle colliders,''
  Phys.\ Rept.\  {\bf 641} (2016) 1
  doi:10.1016/j.physrep.2016.06.001
  [arXiv:1603.04993 [hep-ph]].
  %%CITATION = doi:10.1016/j.physrep.2016.06.001;%%
  %129 citations counted in INSPIRE as of 17 Jul 2018



\bibitem{leptoquarks}
%\cite{Crivellin:2017zlb}
%\bibitem{Crivellin:2017zlb}
  A.~Crivellin, D.~MÃŒller and T.~Ota,
  %``Simultaneous explanation of R(D$^{(?)}$) and b?s?$^{+}$ ?$^{?}$: the last scalar leptoquarks standing,''
  JHEP {\bf 1709} (2017) 040
  doi:10.1007/JHEP09(2017)040
  [arXiv:1703.09226 [hep-ph]].
  %%CITATION = doi:10.1007/JHEP09(2017)040;%%
  %48 citations counted in INSPIRE as of 01 Feb 2018
  


\bibitem{RK-models}
%\cite{Hiller:2017bzc}
%\bibitem{Hiller:2017bzc}
  G.~Hiller and I.~Nisandzic,
  %``$R_K$ and $R_{K^{\ast}}$ beyond the standard model,''
  Phys.\ Rev.\ D {\bf 96} (2017) no.3,  035003
  doi:10.1103/PhysRevD.96.035003
  [arXiv:1704.05444 [hep-ph]];
  %%CITATION = doi:10.1103/PhysRevD.96.035003;%%
  %64 citations counted in INSPIRE as of 01 Feb 2018


  
  
\bibitem{LQ-fit}  
%\cite{Buttazzo:2017ixm}
%\bibitem{Buttazzo:2017ixm}
  D.~Buttazzo, A.~Greljo, G.~Isidori and D.~Marzocca,
  %``B-physics anomalies: a guide to combined explanations,''
  JHEP {\bf 1711} (2017) 044
  doi:10.1007/JHEP11(2017)044
  [arXiv:1706.07808 [hep-ph]].
  %%CITATION = doi:10.1007/JHEP11(2017)044;%%
  %28 citations counted in INSPIRE as of 01 Feb 2018





\bibitem{LQ-early}
%\cite{Bauer:2015knc}
%\bibitem{Bauer:2015knc}
  M.~Bauer and M.~Neubert,
  %``Minimal Leptoquark Explanation for the R$_{D^{(*)}}$ , R$_K$ , and $(g-2)_g$ Anomalies,''
  Phys.\ Rev.\ Lett.\  {\bf 116} (2016) no.14,  141802
  doi:10.1103/PhysRevLett.116.141802
  [arXiv:1511.01900 [hep-ph]].
  %%CITATION = doi:10.1103/PhysRevLett.116.141802;%%
  %172 citations counted in INSPIRE as of 17 Jul 2018


\bibitem{nomura}
%\cite{Chen:2017hir}
%\bibitem{Chen:2017hir}
  C.~H.~Chen, T.~Nomura and H.~Okada,
  %``Excesses of muon $g-2$, $R_{D^{(\ast)}}$, and $R_K$ in a leptoquark model,''
  Phys.\ Lett.\ B {\bf 774} (2017) 456
  doi:10.1016/j.physletb.2017.10.005
  [arXiv:1703.03251 [hep-ph]].
  %%CITATION = doi:10.1016/j.physletb.2017.10.005;%%
  %32 citations counted in INSPIRE as of 17 Jul 2018


\bibitem{kenji}
%\cite{Matsuzaki:2017bpp}
%\bibitem{Matsuzaki:2017bpp}
  S.~Matsuzaki, K.~Nishiwaki and R.~Watanabe,
  %``Phenomenology of flavorful composite vector bosons in light of $B$ anomalies,''
  JHEP {\bf 1708} (2017) 145
  doi:10.1007/JHEP08(2017)145
  [arXiv:1706.01463 [hep-ph]].
  %%CITATION = doi:10.1007/JHEP08(2017)145;%%
  %13 citations counted in INSPIRE as of 17 Jul 2018



\bibitem{watanabe}
%\cite{Kumar:2018kmr}
%\bibitem{Kumar:2018kmr}
  J.~Kumar, D.~London and R.~Watanabe,
  %``Combined Explanations of the $b \to s \mu^+ \mu^-$ and $b \to c \tau^- {\bar\nu}$ Anomalies: a General Model Analysis,''
  arXiv:1806.07403 [hep-ph].
  %%CITATION = ARXIV:1806.07403;%%
  %1 citations counted in INSPIRE as of 17 Jul 2018




\bibitem{unification}
%\cite{Assad:2017iib}
%\bibitem{Assad:2017iib}
  N.~Assad, B.~Fornal and B.~Grinstein,
  %``Baryon Number and Lepton Universality Violation in Leptoquark and Diquark Models,''
  Phys.\ Lett.\ B {\bf 777} (2018) 324
  doi:10.1016/j.physletb.2017.12.042
  [arXiv:1708.06350 [hep-ph]];
  %%CITATION = doi:10.1016/j.physletb.2017.12.042;%%
  %21 citations counted in INSPIRE as of 20 Jul 2018
%\cite{DiLuzio:2017vat}
%\bibitem{DiLuzio:2017vat}
  L.~Di Luzio, A.~Greljo and M.~Nardecchia,
  %``Gauge leptoquark as the origin of B-physics anomalies,''
  Phys.\ Rev.\ D {\bf 96} (2017) no.11,  115011
  doi:10.1103/PhysRevD.96.115011
  [arXiv:1708.08450 [hep-ph]];
  %%CITATION = doi:10.1103/PhysRevD.96.115011;%%
  %15 citations counted in INSPIRE as of 01 Feb 2018
  %\cite{Calibbi:2017qbu}
%\bibitem{Calibbi:2017qbu}
  L.~Calibbi, A.~Crivellin and T.~Li,
  %``A model of vector leptoquarks in view of the $B$-physics anomalies,''
  arXiv:1709.00692 [hep-ph];
  %%CITATION = ARXIV:1709.00692;%%
  %41 citations counted in INSPIRE as of 20 Jul 2018
%\cite{Bordone:2017bld}
%\bibitem{Bordone:2017bld}
  M.~Bordone, C.~Cornella, J.~Fuentes-Martin and G.~Isidori,
  %``A three-site gauge model for flavor hierarchies and flavor anomalies,''
  arXiv:1712.01368 [hep-ph];
  %%CITATION = ARXIV:1712.01368;%%
  %7 citations counted in INSPIRE as of 01 Feb 2018
  %\cite{Blanke:2018sro}
%\bibitem{Blanke:2018sro}
  M.~Blanke and A.~Crivellin,
  %``$B$ Meson Anomalies in a Pati-Salam Model within the Randall-Sundrum Background,''
  Phys.\ Rev.\ Lett.\  {\bf 121} (2018) no.1,  011801
  doi:10.1103/PhysRevLett.121.011801
  [arXiv:1801.07256 [hep-ph]];
  %%CITATION = doi:10.1103/PhysRevLett.121.011801;%%
  %23 citations counted in INSPIRE as of 20 Jul 2018
%\cite{Becirevic:2018afm}
%\bibitem{Becirevic:2018afm}
  D.~Becirevic, I.~Dorsner, S.~Fajfer, D.~A.~Faroughy, N.~Kosnik and O.~Sumensari,
  %``Scalar leptoquarks from GUT to accommodate the $B$-physics anomalies,''
  arXiv:1806.05689 [hep-ph].
  %%CITATION = ARXIV:1806.05689;%%
  %6 citations counted in INSPIRE as of 17 Jul 2018




\bibitem{crivellin}
%\cite{Capdevila:2017bsm}
%\bibitem{Capdevila:2017bsm}
  B.~Capdevila, A.~Crivellin, S.~Descotes-Genon, J.~Matias and J.~Virto,
  %``Patterns of New Physics in $b\to s\ell^+\ell^-$ transitions in the light of recent data,''
  arXiv:1704.05340 [hep-ph].
  %%CITATION = ARXIV:1704.05340;%%
  %40 citations counted in INSPIRE as of 08 Jul 2017



%RD combination

\bibitem{hflav}
%\cite{Amhis:2016xyh}
%\bibitem{Amhis:2016xyh}
  Y.~Amhis {\it et al.} [HFLAV Collaboration],
  %``Averages of $b$-hadron, $c$-hadron, and $\tau$-lepton properties as of summer 2016,''
  Eur.\ Phys.\ J.\ C {\bf 77} (2017) no.12,  895
  doi:10.1140/epjc/s10052-017-5058-4
  [arXiv:1612.07233 [hep-ex]].
  %%CITATION = doi:10.1140/epjc/s10052-017-5058-4;%%
  %246 citations counted in INSPIRE as of 12 Feb 2018


%RD lattice

\bibitem{RD-lattice}
%\cite{Na:2015kha}
%\bibitem{Na:2015kha}
  H.~Na {\it et al.} [HPQCD Collaboration],
  %``$B \rightarrow D l \nu$ form factors at nonzero recoil and extraction of $|V_{cb}|$,''
  Phys.\ Rev.\ D {\bf 92} (2015) no.5,  054510
   Erratum: [Phys.\ Rev.\ D {\bf 93} (2016) no.11,  119906]
  doi:10.1103/PhysRevD.93.119906, 10.1103/PhysRevD.92.054510
  [arXiv:1505.03925 [hep-lat]].
  %%CITATION = doi:10.1103/PhysRevD.93.119906, 10.1103/PhysRevD.92.054510;%%
  %130 citations counted in INSPIRE as of 12 Feb 2018



\bibitem{RD-SM}
%\cite{Fajfer:2012vx}
%\bibitem{Fajfer:2012vx}
  S.~Fajfer, J.~F.~Kamenik and I.~Nisandzic,
  %``On the $B \to D^* \tau \bar \nu_{\tau}$ Sensitivity to New Physics,''
  Phys.\ Rev.\ D {\bf 85} (2012) 094025
  doi:10.1103/PhysRevD.85.094025
  [arXiv:1203.2654 [hep-ph]];
  %%CITATION = doi:10.1103/PhysRevD.85.094025;%%
%\cite{Bernlochner:2017jka}
%\bibitem{Bernlochner:2017jka}
  F.~U.~Bernlochner, Z.~Ligeti, M.~Papucci and D.~J.~Robinson,
  %``Combined analysis of semileptonic $B$ decays to $D$ and $D^*$: $R(D^{(*)})$, $|V_{cb}|$, and new physics,''
  Phys.\ Rev.\ D {\bf 95} (2017) no.11,  115008
  doi:10.1103/PhysRevD.95.115008
  [arXiv:1703.05330 [hep-ph]].
  %%CITATION = doi:10.1103/PhysRevD.95.115008;%%
  %39 citations counted in INSPIRE as of 12 Feb 2018


\bibitem{RD-SMetc}
%\cite{Bigi:2016mdz}
%\bibitem{Bigi:2016mdz}
  D.~Bigi and P.~Gambino,
  %``Revisiting $B\to D \ell \nu$,''
  Phys.\ Rev.\ D {\bf 94} (2016) no.9,  094008
  doi:10.1103/PhysRevD.94.094008
  [arXiv:1606.08030 [hep-ph]];
  %%CITATION = doi:10.1103/PhysRevD.94.094008;%%
  %54 citations counted in INSPIRE as of 12 Feb 2018
%\cite{Lattice:2015rga}
%\bibitem{Lattice:2015rga}
  J.~A.~Bailey {\it et al.} [MILC Collaboration],
  %``BâDâÎœ form factors at nonzero recoil and |V$_{cb}$| from 2+1-flavor lattice QCD,''
  Phys.\ Rev.\ D {\bf 92} (2015) no.3,  034506
  doi:10.1103/PhysRevD.92.034506
  [arXiv:1503.07237 [hep-lat]];
  %%CITATION = doi:10.1103/PhysRevD.92.034506;%%
  %115 citations counted in INSPIRE as of 12 Feb 2018
  %\cite{Bigi:2017jbd}
%\bibitem{Bigi:2017jbd}
  D.~Bigi, P.~Gambino and S.~Schacht,
  %``$R(D^*)$, $|V_{cb}|$, and the Heavy Quark Symmetry relations between form factors,''
  JHEP {\bf 1711} (2017) 061
  doi:10.1007/JHEP11(2017)061
  [arXiv:1707.09509 [hep-ph]];
  %%CITATION = doi:10.1007/JHEP11(2017)061;%%
  %22 citations counted in INSPIRE as of 12 Feb 2018
%\cite{Jaiswal:2017rve}
%\bibitem{Jaiswal:2017rve}
  S.~Jaiswal, S.~Nandi and S.~K.~Patra,
  %``Extraction of $|V_{cb}|$ from $B\to D^{(*)}\ell\nu_\ell$ and the Standard Model predictions of $R(D^{(*)})$,''
  JHEP {\bf 1712} (2017) 060
  doi:10.1007/JHEP12(2017)060
  [arXiv:1707.09977 [hep-ph]].
  %%CITATION = doi:10.1007/JHEP12(2017)060;%%
  %15 citations counted in INSPIRE as of 12 Feb 2018


\bibitem{RD-bestfit}
%\cite{Altmannshofer:2017poe}
%\bibitem{Altmannshofer:2017poe}
  W.~Altmannshofer, P.~S.~Bhupal Dev and A.~Soni,
  %``$R_{D^{(*)}}$ anomaly: A possible hint for natural supersymmetry with $R$-parity violation,''
  Phys.\ Rev.\ D {\bf 96} (2017) no.9,  095010
  doi:10.1103/PhysRevD.96.095010
  [arXiv:1704.06659 [hep-ph]].
  %%CITATION = doi:10.1103/PhysRevD.96.095010;%%
  %18 citations counted in INSPIRE as of 12 Feb 2018





\bibitem{cutoff}
%\cite{DiLuzio:2017chi}
%\bibitem{DiLuzio:2017chi}
  L.~Di Luzio and M.~Nardecchia,
  %``What is the scale of new physics behind the $B$-flavour anomalies?,''
  Eur.\ Phys.\ J.\ C {\bf 77} (2017) no.8,  536
  doi:10.1140/epjc/s10052-017-5118-9
  [arXiv:1706.01868 [hep-ph]].
  %%CITATION = doi:10.1140/epjc/s10052-017-5118-9;%%
  %12 citations counted in INSPIRE as of 01 Feb 2018



%LQ bounds  
  \bibitem{2bMET}
  %\cite{Sirunyan:2017kiw}
%\bibitem{Sirunyan:2017kiw}
  A.~M.~Sirunyan {\it et al.} [CMS Collaboration],
  %``Search for the pair production of third-generation squarks with two-body decays to a bottom or charm quark and a neutralino in proton?proton collisions at $\sqrt{s}$ = 13 TeV,''
  Phys.\ Lett.\ B {\bf 778} (2018) 263
  doi:10.1016/j.physletb.2018.01.012
  [arXiv:1707.07274 [hep-ex]];
  %%CITATION = doi:10.1016/j.physletb.2018.01.012;%%
  %17 citations counted in INSPIRE as of 11 Jul 2018
ATLAS Collaboration, ATLAS-CONF-2017-038.


\bibitem{2jetMET}
%\cite{Sirunyan:2017cwe}
%\bibitem{Sirunyan:2017cwe}
  A.~M.~Sirunyan {\it et al.} [CMS Collaboration],
  %``Search for supersymmetry in multijet events with missing transverse momentum in proton-proton collisions at 13 TeV,''
  Phys.\ Rev.\ D {\bf 96} (2017) no.3,  032003
  doi:10.1103/PhysRevD.96.032003
  [arXiv:1704.07781 [hep-ex]];
  %%CITATION = doi:10.1103/PhysRevD.96.032003;%%
  %64 citations counted in INSPIRE as of 11 Jul 2018
ATLAS Collaboration, ATLAS-CONF-2017-022.


\bibitem{2b2mu}
ATLAS Collaboration, ATLAS-CONF-2017-036.



\bibitem{2mu2jet}
%\cite{Khachatryan:2015vaa}
%\bibitem{Khachatryan:2015vaa}
  V.~Khachatryan {\it et al.} [CMS Collaboration],
  %``Search for pair production of first and second generation leptoquarks in proton-proton collisions at $\sqrt{s}$ = 8??TeV,''
  Phys.\ Rev.\ D {\bf 93} (2016) no.3,  032004
  doi:10.1103/PhysRevD.93.032004
  [arXiv:1509.03744 [hep-ex]].
  %%CITATION = doi:10.1103/PhysRevD.93.032004;%%
  %57 citations counted in INSPIRE as of 11 Jul 2018




\bibitem{2t2mu}  
 CMS Collaboration, CMS-PAS-B2G-16-027. 
  
  
 
 \bibitem{mu2jetMET}
 %\cite{Aad:2015caa}
%\bibitem{Aad:2015caa}
  G.~Aad {\it et al.} [ATLAS Collaboration],
  %``Searches for scalar leptoquarks in pp collisions at $\sqrt{s}$ = 8 TeV with the ATLAS detector,''
  Eur.\ Phys.\ J.\ C {\bf 76} (2016) no.1,  5
  doi:10.1140/epjc/s10052-015-3823-9
  [arXiv:1508.04735 [hep-ex]].
  %%CITATION = doi:10.1140/epjc/s10052-015-3823-9;%%
  %95 citations counted in INSPIRE as of 11 Jul 2018

 
 
 
 \bibitem{2tMET}
   %\cite{Sirunyan:2017xse}
%\bibitem{Sirunyan:2017xse}
  A.~M.~Sirunyan {\it et al.} [CMS Collaboration],
  %``Search for top squark pair production in pp collisions at $ \sqrt{s}=13 $ TeV using single lepton events,''
  JHEP {\bf 1710} (2017) 019
  doi:10.1007/JHEP10(2017)019
  [arXiv:1706.04402 [hep-ex]];
  %%CITATION = doi:10.1007/JHEP10(2017)019;%%
  %27 citations counted in INSPIRE as of 11 Jul 2018
%\cite{Aaboud:2017aeu}
%\bibitem{Aaboud:2017aeu}
  M.~Aaboud {\it et al.} [ATLAS Collaboration],
  %``Search for top-squark pair production in final states with one lepton, jets, and missing transverse momentum using 36 fb$^{-1}$ of $\sqrt{s}=13$ TeV pp collision data with the ATLAS detector,''
  JHEP {\bf 1806} (2018) 108
  doi:10.1007/JHEP06(2018)108
  [arXiv:1711.11520 [hep-ex]].
  %%CITATION = doi:10.1007/JHEP06(2018)108;%%
  %20 citations counted in INSPIRE as of 11 Jul 2018
 
   



\bibitem{Zprime}
%\cite{Bian:2017xzg}
%\bibitem{Bian:2017xzg}
  L.~Bian, H.~M.~Lee and C.~B.~Park,
  %``$B$-meson anomalies and Higgs physics in flavored $U(1)'$ model,''
  arXiv:1711.08930 [hep-ph];
  %%CITATION = ARXIV:1711.08930;%%
%\cite{Bian:2017rpg}
%\bibitem{Bian:2017rpg}
  L.~Bian, S.~M.~Choi, Y.~J.~Kang and H.~M.~Lee,
  %``A minimal flavored $U(1)'$ for $B$-meson anomalies,''
  Phys.\ Rev.\ D {\bf 96} (2017) no.7,  075038
  doi:10.1103/PhysRevD.96.075038
  [arXiv:1707.04811 [hep-ph]].
  %%CITATION = doi:10.1103/PhysRevD.96.075038;%%
  %6 citations counted in INSPIRE as of 02 Feb 2018


%RPV


%BKnunu

\bibitem{BKnunu}
%\cite{Alok:2017jgr}
%\bibitem{Alok:2017jgr}
  A.~K.~Alok, B.~Bhattacharya, D.~Kumar, J.~Kumar, D.~London and S.~U.~Sankar,
  %``New physics in $b \rightarrow s \mu^+ \mu^-$: Distinguishing models through CP-violating effects,''
  Phys.\ Rev.\ D {\bf 96} (2017) no.1,  015034
  doi:10.1103/PhysRevD.96.015034
  [arXiv:1703.09247 [hep-ph]].
  %%CITATION = doi:10.1103/PhysRevD.96.015034;%%
  %18 citations counted in INSPIRE as of 15 Jul 2018


\bibitem{BKnunu-exp}
%\cite{Grygier:2017tzo}
%\bibitem{Grygier:2017tzo}
  J.~Grygier {\it et al.} [Belle Collaboration],
  %``Search for $\boldsymbol{B\to h\nu\bar{\nu}}$ decays with semileptonic tagging at Belle,''
  Phys.\ Rev.\ D {\bf 96} (2017) no.9,  091101
   Addendum: [Phys.\ Rev.\ D {\bf 97} (2018) no.9,  099902]
  doi:10.1103/PhysRevD.97.099902, 10.1103/PhysRevD.96.091101
  [arXiv:1702.03224 [hep-ex]].
  %%CITATION = doi:10.1103/PhysRevD.97.099902, 10.1103/PhysRevD.96.091101;%%
  %33 citations counted in INSPIRE as of 15 Jul 2018



\bibitem{BKnunu-SM}
%\cite{Buras:2014fpa}
%\bibitem{Buras:2014fpa}
  A.~J.~Buras, J.~Girrbach-Noe, C.~Niehoff and D.~M.~Straub,
  %``$ B\to {K}^{\left(\ast \right)}\nu \overline{\nu} $ decays in the Standard Model and beyond,''
  JHEP {\bf 1502} (2015) 184
  doi:10.1007/JHEP02(2015)184
  [arXiv:1409.4557 [hep-ph]].
  %%CITATION = doi:10.1007/JHEP02(2015)184;%%
  %125 citations counted in INSPIRE as of 15 Jul 2018


%LQ searches




\bibitem{LQS2}
%\cite{Faroughy:2016osc}
%\bibitem{Faroughy:2016osc}
  D.~A.~Faroughy, A.~Greljo and J.~F.~Kamenik,
  %``Confronting lepton flavor universality violation in B decays with high-$p_T$ tau lepton searches at LHC,''
  Phys.\ Lett.\ B {\bf 764} (2017) 126
  doi:10.1016/j.physletb.2016.11.011
  [arXiv:1609.07138 [hep-ph]];
  %%CITATION = doi:10.1016/j.physletb.2016.11.011;%%
  %52 citations counted in INSPIRE as of 01 Feb 2018
%\cite{Diaz:2017lit}
%\bibitem{Diaz:2017lit}
  B.~Diaz, M.~Schmaltz and Y.~M.~Zhong,
  %``The leptoquark Hunter?s guide: Pair production,''
  JHEP {\bf 1710} (2017) 097
  doi:10.1007/JHEP10(2017)097
  [arXiv:1706.05033 [hep-ph]];
  %%CITATION = doi:10.1007/JHEP10(2017)097;%%
  %14 citations counted in INSPIRE as of 01 Feb 2018
%\cite{Hiller:2018wbv}
%\bibitem{Hiller:2018wbv}
  G.~Hiller, D.~Loose and I.~Nisandzic,
  %``Flavorful leptoquarks at hadron colliders,''
  arXiv:1801.09399 [hep-ph];
  %%CITATION = ARXIV:1801.09399;%%
%\cite{Dorsner:2018ynv}
%\bibitem{Dorsner:2018ynv}
  I.~DorÅ¡ner and A.~Greljo,
  %``Leptoquark toolbox for precision collider studies,''
  arXiv:1801.07641 [hep-ph].
  %%CITATION = ARXIV:1801.07641;%%
  %1 citations counted in INSPIRE as of 01 Feb 2018

  

%(g-2)mu, LFV

\bibitem{amu}
%\cite{Bennett:2006fi}
%\bibitem{Bennett:2006fi}
  G.~W.~Bennett {\it et al.} [Muon g-2 Collaboration],
  %``Final Report of the Muon E821 Anomalous Magnetic Moment Measurement at BNL,''
  Phys.\ Rev.\ D {\bf 73} (2006) 072003
  doi:10.1103/PhysRevD.73.072003
  [hep-ex/0602035].
  %%CITATION = doi:10.1103/PhysRevD.73.072003;%%
  %1416 citations counted in INSPIRE as of 09 Jul 2017



\bibitem{pdg}
%\cite{Olive:2016xmw}
%\bibitem{Olive:2016xmw}
  C.~Patrignani {\it et al.} [Particle Data Group],
  %``Review of Particle Physics,''
  Chin.\ Phys.\ C {\bf 40} (2016) no.10,  100001.
  doi:10.1088/1674-1137/40/10/100001
  %%CITATION = doi:10.1088/1674-1137/40/10/100001;%%
  %1147 citations counted in INSPIRE as of 09 Jul 2017
  




\bibitem{taumu}
  %\cite{Aubert:2009ag}
%\bibitem{Aubert:2009ag}
  B.~Aubert {\it et al.} [BaBar Collaboration],
  %``Searches for Lepton Flavor Violation in the Decays tau+- ---> e+- gamma and tau+- ---> mu+- gamma,''
  Phys.\ Rev.\ Lett.\  {\bf 104} (2010) 021802
  doi:10.1103/PhysRevLett.104.021802
  [arXiv:0908.2381 [hep-ex]].
  %%CITATION = doi:10.1103/PhysRevLett.104.021802;%%
  %364 citations counted in INSPIRE as of 25 Apr 2018
  
  
\bibitem{LQ-DM}  
%\cite{Baker:2015qna}
%\bibitem{Baker:2015qna}
  M.~J.~Baker {\it et al.},
  %``The Coannihilation Codex,''
  JHEP {\bf 1512} (2015) 120
  doi:10.1007/JHEP12(2015)120
  [arXiv:1510.03434 [hep-ph]];
  %%CITATION = doi:10.1007/JHEP12(2015)120;%%
  %39 citations counted in INSPIRE as of 19 Jul 2018
%\cite{Bauer:2015boy}
%\bibitem{Bauer:2015boy}
  M.~Bauer and M.~Neubert,
  %``Flavor anomalies, the 750 GeV diphoton excess, and a dark matter candidate,''
  Phys.\ Rev.\ D {\bf 93} (2016) no.11,  115030
  doi:10.1103/PhysRevD.93.115030
  [arXiv:1512.06828 [hep-ph]].
  %%CITATION = doi:10.1103/PhysRevD.93.115030;%%
  %128 citations counted in INSPIRE as of 17 Jul 2018  
  
  
\bibitem{LQ-DM2}    
%\cite{Queiroz:2014pra}
%\bibitem{Queiroz:2014pra}
  F.~S.~Queiroz, K.~Sinha and A.~Strumia,
  %``Leptoquarks, Dark Matter, and Anomalous LHC Events,''
  Phys.\ Rev.\ D {\bf 91} (2015) no.3,  035006
  doi:10.1103/PhysRevD.91.035006
  [arXiv:1409.6301 [hep-ph]];
  %%CITATION = doi:10.1103/PhysRevD.91.035006;%%
  %69 citations counted in INSPIRE as of 19 Jul 2018  
%\cite{Allanach:2015ria}
%\bibitem{Allanach:2015ria}
  B.~Allanach, A.~Alves, F.~S.~Queiroz, K.~Sinha and A.~Strumia,
  %``Interpreting the CMS $\ell^+\ell^- jj E\!\!\!\!/_{\rm T}$ Excess with a Leptoquark Model,''
  Phys.\ Rev.\ D {\bf 92} (2015) no.5,  055023
  doi:10.1103/PhysRevD.92.055023
  [arXiv:1501.03494 [hep-ph]].
  %%CITATION = doi:10.1103/PhysRevD.92.055023;%%
  %40 citations counted in INSPIRE as of 19 Jul 2018  

  
  
  
\bibitem{TBP}
S.~M.~Choi, Y.~J.~Kang, H.~M.~Lee and T.~G.~Ro, To appear.   
  
  
\bibitem{cline}
%\cite{Cline:2012nw}
%\bibitem{Cline:2012nw}
  J.~M.~Cline,
  %``130 GeV dark matter and the Fermi gamma-ray line,''
  Phys.\ Rev.\ D {\bf 86} (2012) 015016
  doi:10.1103/PhysRevD.86.015016
  [arXiv:1205.2688 [hep-ph]].
  %%CITATION = doi:10.1103/PhysRevD.86.015016;%%
  %81 citations counted in INSPIRE as of 15 Jul 2018
  
  
\bibitem{loops}  
%\cite{Lee:2012bq}
%\bibitem{Lee:2012bq}
  H.~M.~Lee, M.~Park and W.~I.~Park,
  %``Fermi Gamma Ray Line at 130 GeV from Axion-Mediated Dark Matter,''
  Phys.\ Rev.\ D {\bf 86} (2012) 103502
  [arXiv:1205.4675 [hep-ph]];
  %%CITATION = ARXIV:1205.4675;%%
  %63 citations counted in INSPIRE as of 18 Aug 2015
%\cite{Lee:2012wz}
%\bibitem{Lee:2012wz}
  H.~M.~Lee, M.~Park and W.~I.~Park,
  %``Axion-mediated dark matter and Higgs diphoton signal,''
  JHEP {\bf 1212} (2012) 037
  [arXiv:1209.1955 [hep-ph]];
  %%CITATION = ARXIV:1209.1955;%%
  %42 citations counted in INSPIRE as of 18 Aug 2015
%\cite{Lee:2012ph}
%\bibitem{Lee:2012ph}
  H.~M.~Lee, M.~Park and V.~Sanz,
  %``Interplay between Fermi gamma-ray lines and collider searches,''
  JHEP {\bf 1303} (2013) 052
  [arXiv:1212.5647 [hep-ph]];
  %%CITATION = ARXIV:1212.5647;%%
  %22 citations counted in INSPIRE as of 18 Aug 2015  
%\cite{Choi:2016cic}
%\bibitem{Choi:2016cic}
  S.~M.~Choi, Y.~J.~Kang and H.~M.~Lee,
  %``Diphoton resonance confronts dark matter,''
  JHEP {\bf 1607} (2016) 030
  doi:10.1007/JHEP07(2016)030
  [arXiv:1605.04804 [hep-ph]].
  %%CITATION = doi:10.1007/JHEP07(2016)030;%%
  %10 citations counted in INSPIRE as of 17 Jun 2018

  
  
  
%Direct detection  


  
  
\bibitem{hisano}
%\cite{Hisano:2010yh}
%\bibitem{Hisano:2010yh}
  J.~Hisano, K.~Ishiwata, N.~Nagata and M.~Yamanaka,
  %``Direct Detection of Vector Dark Matter,''
  Prog.\ Theor.\ Phys.\  {\bf 126} (2011) 435
  doi:10.1143/PTP.126.435
  [arXiv:1012.5455 [hep-ph]].
  %%CITATION = doi:10.1143/PTP.126.435;%%
  %34 citations counted in INSPIRE as of 05 Aug 2017
  
  
% Loop-induced annihilations
  

% Indirect detections

\bibitem{Fermiline}
%\cite{Ackermann:2015lka}
%\bibitem{Ackermann:2015lka}
  M.~Ackermann {\it et al.} [Fermi-LAT Collaboration],
  %``Updated search for spectral lines from Galactic dark matter interactions with pass 8 data from the Fermi Large Area Telescope,''
  Phys.\ Rev.\ D {\bf 91} (2015) no.12,  122002
  doi:10.1103/PhysRevD.91.122002
  [arXiv:1506.00013 [astro-ph.HE]].
  %%CITATION = doi:10.1103/PhysRevD.91.122002;%%
  %173 citations counted in INSPIRE as of 16 Jul 2018



\bibitem{HESSline}
%\cite{Abdallah:2018qtu}
%\bibitem{Abdallah:2018qtu}
  H.~Abdallah {\it et al.} [HESS Collaboration],
  %``Search for $\gamma$-Ray Line Signals from Dark Matter Annihilations in the Inner Galactic Halo from 10 Years of Observations with H.E.S.S.,''
  Phys.\ Rev.\ Lett.\  {\bf 120} (2018) no.20,  201101
  doi:10.1103/PhysRevLett.120.201101
  [arXiv:1805.05741 [astro-ph.HE]];
  %%CITATION = doi:10.1103/PhysRevLett.120.201101;%%
  %1 citations counted in INSPIRE as of 16 Jul 2018
%\cite{Abdalla:2016olq}
%\bibitem{Abdalla:2016olq}
  H.~Abdalla {\it et al.} [H.E.S.S. Collaboration],
  %``H.E.S.S. Limits on Linelike Dark Matter Signatures in the 100 GeV to 2 TeV Energy Range Close to the Galactic Center,''
  Phys.\ Rev.\ Lett.\  {\bf 117} (2016) no.15,  151302
  doi:10.1103/PhysRevLett.117.151302
  [arXiv:1609.08091 [astro-ph.HE]].
  %%CITATION = doi:10.1103/PhysRevLett.117.151302;%%
  %27 citations counted in INSPIRE as of 16 Jul 2018



\bibitem{Fermi}
%\cite{Ackermann:2015zua}
%\bibitem{Ackermann:2015zua}
  M.~Ackermann {\it et al.} [Fermi-LAT Collaboration],
  %``Searching for Dark Matter Annihilation from Milky Way Dwarf Spheroidal Galaxies with Six Years of Fermi Large Area Telescope Data,''
  Phys.\ Rev.\ Lett.\  {\bf 115} (2015) no.23,  231301
  doi:10.1103/PhysRevLett.115.231301
  [arXiv:1503.02641 [astro-ph.HE]].
  %%CITATION = doi:10.1103/PhysRevLett.115.231301;%%
  %610 citations counted in INSPIRE as of 30 Jun 2018


\bibitem{HESS}
%\cite{Abdallah:2016ygi}
%\bibitem{Abdallah:2016ygi}
  H.~Abdallah {\it et al.} [H.E.S.S. Collaboration],
  %``Search for dark matter annihilations towards the inner Galactic halo from 10 years of observations with H.E.S.S,''
  Phys.\ Rev.\ Lett.\  {\bf 117} (2016) no.11,  111301
  doi:10.1103/PhysRevLett.117.111301
  [arXiv:1607.08142 [astro-ph.HE]].
  %%CITATION = doi:10.1103/PhysRevLett.117.111301;%%
  %92 citations counted in INSPIRE as of 30 Jun 2018



\bibitem{AMS}
%\cite{Aguilar:2016kjl}
%\bibitem{Aguilar:2016kjl}
  M.~Aguilar {\it et al.} [AMS Collaboration],
  %``Antiproton Flux, Antiproton-to-Proton Flux Ratio, and Properties of Elementary Particle Fluxes in Primary Cosmic Rays Measured with the Alpha Magnetic Spectrometer on the International Space Station,''
  Phys.\ Rev.\ Lett.\  {\bf 117} (2016) no.9,  091103.
  doi:10.1103/PhysRevLett.117.091103;
  %%CITATION = doi:10.1103/PhysRevLett.117.091103;%%
  %141 citations counted in INSPIRE as of 30 Jun 2018
%\cite{Cuoco:2017iax}
%\bibitem{Cuoco:2017iax}
  A.~Cuoco, J.~Heisig, M.~Korsmeier and M.~Krämer,
  %``Constraining heavy dark matter with cosmic-ray antiprotons,''
  JCAP {\bf 1804} (2018) no.04,  004
  doi:10.1088/1475-7516/2018/04/004
  [arXiv:1711.05274 [hep-ph]].
  %%CITATION = doi:10.1088/1475-7516/2018/04/004;%%
  %14 citations counted in INSPIRE as of 30 Jun 2018


%Higgs data


  \bibitem{Higgsinv}
  %\cite{Aad:2015pla}
%\bibitem{Aad:2015pla}
  G.~Aad {\it et al.} [ATLAS Collaboration],
  %``Constraints on new phenomena via Higgs boson couplings and invisible decays with the ATLAS detector,''
  JHEP {\bf 1511} (2015) 206
  doi:10.1007/JHEP11(2015)206
  [arXiv:1509.00672 [hep-ex]];
  %%CITATION = doi:10.1007/JHEP11(2015)206;%%
  %210 citations counted in INSPIRE as of 29 Jun 2018
  %\cite{Khachatryan:2016whc}
%\bibitem{Khachatryan:2016whc}
  V.~Khachatryan {\it et al.} [CMS Collaboration],
  %``Searches for invisible decays of the Higgs boson in pp collisions at $\sqrt{s}$ = 7, 8, and 13 TeV,''
  JHEP {\bf 1702} (2017) 135
  doi:10.1007/JHEP02(2017)135
  [arXiv:1610.09218 [hep-ex]].
  %%CITATION = doi:10.1007/JHEP02(2017)135;%%
  %94 citations counted in INSPIRE as of 29 Jun 2018
  
 
\bibitem{Combi} 
 %\cite{Khachatryan:2016vau}
%\bibitem{Khachatryan:2016vau}
  G.~Aad {\it et al.} [ATLAS and CMS Collaborations],
  %``Measurements of the Higgs boson production and decay rates and constraints on its couplings from a combined ATLAS and CMS analysis of the LHC pp collision data at $ \sqrt{s}=7 $ and 8 TeV,''
  JHEP {\bf 1608} (2016) 045
  doi:10.1007/JHEP08(2016)045
  [arXiv:1606.02266 [hep-ex]].
  %%CITATION = doi:10.1007/JHEP08(2016)045;%%
  %683 citations counted in INSPIRE as of 16 Jul 2018
 
 

\bibitem{ATLAS}
%\cite{Aaboud:2018xdt}
%\bibitem{Aaboud:2018xdt}
  M.~Aaboud {\it et al.} [ATLAS Collaboration],
  %``Measurements of Higgs boson properties in the diphoton decay channel with 36 fb$^{-1}$ of $pp$ collision data at $\sqrt{s} = 13$ TeV with the ATLAS detector,''
  arXiv:1802.04146 [hep-ex].
  %%CITATION = ARXIV:1802.04146;%%
  %28 citations counted in INSPIRE as of 16 Jul 2018

  
  
  \bibitem{CMS}
  %\cite{Sirunyan:2018ouh}
%\bibitem{Sirunyan:2018ouh}
  A.~M.~Sirunyan {\it et al.} [CMS Collaboration],
  %``Measurements of Higgs boson properties in the diphoton decay channel in proton-proton collisions at $\sqrt{s} =$ 13 TeV,''
  arXiv:1804.02716 [hep-ex].
  %%CITATION = ARXIV:1804.02716;%%
  %10 citations counted in INSPIRE as of 30 Jun 2018
  
  
  
\bibitem{RPV}
%\cite{Chun:2014jha}
%\bibitem{Chun:2014jha}
  E.~J.~Chun, S.~Jung, H.~M.~Lee and S.~C.~Park,
  %``Stop and Sbottom LSP with R-parity Violation,''
  Phys.\ Rev.\ D {\bf 90} (2014) 115023
  doi:10.1103/PhysRevD.90.115023
  [arXiv:1408.4508 [hep-ph]].
  %%CITATION = doi:10.1103/PhysRevD.90.115023;%%
  %14 citations counted in INSPIRE as of 01 Feb 2018
  
  
  \bibitem{RK-early}
%\cite{Hiller:2014yaa}
%\bibitem{Hiller:2014yaa}
  G.~Hiller and M.~Schmaltz,
  %``$R_K$ and future $b \to s \ell \ell$ physics beyond the standard model opportunities,''
  Phys.\ Rev.\ D {\bf 90} (2014) 054014
  doi:10.1103/PhysRevD.90.054014
  [arXiv:1408.1627 [hep-ph]];
  %%CITATION = doi:10.1103/PhysRevD.90.054014;%%
  %199 citations counted in INSPIRE as of 01 Feb 2018
  
  
  \bibitem{SLee}
 %\bibitem{Biswas:2014gga}
  S.~Biswas, D.~Chowdhury, S.~Han and S.~J.~Lee,
  %``Explaining the lepton non-universality at the LHCb and CMS within a unified framework,''
  JHEP {\bf 1502} (2015) 142
  doi:10.1007/JHEP02(2015)142
  [arXiv:1409.0882 [hep-ph]];
  %%CITATION = doi:10.1007/JHEP02(2015)142;%%
  %57 citations counted in INSPIRE as of 13 Jul 2017

  
  
\bibitem{RPV2}
  %\cite{Deshpand:2016cpw}
%\bibitem{Deshpand:2016cpw}
  N.~G.~Deshpande and X.~G.~He,
  %``Consequences of R-parity violating interactions for anomalies in $\bar B\to D^{(*)} \tau \bar \nu$ and $b\to s \mu^+\mu^-$,''
  Eur.\ Phys.\ J.\ C {\bf 77} (2017) no.2,  134
  doi:10.1140/epjc/s10052-017-4707-y
  [arXiv:1608.04817 [hep-ph]];
  %%CITATION = doi:10.1140/epjc/s10052-017-4707-y;%%
  %33 citations counted in INSPIRE as of 15 Jul 2018
%\cite{Das:2017kfo}
%\bibitem{Das:2017kfo}
  D.~Das, C.~Hati, G.~Kumar and N.~Mahajan,
  %``Scrutinizing $R$-parity violating interactions in light of $R_{K^{(\ast)}}$ data,''
  arXiv:1705.09188 [hep-ph];
  %%CITATION = ARXIV:1705.09188;%%
  %5 citations counted in INSPIRE as of 13 Jul 2017 



  
  

\end{thebibliography}
\end{document}